\documentclass[11pt]{article}

\usepackage[normalem]{ulem} 

\usepackage[utf8]{inputenc}
\usepackage[T1]{fontenc}

\newif\ifcameraready
\camerareadyfalse

\usepackage{amsfonts,amsthm,amsmath,amssymb}

\usepackage{libertine}
\usepackage[libertine]{newtxmath}
\usepackage[T1]{fontenc}

\usepackage{anyfontsize}

\usepackage[letterpaper, margin=1in]{geometry} 
\usepackage{setspace}

\usepackage[hyphens,spaces,obeyspaces]{url}
\usepackage[hidelinks]{hyperref}
\hypersetup{
    colorlinks=true,
    linkcolor=NavyBlue,
    filecolor=NavyBlue,
    citecolor=NavyBlue,
    urlcolor=NavyBlue,
}

\usepackage{preamble}       

\usepackage[style=ieee-alphabetic,
    doi=false,
    url=false]{biblatex}
\DeclareFieldFormat[article,inproceedings,incollection]{title}{%
  \mkbibquote{%
    \iffieldundef{doi}
      {%
        \iffieldundef{url}
          {#1}
          {\href{\thefield{url}}{#1}}%
      }
      {\href{https://doi.org/\thefield{doi}}{#1}}%
  }%
}

\title{The Proof Analysis Problem for Constant-Depth Frege\footnote{An extended abstract of this paper will appear at the 67th IEEE Symposium on Foundations of Computer Science (FOCS 2026).}}

\author{    Noel Arteche\footnote{Lund University and University of Copenhagen, \email{noel.arteche@cs.lth.se}} 
        \and Susanna F. de Rezende\footnote{Lund University, \email{susanna.rezende@cs.lth.se}}
        \and Erfan Khaniki\footnote{University of Oxford, \email{erfan.khaniki@cs.ox.ac.uk}}
        \and Kilian Risse\footnote{ETH Zürich, \email{krisse@ethz.ch}}
    }

\date{}

\begin{document}
\maketitle

\thispagestyle{empty}

\begin{abstract}
We give a parallel algorithm for the \emph{Proof Analysis Problem} of Arteche, Atserias, de Rezende, and Khaniki (FOCS~2025). The problem asks, given a CNF formula~$\varphi$ on $n$ variables and a proof $\pi$ of a size-$s$ Resolution lower bound on~$\varphi$, encoded as $\neg \Reff_s(\varphi)$, to extract a satisfying assignment for~$\varphi$ when one exists. The original work gave a polynomial-time extraction algorithm for Resolution refutations that enabled a formalization of the Atserias--Müller lower bound (FOCS~2019) in Extended Frege, but it appeared inherently sequential. We devise a new algorithm that performs assignment extraction in low-depth circuit classes. 

Specifically, we give an assignment extraction algorithm computable by polynomial-size $\TC^0$ circuits and by quasipolynomial-size $\ACZ$ circuits, and we prove the correctness of the latter within quasipolynomial-size $\ACZ$-Frege. This is achieved thanks to a new formalization of the Atserias--Müller lower bound in the relativized theory $\BussTfourtwo(\alpha)$ of bounded arithmetic, which corresponds to \hbox{depth-$4$} Frege systems. One of the main technical ingredients in this formalization is a derandomization of the random restriction argument of Atserias and Müller using limited independence and tools from unconditional derandomization, combined with \citeauthor{jerabek:phd-thesis}'s approximate counting framework.

This yields quasipolynomial-size $\ACZ$-Frege proofs for a broad class of Resolution lower bounds, making progress on a question of Pudlák. A key feature of our approach is an improved parameter regime: unlike prior $\ACZ$-Frege formalizations of Resolution lower bounds by Li, Li, and Ren (STOC~2026), which only proved $\neg \Reff_s(\varphi)$ formulas for $s = 2^{\Omega(n)}$, our approach applies to all $s$, including $s = n^{O(1)}$.

With these improved parameters we show that the Proof Analysis Problem for $\ACZ$-Frege is $\NP$-hard under quasipolynomial-time reductions, which rules out certain Atserias--Müller-type approaches to the hardness of automating constant-depth Frege. Furthermore, automating $\ACZ$-Frege on Resolution lower bound formulas suffices to automate it on all formulas.

\end{abstract}

\newpage
\thispagestyle{empty}
\tableofcontents

\newpage
\setcounter{page}{1}

\newpage
\section{Introduction}
\label{sec:introduction}


The computational complexity of automated theorem proving has been a central concern of theoretical computer science since the birth of the field and the seminal works on $\NP$-completeness \autocite{Cook71, Cook75}. In the context of proof complexity, this question is captured by the notion of \emph{automatability}, introduced by Bonet, Pitassi, and Raz \autocite{BPR00}: a proof system $S$ for propositional logic is automatable if there exists an algorithm that finds a proof in time $\poly(s)$ whenever a proof of size $s$ exists. Despite early hardness results based on cryptographic assumptions \autocite{KP98, BPR00, BDGMP04}, establishing the non-automatability of standard systems remained elusive for nearly two decades after the work of Bonet, Pitassi, and Raz.

In 2019, Atserias and Müller \autocite{AM20} proved that Resolution---the most extensively studied propositional proof system---cannot be automated unless $\P = \NP$, an optimal hardness result. At the heart of their argument lies a clever meta-mathematical reduction that forces Resolution to reason about itself. Their reduction maps a CNF formula $\varphi$ to a new formula $\Reff_s(\varphi)$, which asserts that $\varphi$ admits a Resolution refutation of length $s$. When negated, this is the natural propositional encoding of a Resolution size lower bound. If $\varphi$ is satisfiable, then $\Reff_s(\varphi)$ is unsatisfiable, as it encodes a \emph{trivial} lower bound. For such formulas, Pudlák \autocite{Pudlak03} had already shown how to efficiently construct a Resolution refutation from an explicit satisfying assignment for $\varphi$. In contrast, Atserias and Müller proved that if $\varphi$ is unsatisfiable, then $\Reff_s(\varphi)$ requires exponential size to be refuted in Resolution. This creates an exponential gap between the satisfiable and unsatisfiable cases, yielding the $\NP$-hardness reduction.

Essential to this and similar arguments for other weak proof systems~\autocite{B20, G20, GKSMP20, dRGNPRS21, IR22, P23} is whether propositional proof systems can prove their own non-trivial size lower bounds. Pudlák soon identified this question as \say{more fundamental than automatability} \autocite{Pudlak20}, and called for a systematic study of when concrete systems can establish proof complexity lower bounds. Recently, Arteche, Atserias, de Rezende, and Khaniki \autocite{PAP} approached these questions from a computational perspective by introducing the \emph{Proof Analysis Problem}. For a fixed proof system $S$, the Proof Analysis Problem for $S$ (denoted $\PAP_S$) asks, given a CNF formula $\varphi$ and an $S$-proof $\pi$ of the Resolution lower bound statement $\neg \Reff_s(\varphi)$, to decide whether $\varphi$ is satisfiable---and hence whether $\pi$ is proving a trivial lower bound. If the problem $\PAP_S$ is decidable in polynomial time, we say that the proof system $S$ is \emph{analyzable}. Due to the lower bound on $\Reff$ formulas, Resolution is readily seen to be analyzable: if the refutation of $\Reff_s(\varphi)$ is short and correct, then it must be for a satisfiable $\varphi$. It is not hard to see that for any system $S$ at least as strong as Resolution, if $S$ is analyzable, then the system cannot be automated unless $\P = \NP$, thereby generalizing the Atserias--Müller reduction.

\paragraph{The Proof Analysis Problem for constant-depth Frege: easy or hard?}
As discussed above, if a proof system is
analyzable, then automating it is $\NP$-hard. Conversely, if the
system is \emph{not} analyzable, an argument in the style of Atserias
and Müller 
does not suffice
to prove the
hardness of automating it. 
Previous work~\autocite{PAP}  
showed that the 
Proof
Analysis
Problem for Extended Frege ($\PAP_\EF$) is $\NP$-hard, implying that
new techniques are needed to attack the non-automatability of strong
proof systems. However, the status of analyzability remained open for
proof systems in between Resolution and Extended Frege.

Of central interest are constant-depth Frege systems, where only partial hardness results are known for automatability \autocite{BDGMP04, ACG24, P23} and $\NP$-hardness remains unknown over CNF formulas. The Atserias--Müller techniques appear insufficient to settle this, so proving that the analysis problem for these systems is hard would suggest that new ideas are needed. To prove that $\PAP_S$ is $\NP$-hard it is enough for $S$ to derive non-trivial Resolution lower bounds. For constant-depth Frege, the question of whether Resolution lower bounds are provable in polynomial size remains an open problem \autocite[Problem 3]{Pudlak20}. A first step in this direction was recently obtained by Li, Li, and Ren \autocite{LLR24} who established some formalizations in quasipolynomial-size constant-depth Frege. 

\myparagraph{The assignment extraction problem}
While the Proof Analysis Problem asks to \emph{decide} whether the
formula $\varphi$ in the input is satisfiable, the natural
\emph{search problem} associated with it asks for a satisfying
assignment to be extracted from the proof of $\neg
\Reff_s(\varphi)$. In \autocite{PAP}, it was shown that a polynomial-time
extraction algorithm exists when the proofs being analyzed are
themselves Resolution refutations. The design of this algorithm is intimately connected 
to the complexity of the Atserias--Müller lower bound. Ultimately, the
extraction algorithm amounts to a new constructive proof of the lower
bound on $\Reff$ formulas: given a small Resolution refutation of
$\Reff_s(\varphi)$, the algorithm always succeeds in finding a
satisfying assignment for $\varphi$, implying that if $\varphi$ is
unsatisfiable, then no small Resolution refutation of
$\Reff_s(\varphi)$ exists.

Algorithms for the assignment extraction problem and constructive proofs of the lower bound for $\Reff$ formulas \autocite{AM20, Garlik19} are thus two sides of the same coin. 
\citeauthor{PAP}~\autocite{PAP}
further showed that this constructive proof can be formalized in Cook's theory $\PV_1$ of bounded arithmetic, which captures polynomial-time reasoning. Applying Buss's witnessing theorem, one immediately recovers the extraction algorithm from the formalized $\PVO$ proof. Moreover, by Cook's propositional translation, it follows that Extended Frege proves the Atserias--Müller lower bound in its propositional form, yielding a series of interesting corollaries for proof complexity.

One of the main questions left open was then to determine whether the lower bound could be formalized in weaker theories and proof systems---and, 
consequently,
whether more efficient extraction algorithms are possible. The known algorithm appeared inherently sequential in multiple aspects and difficult to parallelize or implement in sublinear space, suggesting that the extraction problem could even be complete for $\FP$. Whether complete for $\FP$ or efficiently parallelizable, either outcome would be significant. On the one hand, if the problem turned out to be $\FP$-complete, that would imply that no theory below $\PVO$ would prove the Atserias--Müller lower bound: if, for instance, the theory $\VZ$ for $\ACZ$ reasoning proved such a statement, witnessing theorems would imply the false inclusion $\P \subseteq \ACZ$. On the other hand, the alternative outcome of a new, more efficient algorithm in, say, $\NC$, $\ComplexityFont{L}$ or $\TC$, would suggest that a more efficient proof of the lower bound might be possible.

\myparagraph{This paper}
In this work, we show that the latter scenario turns out to be the
right one: we devise a new algorithm capable of performing assignment
extraction by polynomial-size $\TC^0$ circuits and
quasipolynomial-size $\ACZ$ circuits. Inspired by this new algorithm,
we obtain a new proof of the Atserias--Müller lower bound in
quasipolynomial-size constant-depth Frege systems.  This, in turn,
gives constant-depth Frege proofs of a large class of Resolution lower
bounds in all parameter regimes. Equipped with these, we can
extend the $\NP$-hardness reduction for $\PAP_{\EF}$ to $\ACZFnodepth$, under quasipolynomial-time reductions---concluding that the hardness of automating such systems requires new ideas, or at least stronger $\Reff$ formulas.

\subsection{Contributions}
\label{subsec:intro-contributions}

We continue the systematic study of the Proof Analysis Problem initiated by 
\citeauthor{PAP}~\autocite{PAP}
and address their two main open problems:

\begin{enumerate}
	\item Is the assignment extraction problem for Resolution refutations computable in some subclass of $\FP$? And, if so, can this be formalized in a weaker theory of bounded arithmetic than $\PVO$?
	\item Are constant-depth Frege systems analyzable? That is, is $\PAP_S$ in $\P$ for $S = \ACZFnodepth$?
\end{enumerate}

\subsubsection{A new extraction algorithm for Resolution}
\label{subsubsec:intro-contributions-new-algo}
Our first contribution is a new parallel algorithm for the assignment extraction problem for Resolution, which can be implemented in low-depth circuit classes. Recall that $\TC^0$ is the class of problems computable by polynomial-size constant-depth Boolean circuits with unbounded fan-in and threshold gates.

\begin{theorem}[Assignment extraction in $\TC^0$, informal]
    \label{thm:extraction-TC0-informal}
    The assignment extraction problem for Resolution is computable by uniform multi-output $\TC^0$ circuits of polynomial size. 
\end{theorem}

This algorithm inspects each clause of the Resolution refutation in parallel in search for a satisfying assignment, and we show that there always exists a clause in the refutation that \emph{hides} the assignment. While extracting the assignment is highly non-trivial, the actual implementation of the algorithm is fully parallelizable. More precisely, we show that there exists a polynomial-size $\TC^0$ circuit $E$ such that for every Resolution refutation $\pi$ of $\Reff_s(\varphi)$ for satisfiable $\varphi$, there exists at least one clause $C \in \pi$ such that $E(C)$ outputs a satisfying assignment. The final procedure then evaluates $E(C)$ in parallel for every $C \in \pi$.

The algorithm appears somewhat surprising for two reasons. First, the original polynomial-time algorithm consisted of two subroutines, both of which appear very hard to parallelize.
Second, and more surprisingly, 
the natural proof of correctness of this new algorithm invokes the correctness of the original algorithm. We discuss this at length in the technical overview below (see \Cref{subsubsec:intro-techniques-new-algo}).

Furthermore, we show that a quasipolynomial increase in size suffices to remove the threshold gates.

\begin{theorem}[Assignment extraction in quasipolynomial $\ACZ$, informal]
    \label{thm:extraction-AC0-informal}
    The assignment extraction problem for Resolution is computable by uniform multi-output $\ACZ$ circuits of quasipolynomial size.
\end{theorem}

The formal statements can be found in \Cref{subsec:new-algo}.

\subsubsection{Formalization of the Atserias--Müller lower bound in $\BussTfourtwo(\alpha)$ and constant-depth Frege}
\label{subsubsec:intro-contributions-new-AM-proof}

As discussed earlier,
proofs of correctness of the
extraction algorithms and proofs of the Atserias--Müller lower bound
are two sides of the same coin. Inspired by the new algorithms above,
the next question is whether they can be proven correct in systems weaker than Extended Frege. For the sake of completeness, let us state the lower bound and the previously known formalization in $\PVO$, which proves a slightly weaker version of the statement.

\begin{theorem}[The Atserias--Müller lower bound in $\PVO$ \autocite{AM20, PAP}]
  \label{thm:AM-intro}
    For every $n$-variate unsatisfiable
    CNF formula~$\varphi$ and every~$s \in \bbN$, it holds that any
    Resolution refutation of the formula~$\Reff_s(\varphi)$ is of
    size $2^{\Omega(s/n)}$. Furthermore, the theory $\PVO$ proves that
    any such refutation is of size~$2^{\Omega(s/n^2)}$.
  %
\end{theorem}

The target theory for a new formalization of the proof using the $\TC^0$ algorithm (\Cref{thm:extraction-TC0-informal}) is the two-sorted theory $\VTC^0$ of Cook and Nguyen \autocite{CN10}. However, the natural proof of correctness invokes the correctness of the original algorithm, only available in $\PVO$. We find a way around this via an alternative proof, which reasons about objects of quasipolynomial size. Due to these size bounds, the proof cannot go through in $\VTC^0$, but can be carried out in the two-sorted theory $\VZT$ corresponding to quasipolynomial $\ACZ$ reasoning. Effectively, this formalizes the correctness of the $\ACZ$ algorithm (\Cref{thm:extraction-AC0-informal}).

While $\VZT$ would be enough for quasipolynomial $\ACZ$ reasoning, we choose to work instead over specific levels $\BussTitwo(\alpha)$ of the relativized hierarchy of Buss \autocite{Buss86}. This approach still translates into proofs in constant-depth Frege systems via the Paris--Wilkie translation, and it has the added benefit that we can easily control the specific depth we are translating into. Since we work in very low theories of the hierarchy, this appears more informative and worth making explicit.

In this setting, the Resolution refutations are encoded by an external
relation symbol, denoted $\alpha$ in this introduction. Since the new algorithm analyzes each clause of
the refutation in parallel to extract an assignment from one of them,
we can state the result as follows.
\ifcameraready
\else
(See \Cref{thm:poly-regime-ref-lwb-all-CNF} for more details about the encoding of the statement.)
\fi
Note that this new formalization recovers the better $2^{\Omega({s/n})}$ bound of \autocite{AM20}.

\begin{theorem}[The Atserias--Müller lower bound in $\BussTfourtwo(\alpha)$, informal]
    \label{thm:AM-new-proof-informal}    
    The theory $\BussTfourtwo(\alpha)$ proves
    that for every $n$-variate
    CNF formula $\varphi$, every $s \geq n^2$,
    and every Resolution refutation~$\pi$ of $\Reff_s(\varphi)$, if none of the clauses of $\pi$ encodes a satisfying assignment for $\varphi$, then $|\pi| > 2^{\Omega(s/n)}$.
\end{theorem}

Formalizing arguments in bounded arithmetic is particularly exciting
when the more constructive reasoning leads to alternative proofs of known
statements.
This has happened, for example, with Razborov's now\nobreakdash-famous
\say{encoding argument} for Håstad's switching lemma
\autocite{Razborov95, Beame94}. Remarkably, 
Razborov's new proof of the switching lemma
  is recognized for enabling proofs of at least
two major conjectures in combinatorics \autocite{AlweissLovettWuZhang2021,
  ParkPham2024}.  
  Another example where formalization in bounded arithmetic led to alternative proofs
  is a recent hybrid argument proving the
Schwartz--Zippel lemma in $\SOT$ \autocite{AT24}. 

Our formalization also relies on an alternative proof of the lower bound. At the core of our proof is a new derandomization of
the random restriction argument used in the lower bound.
While the earlier work~\autocite{PAP} derandomized this by a greedy algorithm, we instead rely on
$k$-wise independent distributions within the proof system itself. The
technical overview in \Cref{subsubsec:intro-techniques-new-proof-AM} delves further into this.

Overall, this new formalization shows that the reasoning power needed to establish one of the most celebrated recent lower bound results of proof complexity can be made extremely efficient and constructive,
a core concern for ongoing work on the reverse mathematics of computational complexity (see \Cref{subsec:intro-related-work} for a discussion about related work in this direction).  

\myparagraph{Upper bounds in the propositional setting}
Now, via the Paris--Wilkie translation, we can get the following
propositional upper bounds. For the sake of precision, all of our statements about propositional systems use the Tait-style cedent calculus $\PK$
\autocite{Krajicek94} (see \Cref{subsec:prelim-Frege} for the
definitions). In this setting, the depth-$0$ subsystem $\PK_0$ is
precisely Resolution, and we consider \say{half-depths}, like
$\PK_{d+\nicefrac{1}{2}}$, which refers to the subsystem of $\PK_{d+1}$ where all
formulas have depth $d+1$ and the connectives have unbounded fan-in
but the bottom-most connectives have logarithmic fan-in in the size of the proof. In this
notation, $\Res(\log)$ corresponds to $\PK_{0+\nicefrac{1}{2}}$. These
\say{half-depths} arise naturally when translating from bounded
arithmetic. As a general rule, $\mathsf{T}_2^{d+2}(\alpha)$ produces
quasipolynomial-size proofs in $\PK_{d+\nicefrac{1}{2}}$---although the final
depth also depends on the complexity of the statement being proven,
which does affect our translations
\ifcameraready
(see
\Cref{thm:Paris-Wilkie-translation} for the details and the full version for the complete formal statement).
\else
(see
\Cref{thm:Paris-Wilkie-translation} for the details and \Cref{thm:fragment-characterization} for the complete formal statement).
\fi

\begin{corollary}[The Atserias--Müller lower bound in con\-stant-depth
  Frege, informal]
    \label{thm:AM-propositional-informal}
    For every $d\in\bbN$ and every CNF formula $\varphi$, if $\PK_{(d+3)+\nicefrac{1}{2}}$ refutes $\varphi$ in size $\ell$,
    then $\PK_{{(d+3)+\nicefrac{1}{2}}}$ refutes $\Reff(\Reff(\varphi))$ in
    size $\quasipoly(\ell)$.
\end{corollary}

\subsubsection{The Proof Analysis Problem for bounded-depth Frege is
  $\NP$-hard}
\label{subsubsec:intro-contributions-PAP-NP-hard}

We recover the
$\NP$-hardness reduction used to show that the Proof Analysis Problem
for Extended Frege is $\NP$-hard, and extend it all the way to
constant-depth Frege, albeit under quasipolynomial-time reductions. The formal statement is in \Cref{thm:NP-hardness-PAP-PK}. This could be done using \Cref{thm:AM-propositional-informal}, but we get an even better bound on depth via an alternative route based on \cite{LLR24}. \Cref{sec:PAP-NP-hard} outlines the different approaches.

\begin{theorem}[$\NP$-hardness of $\PAP$, informal]
    \label{thm:NP-hardness-informal}
  For every
  propositional proof system $S$ 
  that p-simulates
  depth-2 Frege, 
  it holds that
  the Proof Analysis Problem for $S$ is
  $\NP$-hard under quasipolynomial-time reductions.
\end{theorem}

This immediately implies that the basic Atserias--Müller approach of using the $\Reff$ formula for Resolution for hardness of automatability does not work for constant-depth systems. This is because any proof along those lines using the $\Reff$ formulas works by showing that the system in question is analyzable, and then argues that if a system is both analyzable and automatable, then $\P = \NP$. Of course, this result does not rule out the possibility of using $\Reff^Q$ formulas where $Q$ is a system stronger than Resolution. To make the existing reduction in \Cref{thm:NP-hardness-informal} polynomial-time computable, it would suffice to solve Pudlák's question~\autocite[Problem
3]{Pudlak20} of formalizing Resolution lower bounds in \emph{polynomial-size} $\ACZ$-Frege.

\subsubsection{The $\Reff$ formulas are complete for automatability}
\label{subsubsec:intro-contributions-Reff-complete}

Our last result, also a byproduct of the propositional translations,
is that efficient proof search over Resolution lower bounds suffices
to automate proof systems over all formulas. 
\citeauthor{PAP}~\autocite{PAP}
used their formalizations
in $\PVO$ to show this result for any system that p-simulates Extended Frege. In this work, we extend this to constant-depth systems
with quasipolynomial time bounds.
\ifcameraready
The formal statement and proof can be found in the full version.
\else
The formal statement is in \Cref{theorem:eqivalence-automatability}.
\fi
\begin{theorem}[Automatability in terms of $\Reff$ formulas, informal]
  \label{thm:automatability-informal}
  For every $d \geq 3$, the proof system $\PK_{d+\nicefrac{1}{2}}$ is automatable
  on $\Reff$ formulas in quasipolynomial time if, and only if, it is
  automatable on all CNF formulas in quasipolynomial time.
\end{theorem}

The idea in all of these results is to leverage the propositional
translation as follows: if the system $S$ is strong enough to prove
the Atserias--Müller statement, then a size-$\ell$ proof of
$\neg \varphi$ implies a $\quasipoly(\ell)$-size 
proof of
$\neg \Reff(\Reff(\varphi))$. We can now use the automating algorithm on
this $\Reff$ formula to find a proof, and finally, use the
construction of Pudlák \autocite{Pudlak03} to recover a proof of
$\neg \varphi$ from the proof of $\neg \Reff(\Reff(\varphi))$. The
latter step was formalized in Resolution by \citeauthor{PAP}~\autocite[Theorem 7.1]{PAP}, and only increases the size by an additive polynomial.

\ifcameraready
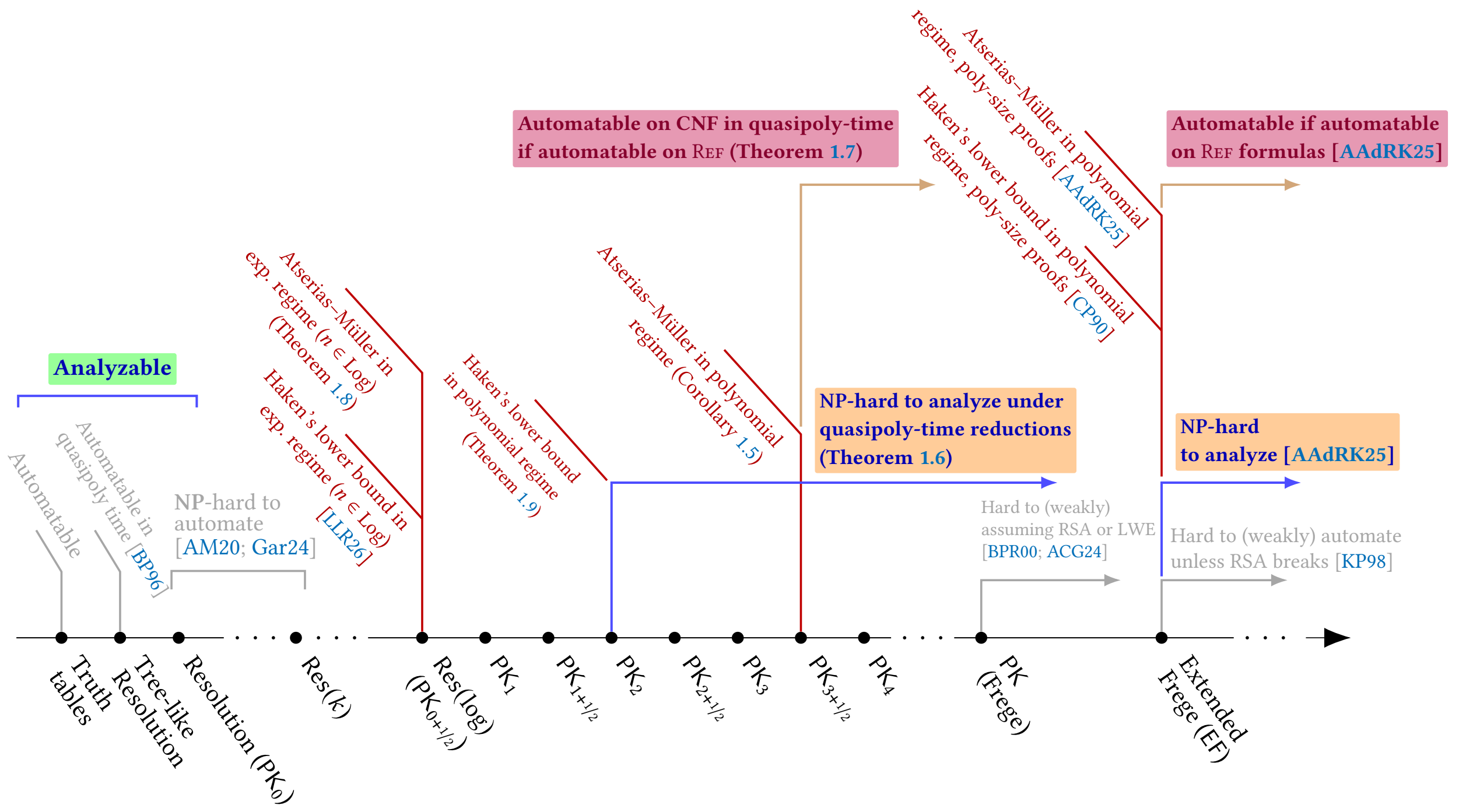
\begin{figure*}[t]
    \centering
    \begin{tikzpicture}[
    x=1.35cm, y=1.15cm,
    >={Latex[length=3mm]},
    axis/.style={line width=0.5pt},
    dot/.style={circle, fill=black, inner sep=1.8pt},
    tinydot/.style={circle, fill=black, inner sep=0.5pt},
    redseg/.style={red!75!black, line width=0.9pt},
    blueann/.style={blue!70, line width=1pt, -{Latex[length=2.5mm]}},
    autoann/.style={brown!70, line width=1pt, -{Latex[length=2.5mm]}},
    grayann/.style={gray!70, line width=0.9pt},
    grayann2/.style={gray!70, line width=0.9pt, -{Latex[length=2.5mm]}},
    lab/.style={font=\small, rotate=-55, anchor=north west,align=left,below=30pt,text height=2.5ex,
    text depth=0.25ex},
]

\tikzset{
  rednote/.style={
    text=red!70!black,
    font=\footnotesize,
    align=right,
    rotate=-48,
    anchor=west
  },
  bluenote/.style={
    text=blue!70!black,
    font=\bfseries\small,
    align=left,
    anchor=west
  },
  autonote/.style={
    text=purple!70!black,
    font=\bfseries\small,
    align=left,
    anchor=west
},
  graynote/.style={
    text=gray!70,
    font=\footnotesize,
    align=left,
    anchor=west
  },
  hilite/.style={
    fill=orange!40,
    inner sep=2pt,
    rounded corners=1pt
  },
  autolite/.style={
    fill=purple!40,
    inner sep=2pt,
    rounded corners=1pt
  },
  greenhilite/.style={
    fill=green!40,
    inner sep=2pt,
    rounded corners=1pt
  }
}
\draw[axis] (0,0) -- (2.3,0);
\draw[axis] (3.9,0) -- (9.7,0);
\draw[axis] (10.4,0) -- (13.5,0);

\foreach \x in 
{2.45, 3.35, 9.85,13.65}
{\foreach \i in {0,1,2}
\node[tinydot] at (\i*0.2+\x,0) {};
}

\draw[-{Latex[length=4mm]}] (14.3,0) -- ++(0.5,0);

\def\incline{-55}

\foreach \x/\shift/\name in 
{0.5/0.15/{Truth\\tables},
1.15/0.1/{Tree-like\\Resolution},
1.8/0/{Resolution ($\PK_0$)},
3.1/0/{$\Res(k)$},
4.5/0.25/{$\Res(\log)$\\($\PK_{0+\nicefrac{1}{2}}$)}/
}{
  \node[dot] at (\x,0) {};
 \rotlabel{\x-0.10-\shift}{-0.08}{\incline}{}{\name}
}

\foreach \x in {1,...,4}{
\foreach \j in {0,1}{
\ifodd\j\def\plushalf{+\nicefrac{1}{2}}
\else \def\plushalf{} \fi
\pgfmathtruncatemacro{\cond}{(\x==4 && \j==1)}
\ifnum\cond=0
  \node[dot] at ({2*\x+\j)*0.7+3.8},0) {};
   \rotlabel{{2*\x+\j)*0.7+3.7}}{-0.08}{\incline}{}{$\PK_{\x\plushalf}$}
\fi
}
}

\foreach \x/\name in 
{
10.7/{$\PK$\\(Frege)},
12.7/{Extended\\Frege ($\EF$)}/
}{
  \node[dot] at (\x,0) {};
  \rotlabel{\x-0.2}{-0.08}{\incline}{}{\name}
}

\draw[blueann, -] (0.02,3) -- (0.02,3.15) -- (2.0,3.15) -- (2.0,3);
\node[bluenote, greenhilite] at (0.35,3.5) {Analyzable};

\draw[grayann] (0.5,0.08) -- (0.5,0.86) -- (0.22,1.4);
\rotlabel{-0.2}{2.6}{-58}{text=gray!70,
    font=\small}{Automatable}

\draw[grayann] (1.15,0.08) -- (1.15,0.86) -- (0.87,1.40);
\rotlabel{0.3}{3}{-58}{text=gray!70,
    font=\footnotesize}{Automatable in\\quasipoly time~\autocite{BP96}}

\draw[grayann, -] (1.72,0.63) -- (1.72,0.87) -- (3.2,0.87) -- (3.2,0.63);

\ifcameraready
\rotlabel{1.65}{1.6}{0}{text=gray!70,
    font=\footnotesize}{$\NP$-hard to\\automate \autocite{AM20, G20}}
\else
\rotlabel{1.65}{2.0}{0}{text=gray!70,
    font=\small}{$\NP$-hard to\\automate\\ \autocite{AM20, G20}}
\fi

\draw[redseg] (4.50,0.08) -- (4.50,3.45) -- (3.65,4.55);
\draw[redseg] (4.5,1.55) -- (3.65,2.65);

\ifcameraready
\node[rednote] at (2.50,5.1) {Atserias--Müller in  \\ exp. regime ($n\in \Log$) \\ (\Cref{thm:AM-LLR-regime-informal})};
\node[rednote] at (2.5,3.4) {Haken's lower bound in\\ exp. regime ($n \in \Log$) \autocite{LLR24}};
\else
\node[rednote] at (2.50,5.1) {Atserias--Müller in  \\ exp. regime ($n\in \Log$)\\(\Cref{thm:AM-LLR-regime-informal})};
\node[rednote] at (2.5,3.3) {Haken's lower bound in\\ exp. regime ($n \in \Log$) \\ \autocite{LLR24}};
\fi

\draw[redseg] (6.55,2.05) -- (5.75,3.1);
\ifcameraready
\node[rednote] at (4.7,3.55) {\footnotesize Haken's lower bound\\ \footnotesize in  polynomial regime \\ (\footnotesize{\Cref{thm:Haken-PHP-polyregime-informal}})};
\else
\node[rednote] at (4.7,3.5) {\scriptsize Haken's lower bound\\ \scriptsize in  polynomial regime \\ (\scriptsize{\Cref{thm:Haken-PHP-polyregime-informal}})};
\fi


\draw[redseg] (8.7,0.08) -- (8.7,2.65) -- (7.85,3.75);
\node[rednote] at (6.3,5.05) {Atserias--Müller in polynomial \\ regime  (\Cref{thm:AM-propositional-informal})};

\draw[autoann] (8.70,2.75) -- (8.70,5.9) -- (10.2,5.9);
\node[autonote, autolite] at (5.5,6.5) {\footnotesize{Automatable on CNF in quasipoly-time} \\ \footnotesize if automatable on $\Reff$ (\Cref{thm:automatability-informal})};

\draw[blueann] (6.6,0.08) -- (6.6,2.02) -- (11.55,2.02);
\node[bluenote, hilite] at (8.85,2.7) {\footnotesize{$\NP$-hard to analyze under} \\ \footnotesize quasipoly-time reductions \\ 
 \footnotesize (\Cref{thm:NP-hardness-informal})};

\draw[grayann2] (10.70,0.08) -- (10.70,0.75) -- (12.25,0.75);
\ifcameraready
\node[graynote,font=\footnotesize] at (10.5,1.25) 
{Hard to (weakly) \\ automate unless DH \\ and LWE break \autocite{BPR00, ACG24}
};
\else
\node[graynote,font=\scriptsize] at (10.6,1.4) 
{Hard to (weakly) \\ assuming RSA or LWE \\ \autocite{BPR00, ACG24}
};
\fi

\draw[grayann2] (12.7,0.08) -- (12.7,0.75) -- (14.10,0.75);
\node[graynote,font=\footnotesize] at (12.7,1.15) {Hard to (weakly) automate\\ unless RSA breaks \autocite{KP98}};

\draw[blueann] (12.70,0.8) -- (12.70,2.02) -- (14.25,2.02);
\node[bluenote, hilite] at (12.85,2.55) {\footnotesize $\NP$-hard\\ \footnotesize to analyze \autocite{PAP}};

\draw[autoann] (12.70,5.6) -- (12.70,5.9) -- (14.25,5.9);
\node[autonote, autolite] at (12.75,6.5) {\footnotesize Automatable if automatable \\ \footnotesize on $\Reff$ formulas \autocite{PAP}};

\draw[redseg] (12.7,2.10) -- (12.7,5.50) -- (11.85,6.60);
\node[rednote] at (10.0,8.35) {Atserias--Müller in  polynomial \\ regime, poly-size proofs \autocite{PAP}};

\draw[redseg] (12.7,4.0) -- (11.85,5.1);
\node[rednote] at (9.85,7.1) {Haken's lower bound in  polynomial \\ regime, poly-size proofs \autocite{CP90} };

\end{tikzpicture}
    \caption{The landscape of automatability, analyzability and formalized Resolution lower bounds. In \textcolor{red!75!black}{red}, upper bounds for formalized Resolution lower bounds; in \textcolor{gray}{gray}, upper and lower bounds on automatability; in \textcolor{green}{green}, analyzable proof systems; in \textcolor{orange}{orange}, $\NP$-hardness results on analyzability; in \textcolor{purple}{purple}, results bootstrapping automatability on $\Reff$ formulas to automatability on all formulas. All results in \textcolor{red!75!black}{red} for constant-depth systems correspond to quasipolynomial-size proofs.}
    \label{fig:system-metro-line}
\end{figure*}
\else
\begin{figure}
    \centering
    \begin{tikzpicture}[
    x=1.35cm, y=1.15cm,
    >={Latex[length=3mm]},
    axis/.style={line width=0.5pt},
    dot/.style={circle, fill=black, inner sep=1.8pt},
    tinydot/.style={circle, fill=black, inner sep=0.5pt},
    redseg/.style={red!75!black, line width=0.9pt},
    blueann/.style={blue!70, line width=1pt, -{Latex[length=2.5mm]}},
    autoann/.style={brown!70, line width=1pt, -{Latex[length=2.5mm]}},
    grayann/.style={gray!70, line width=0.9pt},
    grayann2/.style={gray!70, line width=0.9pt, -{Latex[length=2.5mm]}},
    lab/.style={font=\small, rotate=-55, anchor=north west,align=left,below=30pt,text height=2.5ex,
    text depth=0.25ex},
]

\tikzset{
  rednote/.style={
    text=red!70!black,
    font=\footnotesize,
    align=right,
    rotate=-48,
    anchor=west
  },
  bluenote/.style={
    text=blue!70!black,
    font=\bfseries\small,
    align=left,
    anchor=west
  },
  autonote/.style={
    text=purple!70!black,
    font=\bfseries\small,
    align=left,
    anchor=west
},
  graynote/.style={
    text=gray!70,
    font=\footnotesize,
    align=left,
    anchor=west
  },
  hilite/.style={
    fill=orange!40,
    inner sep=2pt,
    rounded corners=1pt
  },
  autolite/.style={
    fill=purple!40,
    inner sep=2pt,
    rounded corners=1pt
  },
  greenhilite/.style={
    fill=green!40,
    inner sep=2pt,
    rounded corners=1pt
  }
}
\draw[axis] (0,0) -- (2.3,0);
\draw[axis] (3.9,0) -- (9.7,0);
\draw[axis] (10.4,0) -- (13.5,0);

\foreach \x in 
{2.45, 3.35, 9.85,13.65}
{\foreach \i in {0,1,2}
\node[tinydot] at (\i*0.2+\x,0) {};
}

\draw[-{Latex[length=4mm]}] (14.3,0) -- ++(0.5,0);

\def\incline{-55}

\foreach \x/\shift/\name in 
{0.5/0.15/{Truth\\tables},
1.15/0.1/{Tree-like\\Resolution},
1.8/0/{Resolution ($\PK_0$)},
3.1/0/{$\Res(k)$},
4.5/0.25/{$\Res(\log)$\\($\PK_{0+\nicefrac{1}{2}}$)}/
}{
  \node[dot] at (\x,0) {};
 \rotlabel{\x-0.10-\shift}{-0.08}{\incline}{}{\name}
}

\foreach \x in {1,...,4}{
\foreach \j in {0,1}{
\ifodd\j\def\plushalf{+\nicefrac{1}{2}}
\else \def\plushalf{} \fi
\pgfmathtruncatemacro{\cond}{(\x==4 && \j==1)}
\ifnum\cond=0
  \node[dot] at ({2*\x+\j)*0.7+3.8},0) {};
   \rotlabel{{2*\x+\j)*0.7+3.7}}{-0.08}{\incline}{}{$\PK_{\x\plushalf}$}
\fi
}
}

\foreach \x/\name in 
{
10.7/{$\PK$\\(Frege)},
12.7/{Extended\\Frege ($\EF$)}/
}{
  \node[dot] at (\x,0) {};
  \rotlabel{\x-0.2}{-0.08}{\incline}{}{\name}
}

\draw[blueann, -] (0.02,3) -- (0.02,3.15) -- (2.0,3.15) -- (2.0,3);
\node[bluenote, greenhilite] at (0.35,3.5) {Analyzable};

\draw[grayann] (0.5,0.08) -- (0.5,0.86) -- (0.22,1.4);
\rotlabel{-0.2}{2.6}{-58}{text=gray!70,
    font=\small}{Automatable}

\draw[grayann] (1.15,0.08) -- (1.15,0.86) -- (0.87,1.40);
\rotlabel{0.3}{3}{-58}{text=gray!70,
    font=\footnotesize}{Automatable in\\quasipoly time~\autocite{BP96}}

\draw[grayann, -] (1.72,0.63) -- (1.72,0.87) -- (3.2,0.87) -- (3.2,0.63);

\ifcameraready
\rotlabel{1.65}{1.6}{0}{text=gray!70,
    font=\footnotesize}{$\NP$-hard to\\automate \autocite{AM20, G20}}
\else
\rotlabel{1.65}{2.0}{0}{text=gray!70,
    font=\small}{$\NP$-hard to\\automate\\ \autocite{AM20, G20}}
\fi

\draw[redseg] (4.50,0.08) -- (4.50,3.45) -- (3.65,4.55);
\draw[redseg] (4.5,1.55) -- (3.65,2.65);

\ifcameraready
\node[rednote] at (2.50,5.1) {Atserias--Müller in  \\ exp. regime ($n\in \Log$) \\ (\Cref{thm:AM-LLR-regime-informal})};
\node[rednote] at (2.5,3.4) {Haken's lower bound in\\ exp. regime ($n \in \Log$) \autocite{LLR24}};
\else
\node[rednote] at (2.50,5.1) {Atserias--Müller in  \\ exp. regime ($n\in \Log$)\\(\Cref{thm:AM-LLR-regime-informal})};
\node[rednote] at (2.5,3.3) {Haken's lower bound in\\ exp. regime ($n \in \Log$) \\ \autocite{LLR24}};
\fi

\draw[redseg] (6.55,2.05) -- (5.75,3.1);
\ifcameraready
\node[rednote] at (4.7,3.55) {\footnotesize Haken's lower bound\\ \footnotesize in  polynomial regime \\ (\footnotesize{\Cref{thm:Haken-PHP-polyregime-informal}})};
\else
\node[rednote] at (4.7,3.5) {\scriptsize Haken's lower bound\\ \scriptsize in  polynomial regime \\ (\scriptsize{\Cref{thm:Haken-PHP-polyregime-informal}})};
\fi


\draw[redseg] (8.7,0.08) -- (8.7,2.65) -- (7.85,3.75);
\node[rednote] at (6.3,5.05) {Atserias--Müller in polynomial \\ regime  (\Cref{thm:AM-propositional-informal})};

\draw[autoann] (8.70,2.75) -- (8.70,5.9) -- (10.2,5.9);
\node[autonote, autolite] at (5.5,6.5) {\footnotesize{Automatable on CNF in quasipoly-time} \\ \footnotesize if automatable on $\Reff$ (\Cref{thm:automatability-informal})};

\draw[blueann] (6.6,0.08) -- (6.6,2.02) -- (11.55,2.02);
\node[bluenote, hilite] at (8.85,2.7) {\footnotesize{$\NP$-hard to analyze under} \\ \footnotesize quasipoly-time reductions \\ 
 \footnotesize (\Cref{thm:NP-hardness-informal})};

\draw[grayann2] (10.70,0.08) -- (10.70,0.75) -- (12.25,0.75);
\ifcameraready
\node[graynote,font=\footnotesize] at (10.5,1.25) 
{Hard to (weakly) \\ automate unless DH \\ and LWE break \autocite{BPR00, ACG24}
};
\else
\node[graynote,font=\scriptsize] at (10.6,1.4) 
{Hard to (weakly) \\ assuming RSA or LWE \\ \autocite{BPR00, ACG24}
};
\fi

\draw[grayann2] (12.7,0.08) -- (12.7,0.75) -- (14.10,0.75);
\node[graynote,font=\footnotesize] at (12.7,1.15) {Hard to (weakly) automate\\ unless RSA breaks \autocite{KP98}};

\draw[blueann] (12.70,0.8) -- (12.70,2.02) -- (14.25,2.02);
\node[bluenote, hilite] at (12.85,2.55) {\footnotesize $\NP$-hard\\ \footnotesize to analyze \autocite{PAP}};

\draw[autoann] (12.70,5.6) -- (12.70,5.9) -- (14.25,5.9);
\node[autonote, autolite] at (12.75,6.5) {\footnotesize Automatable if automatable \\ \footnotesize on $\Reff$ formulas \autocite{PAP}};

\draw[redseg] (12.7,2.10) -- (12.7,5.50) -- (11.85,6.60);
\node[rednote] at (10.0,8.35) {Atserias--Müller in  polynomial \\ regime, poly-size proofs \autocite{PAP}};

\draw[redseg] (12.7,4.0) -- (11.85,5.1);
\node[rednote] at (9.85,7.1) {Haken's lower bound in  polynomial \\ regime, poly-size proofs \autocite{CP90} };

\end{tikzpicture}
    \caption{The landscape of automatability, analyzability and formalized Resolution lower bounds. In \textcolor{red!75!black}{red}, upper bounds for formalized Resolution lower bounds; in \textcolor{gray}{gray}, upper and lower bounds on automatability; in \textcolor{green}{green}, analyzable proof systems; in \textcolor{orange}{orange}, $\NP$-hardness results on analyzability; in \textcolor{purple}{purple}, results bootstrapping automatability on $\Reff$ formulas to automatability on all formulas. All results in \textcolor{red!75!black}{red} for constant-depth systems correspond to quasipolynomial-size proofs.}
    \label{fig:system-metro-line}
\end{figure}
\fi

\subsection{Technical intermezzo: the exponential versus polynomial-size lower bound regimes}
\label{subsubsec:intro-intermezzo}

The theory $\BussTfourtwo(\alpha)$ and the proof system $\PK_{3+\nicefrac{1}{2}}$ are slightly stronger than the ones used in previous formalizations of Resolution lower bounds. For example, Li, Li, and Ren \autocite{LLR24} showed, through the language of $\TFNP$, that Haken's~\autocite{Haken85} exponential Resolution lower bound for the pigeonhole principle formulas ($\PHP_n$) is provable in $\LLR \subseteq \BussTtwotwo(\alpha)$, which corresponds to quasipolynomial-size $\Res(\log)$ refutations (i.e., $\PK_{0+\nicefrac{1}{2}}$) of the formulas $\Reff_{2^{\varepsilon n}}(\PHP_n)$ for a suitable constant $\varepsilon$.

Before we can state the rest of our results, a crucial technical discussion is in place. In the metalanguage, the $2^{\Omega(n)}$ lower bound proven by Haken certainly implies a weaker lower bound of, say, $n^k$ for any fixed $k$---but this is not obvious at the propositional level. That is, the formula $\Reff_{n^k}(\PHP_n)$ logically implies $\Reff_{2^{\varepsilon n}}(\PHP_n)$ and the system can perform this derivation, but this incurs an exponential blow-up in size if we measure the proof in terms of the starting formula $\Reff_{n^k}(\PHP_n)$.

This mismatch between the size of the inner formula and the outer lower bound parameter can give unreasonable power to the proof system. In the case of $\PHP_n$, the formula $\Reff_{2^{\varepsilon n}}(\PHP_n)$ has size $2^{O(n)}$, while the inner $\PHP_n$ formula has $O(n^2)$ variables and $O(n^3)$ clauses. In this regime, for instance, the proof system could argue that some formula over $n$ variables is unsatisfiable simply by a brute-force argument---and this would only be a polynomial-size proof in terms of the $\Reff$ formula. Similarly, suppose the lower bound argument relies on analyzing some probabilistic event over the uniform distribution for $n$ bits. When starting from the formula $\Reff_{2^{\varepsilon n}}(\PHP_n)$, the proof system could exactly count the probability of any such event in polynomial-size---while this kind of exact counting would be infeasible if we are proving $\neg \Reff_{n^k}(\PHP_n)$ in polynomial size.

This situation causes two problems. Conceptually, one might cast doubt as to whether a formalization of the lower bound $\neg\Reff_{2^{\varepsilon n}}(\PHP_n)$ as in \cite{LLR24} is satisfactory\footnote{Arguably, the exponential-size regime makes perfect sense from the point of view of \citeauthor{LLR24} \autocite{LLR24}, since in the context of $\TFNP$, analyzing purported Resolution refutations of polynomial size would be a trivial computational task, and formalizing the lower bounds is not their main priority.} in capturing the complexity of the argument if this cannot be easily scaled down to prove $\neg\Reff_{n^k}(\PHP_n)$. Pragmatically, for many applications, like an $\NP$-hardness reduction for automatability, one wants to be able to construct the $\Reff_s(\varphi)$ formula efficiently starting from the inner formula $\varphi$. In this setting, we are interested in fixing~$s$ to be a small polynomial in the size of $\varphi$. 

This mismatch between parameters also arises when working within one-sorted theories of bounded arithmetic to then obtain propositional upper bounds via translations. We distinguish two regimes.

\begin{enumerate}[label=(\alph*)]
    \item In the \emph{polynomial-size regime}, we state the lower bound by saying: \say{for every $n$ and every refutation of $\PHP_n$ of length $t$, if $t$ is small, then the refutation is incorrect}; here, \say{small} can be a fixed polynomial, like $n^k$, or an exponential $2^{\varepsilon n}$, stated as $\log  t >  \varepsilon n$, since the exponential function $2^n$ is not provably total for every $n$.
    \item In the \emph{exponential-size regime}, we state the lower bound by saying: \say{for every $n \in \Log$ and every refutation of $\PHP_n$ of length $t$, if $t < 2^{\varepsilon n}$, then the refutation is incorrect}.
\end{enumerate}

The term \emph{polynomial-size regime} is perhaps a misnomer, since we can also get exponential lower bounds---it is just that the polynomial-size lower bounds are the more interesting ones from the point of view of formalization. The key difference is that in (b) we assume $n \in \Log$. The expression $n \in \Log$ is a common shorthand in bounded arithmetic for the formula $\exists x \ |x| = n$. This is often put in words by saying that \say{$n$ is a \emph{length}}, since it is the length of some string $x$. We think of numbers in $\Log$ as small, meaning that the number $2^n$ exists in the theory and can be stated explicitly.

The advantage of assuming $n \in \Log$ is clear: $2^n$ exists and we can perform a great deal of exact counting. The disadvantage appears when dealing with propositional translations: if $n \in \Log$, then the propositional translation trivializes for $\neg \Reff_{n^k}(\PHP_n)$, as the formula is artificially padded to have size $2^{n}$.

In the exponential-size regime, our proof of the Atserias--Müller statement can be carried out in the theory $\LLR$ used in \autocite{LLR24}. 
\ifcameraready
The full version of this paper contains the formal statement and proof.
\else
\Cref{thm:exp-regime-ref-lwb} contains the formal statement.
\fi
\begin{theorem}[The Atserias--Müller lower bound in $\LLR$, informal]
    \label{thm:AM-LLR-regime-informal}     
    The theory $\LLR$ proves that for every $n\in\Log$ and every $s \in \Log$, if $\varphi$ is an $n$-variate CNF formula, $s \geq n^2$, and $\pi$ is a Resolution refutation of $\Reff_s(\varphi)$ such that none of the clauses of~$\pi$ encodes a satisfying assignment for $\varphi$, then $|\pi| > 2^{\Omega(s/n)}$.
\end{theorem}



The known $\Res(\log)$ upper bounds for $\Reff(\PHP_n)$ coming from the formalizations of Li, Li, and Ren~\autocite{LLR24} are in the exponential-size regime and cannot be directly scaled down within $\Res(\log)$ to obtain fixed-polynomial lower bounds for $\PHP_n$. 
Interestingly, for the particular case of these formulas, we devise a way to adapt the size parameters at the cost of increasing the depth of the proof system by a constant (recall that $\Res(\log)$ is $\PK_{0+\nicefrac{1}{2}}$).

\begin{theorem}[Haken's lower bound in the polynomial-size regime, informal]
    \label{thm:Haken-PHP-polyregime-informal}
    For every $k \in \bbN$, there are quasipolynomial-size $\PK_{2}$-proofs of the formulas $\neg \Reff_{n^k}(\PHP_n)$.
\end{theorem}

The idea here is to use the $\Res(\log)$ refutations of $\Reff_{2^{\varepsilon n}}(\PHP_n)$ and apply a restriction sending $\PHP_n$ to an exponentially smaller subformula. This incurs an increase in depth, so we no longer stay in $\Res(\log)$. This approach, however, works only for $\PHP$ and similar formulas where we can restrict to an exponentially smaller sub-instance of the same formula. In particular, this approach does not seem to work for the $\Reff$ formulas. The formal statement is given in \Cref{thm:RefPHP-poly-regime}.

\subsection{Technical overview}
\label{subsec:intro-techniques}
In what follows, we present a summary of the main technical challenges and the key ideas behind the proofs.

\subsubsection{The new extraction algorithm}
\label{subsubsec:intro-techniques-new-algo}
We begin by describing the original extraction algorithm of~\autocite{PAP} and show how to overcome the two main obstacles to parallelizing it. The original algorithm 
has two steps, mirroring the two
components of the lower bound for $\Reff$ formulas: a restriction
argument followed by a width lower bound. This follows the standard
template for Resolution size lower bounds.

In the formula $\Reff_s(\varphi)$, variables are partitioned into $s$
\emph{blocks}, each encoding one of the $s$ clauses of the purported
refutation. The \emph{block-width} of a clause $C$ over the variables of $\Reff_s(\varphi)$, denoted $\bw(C)$, is the number of distinct blocks mentioned in~$C$, and for a refutation $\pi$, we define $\bw(\pi)$ as the maximum block-width over all clauses in $\pi$.

Given a Resolution refutation $\pi$ of $\Reff_s(\varphi)$, the algorithm proceeds as follows:

\begin{enumerate}
\item \emph{Deterministic restriction.} The algorithm constructs a
  restriction $\rho$ such that
  $\Reff_s(\varphi)_{\restriction \rho} \equiv \Reff_{s/2}(\varphi)$,
  and $\pi_{\restriction \rho}$ remains a valid refutation. Crucially,
  $\bw(\pi_{\restriction \rho}) \leq O\bigl(\sqrt{s \log |\pi|}\bigr)$.

\item \emph{Prover--Delayer game.} We think of the width lower bound of \autocite{AM20} as a Prover--Delayer game \autocite{Pudlak00, AD08}. The Prover issues queries about specific variables or chooses to forget the value of a previously queried variable, and the Delayer responds with truth-assignments to these queries. The Prover wins when the partial assignment in the Prover's memory falsifies one of the clauses of the formula. The formula requires large width to be refuted if and only if the Delayer has a strategy that forces the Prover to remember many variables before winning.

Now, starting from the root $\bot$, we
  traverse $\pi_{\restriction \rho}$ according to the Delayer’s
  strategy used in the Prover--Delayer game that establishes the
  block-width lower bound of \autocite{AM20}. As shown in
  \autocite{PAP}, this traversal must eventually reach either:
	\begin{enumerate}[label=(\alph*)]
	    \item a clause $C$ with block-width $\Omega(s/n)$; or
	    \item a satisfying assignment for $\varphi$.
	\end{enumerate}
	If $\pi$ is small enough---less than $2^{\varepsilon n}$ in size---, then $\bw(\pi_{\restriction \rho}) \leq O\bigl(\sqrt{s \log |\pi|}\bigr)$ implies we must be in case (b) and the traversal thus yields a satisfying assignment. If $\pi$ is too large, then we can perform a brute-force search looking for a satisfying assignment while still running in polynomial time.
\end{enumerate}

Both of these steps resist parallelization.

\myparagraph{The challenge of parallelizing the deterministic restriction}
While the original proof of Atserias and Müller used a random restriction, \citeauthor{PAP}~\autocite{PAP} derandomize it via a greedy procedure. At a high level, the algorithm repeatedly identifies the most frequent block in the current refutation, selects a literal that appears in many clauses involving that block, and assigns it so as to eliminate those clauses. The refutation is simplified and the process repeats. After all blocks are processed, the resulting refutation has block-width $O\bigl(\sqrt{s \log |\pi|}\bigr)$.

Although polynomial-time, this procedure is inherently sequential: 
the next literal can only be chosen after simplifying the formula under the current partial assignment. This dependence appears to preclude parallelization.

\myparagraph{The challenge of parallelizing the Prover--Delayer game}
The second step is also inherently
sequential. The algorithm
traverses the refutation while the Delayer maintains a
\emph{reservation}, a partial assignment together with some auxiliary
structure encoding which blocks have been reserved and how they are
arranged across $n$ conceptual layers of the purported refutation. This layered arrangement is supposed to locally
represent an embedding into the canonical exponential-size tree-like
refutation of $\varphi$. In this structure, as long as the formula is unsatisfiable, the Prover can never find a mistake. The arrangement looks similar to \Cref{fig:prove-delayer-example}, except it has only $n$ layers.

This reservation 
is 
both large (requiring linear space) and
history-dependent. In particular, when $\varphi$ is satisfiable, the
Prover only wins once the reservation specifies a block in the
final layer containing information about a satisfying
assignment. Crucially, this assignment is not determined by the
current clause alone, but by the entire traversal history. It is
therefore unclear how to reconstruct it locally or in parallel.

\myparagraph{A parallel quasipolynomial-time algorithm} 
To avoid the sequential nature of the greedy restriction procedure, we first revert to the original random restriction of Atserias and
Müller~\autocite{AM20}. This argument established, by the probabilistic method, that there \emph{exists}
a restriction that not only reduces block-width but actually \emph{kills} all
clauses of block-width at least $ c\cdot \log |\pi|$ for some $c$---much better than the
$c \cdot \sqrt{s\log|\pi|}$ achieved by the previously mentioned greedy construction. While {we} do not know what the actual restriction looks like, this stronger killing property allows us to discard all wide clauses outright. Crucially, we never actually construct the restriction---we simply look at a clause, check its block-width, and only proceed further if the clause has block-width $c \cdot \log |\pi|$.

Now, looking at some clause $C$ of block-width $c \cdot \log |\pi|$, we consider under what circumstances the original algorithm would have found a satisfying assignment at $C$. This would be the case if the algorithm reached $C$ while having in memory a reservation $R$ that specified a width-$n$ clause in the last layer of the purported refutation
that is not a weakening of any axiom. The difficulty is that we do not have access to such a reservation $R$, so we will try to reconstruct it. A naive brute-force search over all possible reservations will not do: there are more than $n$ variables per block, so even for low block-width clauses, this would take exponential time. 

We instead find a valid reservation \emph{only for the pointers} of the block. A brute-force algorithm can try all pointer reservations in time $s^{O(\bw(C))}$, which is quasi-polynomial if $C$ has block-width $O(\log |\pi|)$. These pointers determine a forest-like structure on the blocks. The next step is to determine whether this pointer reservation is consistent with $C$---for example, whether the variables that $C$ enforces to be present or absent in a block are consistent with the left/right pointers leading to the block; and
whether for any two blocks that are in the same connected component given by the pointers, the variables in these blocks are consistent with a valid reservation. This second stage of a consistency check can be easily implemented in parallel by polynomial-size constant-depth circuits.

After this procedure, we may construct a \emph{partial reservation
  $R_0$} for $C$: it sets all pointers in a way consistent with the
clause, but we do not necessarily know what the \emph{contents} of the
blocks involved are. We will say that $R_0$ \emph{encodes a satisfying
  assignment} if $R_0$ contains information about a block appearing in
the last layer and the partial information available is already a
(partial) satisfying assignment for $\varphi$. Furthermore, for the
purpose of this discussion, we say a clause $C$ is an
\emph{assignment-encoding clause} if there is at least one valid
reservation $R_0$ for $C$ and all valid reservations for $C$ encode a
satisfying assignment.

By the correctness of the original algorithm, and the fact that the Delayer cannot keep playing forever, we can conclude that there must exist a clause $C$ in $\pi$ that is assignment-encoding. It thus suffices to analyze each clause $C$ separately and find a single valid reservation for $C$, if one exists: this reservation either encodes a satisfying assignment, and we are done, or it does not, and it therefore witnesses that $C$ is not an assignment-encoding clause.

\myparagraph{The parallel polynomial-time algorithm}
The quasipolynomial algorithm described above still spends
$s^{\Omega(\bw(C))}$ time guessing the pointer structure of a reservation.
To obtain a polynomial-time algorithm, we modify the virtual grid on which
the Delayer arranges the blocks of the $\Reff(\varphi)$ formula to include $\Delta \log |\pi|$ ``dummy'' layers between the original $n$ layers and the axioms of $\varphi$. \Cref{fig:prove-delayer-example} shows this new arrangement of the blocks with the additional layers. On the first $n$ layers the Delayer behaves as before. On each of the
additional layers, however, a block is simply copied to the next layer:
both of its pointers lead to the same child, carrying exactly the same
width-$n$ clause. Only at layer $\ell$ does the Delayer finally declare
that this clause is a weakening of an axiom of $\varphi$.

The purpose of these extra layers is to isolate the final layer from the
complicated branching that may occur in the first $n$ layers. Suppose that
the current clause $C$ has block-width at most $w = c \cdot \log |\pi|$, and consider a
reservation consistent with $C$. Since only blocks mentioned in $C$ have
their outgoing pointers fixed, a connected component determined by the
reservation can contain at most $w$ such blocks before reaching an
unmentioned endpoint. It therefore cannot cross the $\Delta \log|\pi|$ dummy layers
and reach the first $n$ layers, assuming $\Delta > c$. Moreover, there is no branching inside the
dummy region, since every block there has both pointers leading to the same
child. This is the key
simplification.

Now, when analyzing a low-block-width clause $C$, we do not need
to reconstruct the pointers of a reservation at all. Let
$b_1,\ldots,b_w$ be the blocks mentioned in $C$. The algorithm considers every subset $B$ of these.
Intuitively, $B$ is a guess for the set of blocks of $C$ lying on one of
the chains that reaches the final layer. From the literals of $C$, we can
read the partial information that falsifying $C$ forces about the contents
of each block $b\in B$. We collect all this information into a clause
$A_B$. If $A_B$ is non-tautological, it has a unique partial assignment
$\alpha_B$ that falsifies all of its literals. We simply check whether
$\alpha_B$ already satisfies every clause of $\varphi$. If it does, then
every extension of $\alpha_B$ to the remaining variables is a satisfying
assignment, and the algorithm outputs one. Crucially, there are only $2^{w} = 2^{O(\log |\pi|)} = |\pi|^{O(1)}$
subsets to inspect for each clause, so this procedure is polynomial once we
restrict attention to clauses of block-width at most $w = \Theta(\log |\pi|)$.

The whole procedure can be implemented by polynomial-size $\TC^0$
circuits. Threshold gates are used to compute the block-width of a clause
and thereby discard clauses wider than~$w$; all subsets of the remaining
$O(\log|\pi|)$ blocks can then be inspected in parallel. Removing the
threshold gates yields quasipolynomial-size $\ACZ$ circuits.

\ifcameraready
\begin{figure}[t]
    \centering
    \begin{tikzpicture}[
    font=\Large,
    box/.style={minimum width=4.5cm, minimum height=1cm, rounded corners=12pt},
    plainbox/.style={box,draw,thick},
    orangebox/.style={box,draw=orange!90!black,fill=orange!40,very thick},
    greenbox/.style={box,draw=green!60!black,,fill=green!40,very thick},
    yellowbox/.style={box,draw=yellow!70!black,fill=yellow!50,very thick},
    ovalbox/.style={draw, very thick, blue!70!black, fill=yellow!30,
      minimum width=3.9cm, minimum height=1.2cm,ellipse}
      , arrow/.style = {-{Stealth[length=4mm]}, ultra thick},
      redarrow/.style = {arrow, red!80!black},
      yellowarrow/.style = {arrow, yellow!80!black},
      bluearrow/.style = {arrow, blue!70!black}
]

\newlength{\vsep}
\newlength{\hsep}
\newlength{\bigvsep}
\newlength{\bighsep}

\setlength{\hsep}{1cm}
\setlength{\vsep}{1cm}
\setlength{\bighsep}{3cm}
\setlength{\bigvsep}{2cm}

\node[plainbox] (a1) {};
\node[orangebox, right=\hsep of a1] (a2) {$x_1$};
\node[orangebox, right=\hsep of a2] (a3) {$\neg x_1$};
\node[plainbox, right=\bighsep of a3] (a4) {};
\node at ($(a3)!0.5!(a4) + (0,0)$) {$\scalebox{2}{$\ldots$}$};

\node[circle, red!80!black, draw, very thick, minimum size=1cm, fill=orange!40](bot) at ($(a2.north)!0.5!(a3.north) + (0,\vsep+10)$) {$\bot$};

\draw[redarrow] (bot) -- (a2.north);
\draw[redarrow] (bot) -- (a3.north);

\node[orangebox, below=\vsep of a1] (b1) {$x_1 \vee x_2$};
\node[orangebox, below=\vsep of a2] (b2) {$x_1 \vee \neg x_2$};
\node[plainbox, below=\vsep of a3] (b3) {};
\node[greenbox, below=\vsep of a4] (b4) {$x_1 \vee x_2$};
\node at ($(b3)!0.5!(b4) + (0,0)$) {\scalebox{2}{$\ldots$}};

\draw[redarrow] (a2.south) -- (b2.north);
\draw[redarrow] (a2.south) -- (b1.north);

\node[plainbox, below=\vsep of b1] (c1) {};
\node[orangebox, below=\vsep of b2] (c2) {$x_1 \vee \neg  x_2 \vee x_3$};
\node[orangebox, below=\vsep of b3] (c3) {$x_1 \vee \neg x_2 \vee \neg x_3$};
\node[plainbox, below=\vsep of b4] (c4) {};
\node at ($(c3)!0.5!(c4) + (0,0)$) {\scalebox{2}{$\ldots$}};

\draw[redarrow] (b2.south) -- (c2.north);
\draw[redarrow] (b2.south) -- (c3.north);

\node[plainbox, below=\bigvsep of c1] (d1) {};
\node[plainbox, right=\hsep of d1] (d2) {};
\node[yellowbox, right=\hsep of d2] (d3) {$C \lor \neg x_{n-2}$};
\node[greenbox, right=\bighsep of d3] (d4) {$\bigvee_{i=1}^{n-2} x_i$};
\node at ($(d3)!0.5!(d4) + (0,0)$) {\scalebox{2}{$\ldots$}};

\node[greenbox, below=\vsep of d1, , fill=green!40] (e1) {$\bigvee_{i=1}^{n-1} x_i$};
\node[yellowbox, below=\vsep of d2] (e2) {$C \lor \neg x_{n-2} \lor x_{n-1}$};
\node[yellowbox, below=\vsep of d3] (e3) {$C \lor \neg x_{n-2} \lor \neg x_{n-1}$};
\node[plainbox, below=\vsep of d4] (e4) {};
\node at ($(e3)!0.5!(e4) + (0,0)$) {\scalebox{2}{$\ldots$}};

\draw[yellowarrow] (d3.south) -- (e2.north);
\draw[yellowarrow] (d3.south) -- (e3.north);

\node[plainbox, below=\vsep of e1] (f1) {};
\node[yellowbox, below=\vsep of e2] (f2) {\large{$C \lor \neg x_{n-2} 
\lor \neg x_{n-1} \lor x_n$}};
\node[yellowbox, below=\vsep of e3] (f3) {\large{$C \lor \neg x_{n-2} 
\lor \neg x_{n-1} \lor \neg x_n$}};
\node[plainbox, below=\vsep of e4] (f4) {};
\node at ($(f3)!0.5!(f4) + (0,0)$) {\scalebox{2}{$\ldots$}};

\draw[yellowarrow] (e3.south) -- (f2.north);
\draw[yellowarrow] (e3.south) -- (f3.north);

\node at ($(c1)!0.5!(d1)+ (0,0.2)$) {\scalebox{2}{$\vdots$}};
\node at ($(c2)!0.5!(d2)+ (0,0.2)$) {\scalebox{2}{$\vdots$}};
\node at ($(c3)!0.5!(d3)+ (0,0.2)$) {\scalebox{2}{$\vdots$}};
\node at ($(c4)!0.5!(d4)+ (0,0.2)$) {\scalebox{2}{$\vdots$}};

\node[yellowbox, below=\vsep of f1] (g1) 
{\large{$C \lor \neg x_{n-2} 
\lor \neg x_{n-1} \lor \neg x_n$}};
\node[plainbox, below=\vsep of f2] (g2) {};
\node[yellowbox, below=\vsep of f3] (g3) {\large{$C \lor \neg x_{n-2} 
\lor \neg x_{n-1} \lor x_n$}};
\node[plainbox, below=\vsep of f4] (g4) {};
\node at ($(g3)!0.5!(g4) + (0,0)$) {\scalebox{2}{$\ldots$}};

\draw[yellowarrow] (f2.south) -- (g3.north);
\draw[yellowarrow] (f3.south) -- (g1.north);

\node[plainbox, below=\vsep of g1] (h1) {};
\node[yellowbox, below=\vsep of g2] (h2) {\large{$C \lor \neg x_{n-2} 
\lor \neg x_{n-1} \lor \neg x_n$}};
\node[plainbox, below=\vsep of g3] (h3) {};
\node[yellowbox, below=\vsep of g4] (h4) {\large{$C \lor \neg x_{n-2} 
\lor \neg x_{n-1} \lor x_n$}};
\node at ($(h3)!0.5!(h4) + (0,0)$) {\scalebox{2}{$\ldots$}};

\draw[yellowarrow] (g1.south) -- (h2.north);
\draw[yellowarrow] (g3.south) -- (h4.north);

\node[plainbox, below=\vsep of h1] (i1) {};
\node[yellowbox, below=\vsep of h2] (i2) {\large{$C \lor \neg x_{n-2} 
\lor \neg x_{n-1} \lor \neg x_n$}};
\node[yellowbox, below=\vsep of h3] (i3) 
{\large{$C \lor \neg x_{n-2} 
\lor \neg x_{n-1} \lor x_n$}};
\node[plainbox, below=\vsep of h4] (i4) {};
\node at ($(i3)!0.5!(i4) + (0,0)$) {\scalebox{2}{$\ldots$}};

\draw[yellowarrow] (h2.south) -- (i2.north);
\draw[yellowarrow] (h4.south) -- (i3.north);

\node[ovalbox,
      below=\vsep of i2] (j1)
{$C \lor \neg x_n$};

\node[ovalbox,
      below=\vsep of i3] (j2)
{$\neg x_{n-2} \lor x_n$};

\draw[bluearrow] (i2) -- (j1);
\draw[bluearrow] (i3) -- (j2);

\draw[<->, thick, black] 
($(a4.north east)+(1,0)$) --
node[right] {{$n$}}
($(f4.south east)+(1,0)$);

\draw[<->, thick, black] 
($(g4.north east)+(1,0)$) --
node[right] {{$\Delta \log |\pi|$}}
($(i4.south east)+(1,0)$);

\draw[<->, thick, black] 
($(i1.south west)+(0,-3)$) --
node[below] {{$s/(n+\Delta \log |\pi|)$}}
($(i4.south east)+(0,-3)$);

\end{tikzpicture}
    \caption{ 
    Snapshot of the execution of the new extraction algorithm ({Algorithm~\ref{algorithm:logspace}}), when playing over the extended grid. The Prover--Delayer game is played over the formula $\Reff_s(\varphi)$ for $s$ blocks, arranged in $\ell \coloneq n + \Delta\log |\pi|$ layers with $s/\ell$ blocks per layer. In \textcolor{orange!90!black}{orange} and in \textcolor{yellow!70!black}{yellow} are two partial trees in which the Delayer answers consistently by locally following the canonical exponential-size tree-like refutation. The \textcolor{orange!90!black}{orange} tree starts from the root, while the \textcolor{yellow!70!black}{yellow} one starts from some clause $C$ of width $n-3$ that emerged from previous queries. In \textcolor{green!60!black}{green}, blocks that were queried ``out of context'': at layer $i$, the Delayer answers with the clause $x_1 \lor x_2 \lor x_3 \lor \dots \lor x_i$ where all $i$ literals are positive. At layer $n$ there are width-$n$ clauses that will be trivially weakened in the following layers until reaching layer $\ell = n + \Delta \log |\pi|$, where they must point to an axiom (in \textcolor{blue!70!black}{blue}). For $\Delta =0$, we have $\ell = n$ and we recover the grid used in the original algorithm.}
    \label{fig:prove-delayer-example}
\end{figure}
\else
\begin{figure}
    \centering
    \begin{tikzpicture}[
    font=\Large,
    box/.style={minimum width=4.5cm, minimum height=1cm, rounded corners=12pt},
    plainbox/.style={box,draw,thick},
    orangebox/.style={box,draw=orange!90!black,fill=orange!40,very thick},
    greenbox/.style={box,draw=green!60!black,,fill=green!40,very thick},
    yellowbox/.style={box,draw=yellow!70!black,fill=yellow!50,very thick},
    ovalbox/.style={draw, very thick, blue!70!black, fill=yellow!30,
      minimum width=3.9cm, minimum height=1.2cm,ellipse}
      , arrow/.style = {-{Stealth[length=4mm]}, ultra thick},
      redarrow/.style = {arrow, red!80!black},
      yellowarrow/.style = {arrow, yellow!80!black},
      bluearrow/.style = {arrow, blue!70!black}
]

\newlength{\vsep}
\newlength{\hsep}
\newlength{\bigvsep}
\newlength{\bighsep}

\setlength{\hsep}{1cm}
\setlength{\vsep}{1cm}
\setlength{\bighsep}{3cm}
\setlength{\bigvsep}{2cm}

\node[plainbox] (a1) {};
\node[orangebox, right=\hsep of a1] (a2) {$x_1$};
\node[orangebox, right=\hsep of a2] (a3) {$\neg x_1$};
\node[plainbox, right=\bighsep of a3] (a4) {};
\node at ($(a3)!0.5!(a4) + (0,0)$) {$\scalebox{2}{$\ldots$}$};

\node[circle, red!80!black, draw, very thick, minimum size=1cm, fill=orange!40](bot) at ($(a2.north)!0.5!(a3.north) + (0,\vsep+10)$) {$\bot$};

\draw[redarrow] (bot) -- (a2.north);
\draw[redarrow] (bot) -- (a3.north);

\node[orangebox, below=\vsep of a1] (b1) {$x_1 \vee x_2$};
\node[orangebox, below=\vsep of a2] (b2) {$x_1 \vee \neg x_2$};
\node[plainbox, below=\vsep of a3] (b3) {};
\node[greenbox, below=\vsep of a4] (b4) {$x_1 \vee x_2$};
\node at ($(b3)!0.5!(b4) + (0,0)$) {\scalebox{2}{$\ldots$}};

\draw[redarrow] (a2.south) -- (b2.north);
\draw[redarrow] (a2.south) -- (b1.north);

\node[plainbox, below=\vsep of b1] (c1) {};
\node[orangebox, below=\vsep of b2] (c2) {$x_1 \vee \neg  x_2 \vee x_3$};
\node[orangebox, below=\vsep of b3] (c3) {$x_1 \vee \neg x_2 \vee \neg x_3$};
\node[plainbox, below=\vsep of b4] (c4) {};
\node at ($(c3)!0.5!(c4) + (0,0)$) {\scalebox{2}{$\ldots$}};

\draw[redarrow] (b2.south) -- (c2.north);
\draw[redarrow] (b2.south) -- (c3.north);

\node[plainbox, below=\bigvsep of c1] (d1) {};
\node[plainbox, right=\hsep of d1] (d2) {};
\node[yellowbox, right=\hsep of d2] (d3) {$C \lor \neg x_{n-2}$};
\node[greenbox, right=\bighsep of d3] (d4) {$\bigvee_{i=1}^{n-2} x_i$};
\node at ($(d3)!0.5!(d4) + (0,0)$) {\scalebox{2}{$\ldots$}};

\node[greenbox, below=\vsep of d1, , fill=green!40] (e1) {$\bigvee_{i=1}^{n-1} x_i$};
\node[yellowbox, below=\vsep of d2] (e2) {$C \lor \neg x_{n-2} \lor x_{n-1}$};
\node[yellowbox, below=\vsep of d3] (e3) {$C \lor \neg x_{n-2} \lor \neg x_{n-1}$};
\node[plainbox, below=\vsep of d4] (e4) {};
\node at ($(e3)!0.5!(e4) + (0,0)$) {\scalebox{2}{$\ldots$}};

\draw[yellowarrow] (d3.south) -- (e2.north);
\draw[yellowarrow] (d3.south) -- (e3.north);

\node[plainbox, below=\vsep of e1] (f1) {};
\node[yellowbox, below=\vsep of e2] (f2) {\large{$C \lor \neg x_{n-2} 
\lor \neg x_{n-1} \lor x_n$}};
\node[yellowbox, below=\vsep of e3] (f3) {\large{$C \lor \neg x_{n-2} 
\lor \neg x_{n-1} \lor \neg x_n$}};
\node[plainbox, below=\vsep of e4] (f4) {};
\node at ($(f3)!0.5!(f4) + (0,0)$) {\scalebox{2}{$\ldots$}};

\draw[yellowarrow] (e3.south) -- (f2.north);
\draw[yellowarrow] (e3.south) -- (f3.north);

\node at ($(c1)!0.5!(d1)+ (0,0.2)$) {\scalebox{2}{$\vdots$}};
\node at ($(c2)!0.5!(d2)+ (0,0.2)$) {\scalebox{2}{$\vdots$}};
\node at ($(c3)!0.5!(d3)+ (0,0.2)$) {\scalebox{2}{$\vdots$}};
\node at ($(c4)!0.5!(d4)+ (0,0.2)$) {\scalebox{2}{$\vdots$}};

\node[yellowbox, below=\vsep of f1] (g1) 
{\large{$C \lor \neg x_{n-2} 
\lor \neg x_{n-1} \lor \neg x_n$}};
\node[plainbox, below=\vsep of f2] (g2) {};
\node[yellowbox, below=\vsep of f3] (g3) {\large{$C \lor \neg x_{n-2} 
\lor \neg x_{n-1} \lor x_n$}};
\node[plainbox, below=\vsep of f4] (g4) {};
\node at ($(g3)!0.5!(g4) + (0,0)$) {\scalebox{2}{$\ldots$}};

\draw[yellowarrow] (f2.south) -- (g3.north);
\draw[yellowarrow] (f3.south) -- (g1.north);

\node[plainbox, below=\vsep of g1] (h1) {};
\node[yellowbox, below=\vsep of g2] (h2) {\large{$C \lor \neg x_{n-2} 
\lor \neg x_{n-1} \lor \neg x_n$}};
\node[plainbox, below=\vsep of g3] (h3) {};
\node[yellowbox, below=\vsep of g4] (h4) {\large{$C \lor \neg x_{n-2} 
\lor \neg x_{n-1} \lor x_n$}};
\node at ($(h3)!0.5!(h4) + (0,0)$) {\scalebox{2}{$\ldots$}};

\draw[yellowarrow] (g1.south) -- (h2.north);
\draw[yellowarrow] (g3.south) -- (h4.north);

\node[plainbox, below=\vsep of h1] (i1) {};
\node[yellowbox, below=\vsep of h2] (i2) {\large{$C \lor \neg x_{n-2} 
\lor \neg x_{n-1} \lor \neg x_n$}};
\node[yellowbox, below=\vsep of h3] (i3) 
{\large{$C \lor \neg x_{n-2} 
\lor \neg x_{n-1} \lor x_n$}};
\node[plainbox, below=\vsep of h4] (i4) {};
\node at ($(i3)!0.5!(i4) + (0,0)$) {\scalebox{2}{$\ldots$}};

\draw[yellowarrow] (h2.south) -- (i2.north);
\draw[yellowarrow] (h4.south) -- (i3.north);

\node[ovalbox,
      below=\vsep of i2] (j1)
{$C \lor \neg x_n$};

\node[ovalbox,
      below=\vsep of i3] (j2)
{$\neg x_{n-2} \lor x_n$};

\draw[bluearrow] (i2) -- (j1);
\draw[bluearrow] (i3) -- (j2);

\draw[<->, thick, black] 
($(a4.north east)+(1,0)$) --
node[right] {{$n$}}
($(f4.south east)+(1,0)$);

\draw[<->, thick, black] 
($(g4.north east)+(1,0)$) --
node[right] {{$\Delta \log |\pi|$}}
($(i4.south east)+(1,0)$);

\draw[<->, thick, black] 
($(i1.south west)+(0,-3)$) --
node[below] {{$s/(n+\Delta \log |\pi|)$}}
($(i4.south east)+(0,-3)$);

\end{tikzpicture}
    \caption{ 
    Snapshot of the execution of the new extraction algorithm ({Algorithm~\ref{algorithm:logspace}}), when playing over the extended grid. The Prover--Delayer game is played over the formula $\Reff_s(\varphi)$ for $s$ blocks, arranged in $\ell \coloneq n + \Delta\log |\pi|$ layers with $s/\ell$ blocks per layer. In \textcolor{orange!90!black}{orange} and in \textcolor{yellow!70!black}{yellow} are two partial trees in which the Delayer answers consistently by locally following the canonical exponential-size tree-like refutation. The \textcolor{orange!90!black}{orange} tree starts from the root, while the \textcolor{yellow!70!black}{yellow} one starts from some clause $C$ of width $n-3$ that emerged from previous queries. In \textcolor{green!60!black}{green}, blocks that were queried ``out of context'': at layer $i$, the Delayer answers with the clause $x_1 \lor x_2 \lor x_3 \lor \dots \lor x_i$ where all $i$ literals are positive. At layer $n$ there are width-$n$ clauses that will be trivially weakened in the following layers until reaching layer $\ell = n + \Delta \log |\pi|$, where they must point to an axiom (in \textcolor{blue!70!black}{blue}). For $\Delta =0$, we have $\ell = n$ and we recover the grid used in the original algorithm.}
    \label{fig:prove-delayer-example}
\end{figure}
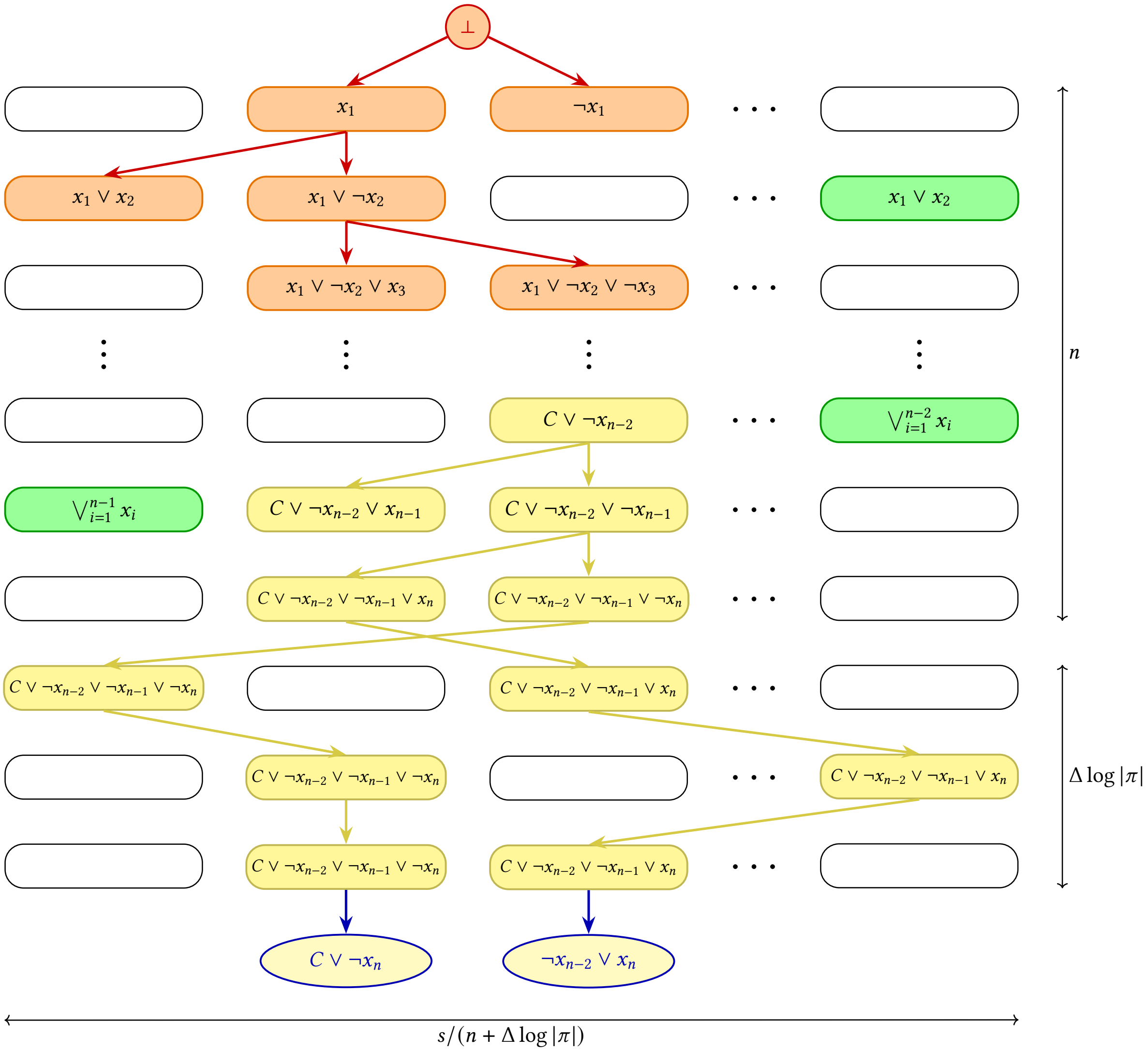
\fi

\subsubsection{Derandomizing the random restriction argument: the new proof of \autocite{AM20}}
\label{subsubsec:intro-techniques-new-proof-AM}
With the new algorithm in hand, we turn to proving its
correctness---and hence the Atserias--Müller lower bound~\autocite{AM20}---within a
suitable theory of bounded arithmetic. We introduce techniques that may be of
independent interest for future formalizations of complexity-theoretic
arguments in bounded arithmetic.

\myparagraph{A new derandomization of the restriction argument}
Our goal is to show that
there is a restriction $\rho$ satisfying all clauses of
block-width at least~{$w = \Theta(\log |\pi|)$}. The original
argument proceeds via the probabilistic method: one defines a suitable
space of restrictions and shows that when sampling uniformly at random over this space, for any
fixed clause $C$ of block-width at least $w$, the probability that the clause
survives is
\begin{equation}
	\label{eq:clause-killed-intro}
	\Pr_{\rho \in \calU} [C_{\restriction \rho} \neq 1] \leq \left(\frac{3}{4}\right)^w.
\end{equation}
A union bound over all clauses
in the refutation $\pi$ yields that the probability that some clause survives is at most $|\pi| \cdot (3/4)^w$, which is strictly
less than $1$ for our choice $w$---so a \say{good} $\rho$ exists.

In bounded arithmetic, we work in the polynomial-size regime, meaning
that the parameters $n$ and~$s$ from the formula $\Reff_s(\varphi)$
are not assumed to be in $\Log$, and hence exact probability
calculations are out of reach. Furthermore, in order to prove the
statement in bounded arithmetic, we must first have a feasibly
constructive procedure to obtain a good restriction---otherwise this
would violate a corresponding witnessing theorem. This means that a
naive use of the probabilistic method alone will not suffice.

Our solution is to use tools from unconditional derandomization. The
key observation is that to establish the bound in
(\ref{eq:clause-killed-intro}), we only require \emph{$w$-wise
  independence}, rather than full independence. We can therefore
replace the uniform distribution $\calU$ by a $w$-wise independent
distribution $\calD$ with much smaller support, following the
classic derandomization paradigm initiated by \citeauthor{ABI86}~\autocite{ABI86} and \citeauthor{Luby86}~\autocite{Luby86}.

The only caveat is quantitative: any $w$-wise independent distribution
over $n$-bit strings has support size at least $n^{\Omega(w)}$
\autocites{Chor85}[Proposition 16.2.3]{AS-BOOK-Probabilistic}, so taking $w = \Theta(\log |\pi|)$ yields a
quasipolynomial-size sample space. Accordingly, we 
formalize
the quasipolynomial-size $\ACZ$ algorithm
(\Cref{thm:extraction-AC0-informal}), rather than the polynomial-size
$\TC^0$ circuits from \Cref{thm:extraction-TC0-informal}.

The observation that limited independence suffices is at the core of the
proof, but we are still left with the task of choosing an explicit construction of
$w$-wise independent distributions computable by constant-depth
circuits and showing that constant-depth Frege systems can reason
about them. We use the algebraic construction of
combinatorial designs due to Nisan and Wigderson~\autocite{NW94}, based on
low-degree polynomials over finite
fields. \citeauthor{jerabek:phd-thesis}~\autocite{jerabek:phd-thesis}
formalized a different (greedy) construction sufficient for the
derandomization of $\BPP$ (i.e., the second construction of Nisan and
Wigderson), but the parameters {of that construction} are
inadequate in our context. As such we formalize
  the algebraic construction in $\PVO$.
  
We then move to the relativized setting, where an external predicate
$\alpha$ encodes the Resolution refutation, and we work in the theory
$\APCtwo(\alpha)$ of \citeauthor{jerabek:phd-thesis}, which extends
$\BussTonetwo(\alpha)$ by the dual weak pigeonhole principle for
$\FP^\NP$ functions ($\dWPHP(\PV_2(\alpha))$). In this theory we have
access to \citeauthor{jerabek:phd-thesis}'s approximate counting
machinery in the relativized setting. We use this to bound the size of
the \say{bad event}, essentially mimicking the bound in
(\ref{eq:clause-killed-intro}), and then simulate the union bound
needed to complete the proof.

We find this derandomization particularly interesting, as it
faithfully captures a probabilistic method argument within a very weak
theory. More broadly, it suggests that techniques from unconditional
derandomization could be systematically imported into bounded
arithmetic, and may prove useful for formalizing other
complexity-theoretic results.

\myparagraph{The block-width lower bound argument}
For the block-width lower bound, the workaround is simpler but a bit
more subtle. Instead of appealing to the correctness of the original
algorithm, we 
rephrase the inductive proof so that it mirrors the
new algorithm. Indeed, the original proof in $\PVO$ worked by
induction on the length of the path constructed by the Delayer to
traverse the refutation. In the relativized context, such a path would
be too large to be encoded by numbers, so we cannot perform this kind
of induction. Instead, our induction is over the indices of the
Resolution refutation.

Let $\pi$ be the Resolution refutation given in reverse order, for convenience.
That is, we assume that $C_1 = \bot$ and every derived clause $C_i$ is derived from clauses $C_j$ and $C_k$ for $j,k > i$. We show that for every clause $C_i \in \pi$, at least one of the following holds:
\begin{enumerate}[label=(\alph*)]
\itemsep=0pt
\item there exists $j \leq i$ such that the algorithm extracts a satisfying assignment from $C_j$;
\item there exists $j \leq i$ such that $C_j$ has large block-width; or
\item there exists $j > i$ such that $C_j$ is \emph{Delayer-consistent}.
\end{enumerate}

The notion of \emph{Delayer-consistency} arises naturally when analyzing the new algorithm. Intuitively, we 
say that a clause~$C$ is Delayer-consistent if there exists a partial reservation~$R$ consistent with $C$ such that the Delayer could have safely continued playing starting from $C$ with reservation~$R$
\ifcameraready
\else
(see \Cref{sec:BW-formalized-Ttwo} for the formal definition).
\fi

The key difference with the original argument is in how the induction is structured. Instead of explicitly constructing the Delayer’s path, we argue locally as follows. At stage $i$, for contradiction, suppose (a)-(c) all fail. Then, by induction hypothesis, $C_i$ must be Delayer-consistent. By analyzing the local structure of the refutation and the definition of consistency, we show that one of the two clauses used to derive $C_i$ is Delayer-consistent, contradicting our assumption.
Thus, rather than building a single path, the argument moves from one Delayer-consistent clause to another, without maintaining the full traversal history. 
Since the refutation is finite, we cannot indefinitely move to further Delayer-consistent clauses, so as long as the refutation has low block-width, the algorithm terminates by finding a satisfying assignment.

\subsubsection{The $\NP$-hardness of the Proof Analysis Problem}
\label{subsubsec:intro-techniques-PAP-NP-hard}
The $\NP$-hardness result in \Cref{thm:NP-hardness-informal} follows the ideas of \citeauthor{PAP}~\autocite{PAP}, in a slightly streamlined form. The key message of the earlier work is that if a propositional proof system $S$ can establish even a single non-trivial Resolution lower bound, then the associated analysis problem for $S$ is $\NP$-hard. We recall this argument in a simplified form.

Suppose that $S$ admits uniform proofs of the lower bounds encoded as $\neg \Reff_{n^k}(\psi_n)$, where $\{\psi_n\}_{n \in \bbN}$ is a fixed family of unsatisfiable CNF formulas over $n$ variables with $\poly(n)$ clauses. For any other formula $\varphi_n$ over $n$ variables one can now efficiently construct a proof $\pi_n$ of the form
\begin{equation}
    \pi_n :  S \vdash \neg \Reff_{n^k}(\varphi_n \lor \psi_n),
\end{equation}
where the disjunction is encoded in CNF using an auxiliary variable, and $\varphi_n$ and $\psi_n$ are over disjoint variable sets.

The derivation reduces the lower bound for $\varphi_n \lor \psi_n$ to that for $\psi_n$. Fix an arbitrary assignment $\alpha$ to the variables of $\varphi_n$. If $\alpha$ satisfies $\varphi_n$, then $\varphi_n \lor \psi_n$ is satisfied, and the corresponding lower bound is trivial. Otherwise, $\varphi_n(\alpha) = \bot$, and under this restriction the formula $\varphi_n \lor \psi_n$ simplifies to $\psi_n$. Provided that~$S$ proves that Resolution is closed under restrictions, it can invoke the lower bound for $\psi_n$ to derive the desired conclusion. Crucial to this reduction is that $\pi_n$ proves a fixed $n^k$ lower bound and not an exponential one, or else we would not be able to construct $\pi_n$ efficiently.

It follows that an algorithm for the analysis problem of~$S$ could be used to decide $\SAT$: given $\varphi_n$ on $n$ variables, construct~$\pi_n$ as described above and run the analysis procedure on $\pi_n$. In~\autocite{PAP}, the reduction for Extended Frege used the fact that Extended Frege can prove Resolution size lower bounds for $\PHP_n$~\autocite{CP90}. To instantiate this argument for constant-depth Frege, we may obtain the proof of a Resolution lower bound from different sources. One such source would be \Cref{thm:AM-propositional-informal}: fixing a formula $\gamma$ such that $\PK_{3+\nicefrac{1}{2}}$ proves $\gamma_n$ unsatisfiable, \Cref{thm:AM-propositional-informal} implies that $\PK_{3+\nicefrac{1}{2}}$ proves $\neg \Reff_{n^k}(\Reff_{n^3}(\gamma_n))$. 
Thus, in the above reduction we may take $\psi_n = \Reff_{n^3}(\gamma_n)$. \Cref{thm:Haken-PHP-polyregime-informal} also gives the proof of a lower bound, even at lower depth, but an even better approach, detailed in \Cref{sec:PAP-NP-hard}, is to rescale the formalizations of \cite{LLR24} and then apply a restriction, to get the reduction to work as low as $\PK_2$. It then remains to verify that constant-depth Frege proves that Resolution is closed under restrictions. This is routine, though somewhat tedious, and completes the proof.

\subsection{Related work}
\label{subsec:intro-related-work}

To the best of our knowledge, the only attempts at formalizing Resolution lower bounds are the article of Cook and Pitassi~\autocite{CP90} on Haken's lower bound in Extended Frege, the earlier work on the Proof Analysis Problem~\autocite{PAP}, and the recent work of Li, Li, and Ren~\autocite{LLR24} in the context of $\TFNP$. The first step towards formalizing constant-depth Frege lower bounds was recently taken by Davis and Robere~\autocite{DR26} in a limited parameter regime, showing that exponential lower bounds on the length of constant-depth Frege proofs of the pigeonhole principle can be proven in $\Res(\log)$ in quasipolynomial size. The result presents a reduction from the weak pigeonhole principle in the style of the $\TFNP$ framework of \autocite{LLR24}. Accordingly, the parameters suffer from some of the same caveats as \autocite{LLR24}: the formalization works only in the exponential-size regime, and it furthermore requires a polynomial bound on the size of the lines of the purported refutations. 

Similar meta-mathematical approaches to proof complexity, heavily featuring variants of the $\Reff$ formulas studied in this paper, appear in the work of \citeauthor{PS19}~\autocite{PS19},  \citeauthor{ST21}~\autocite{ST21}, and \citeauthor{LST26}~\autocite{LST26}. In the context of circuit complexity and communication complexity, more work has been carried out  (see, for example, \autocite{Razborov95, MP20, CLO24, CG25, CLOW26} and references therein). All of these works fit into the general trend of the so-called reverse mathematics of computational complexity. The literature is too vast to review here, so we refer the reader to the survey of Oliveira \autocite{Oliveira24}. In parallel, a growing body of work has used tools and ideas from mathematical logic to prove complexity-theoretic statements (see, e.g., \autocite{azza1,MoritcForcing,khainc,khanw,PS22,Mykyta22,PS23, KraInc,azza2,Kha24,ACG24, AKPS24,KraENS,Mykyta24,azza3, JJ25, TJ26}).

\subsection{Structure of the paper}
\label{subsec:intro-structure-of-the-paper}

\ifcameraready
The rest of the paper is structured as follows. After the preliminaries in \Cref{sec:preliminaries}, \Cref{sec:new-algorithm} introduces the new assignment extraction algorithm, proves its correctness and gives implementations as $\TC^0$ and $\ACZ$ circuits. We then turn to the task of formalizing the correctness in bounded arithmetic: \Cref{sec:restriction-formalized-APC} derandomizes the restriction argument via limited independence in $\APCtwo(\alpha)$. For the formalization of the block-width lower bound, we refer the reader to the full version of the paper, where we also put these two parts together to obtain \Cref{thm:AM-new-proof-informal}, \Cref{thm:AM-propositional-informal}, \Cref{thm:automatability-informal}, and \Cref{thm:AM-LLR-regime-informal}. \Cref{sec:PAP-NP-hard} proves that the Proof Analysis Problem for constant-depth Frege is $\NP$-hard.
\else
The rest of the paper is structured as follows. After the preliminaries in \Cref{sec:preliminaries}, \Cref{sec:new-algorithm} introduces the new assignment extraction algorithm, proves its correctness and gives implementations as $\TC^0$ and $\ACZ$ circuits. We then turn to the task of formalizing the correctness in bounded arithmetic: \Cref{sec:restriction-formalized-APC} derandomizes the restriction argument via limited independence in $\APCtwo(\alpha)$, and \Cref{sec:BW-formalized-Ttwo} formalizes the block-width lower bound. We put these pieces together in \Cref{sec:AM-complete-formal}. In \Cref{sec:propositional-consequence} we apply the Paris--Wilkie translation to get propositional upper bounds and related corollaries. \Cref{sec:PAP-NP-hard} proves that the Proof Analysis Problem for constant-depth Frege is $\NP$-hard.
\fi


\section{Preliminaries}
\label{sec:preliminaries}

We review basic material from proof complexity and bounded arithmetic. For a more comprehensive treatment of proof complexity, we refer to Krajíček \autocite{Krajicek19}. For bounded arithmetic, the survey of \citeauthor{Oliveira24} \autocite{Oliveira24} covers all the necessary material in a style that aligns with our work, but the reader may also look up the classic textbooks on these topics \autocite{Krajicek95, HP93}.

\subsection{Proof complexity}
\label{subsec:proof-complexity}
Following \citeauthor{CR79} \autocite{CR79}, a \emph{propositional proof system} $S$ for the set~$\TAUT$ of propositional tautologies is a polynomial-time function $S : \{ 0,1\}^* \to \TAUT$ whose range is exactly $\TAUT$. If $S(\pi) = \varphi$, then we say $\pi$ is an \emph{$S$-proof} of $\varphi$. If $\varphi$ is an unsatisfiable formula and $\pi$ is an~$S$-proof of the tautology $\neg\varphi$, then we say that $\pi$ is an \emph{$S$-refutation} of~$\varphi$.

For a tautology $\varphi$ and a proof system $S$, we denote by $\size{S}{\varphi} \coloneq \min_{\pi : S(\pi) = \varphi} |\pi| $ the \emph{size} of its smallest $S$-proof. A proof system $S$ is \emph{polynomially bounded} if there exists a constant~$c \in \bbN$ such that for all $\varphi \in \TAUT$ we have $\size{S}{\varphi} \leq |\varphi|^c$. 
\newcommand{\effproves}{\vdash^{\operatorname{poly}}}
For a sequence $\varphi = \{\varphi_n\}_{n\in\bbN}$ of propositional tautologies, we write $S \effproves \varphi$ or simply $S \effproves \varphi_n$ to express that $\size{S}{\varphi_n} = |\varphi_n|^{O(1)}$ as $n$ grows. For a concrete proof $\pi$ achieving this, we write $\pi : S\vdash\varphi_n$. For $s \in \bbN$, we write $S\vdash^{s} \varphi_n$ to express that there exists an~$S$-proof $\pi$ of size $|\pi| \leq s$ such that~$\pi : S \vdash \varphi_n$. For a sequence $\{ \varphi_n \}_{n\in\bbN}$ of formulas or a sequence $\{ \pi_n \}_{n\in\bbN}$ of proofs, we say that the sequence is \emph{$\P$-uniform} if we can construct the $n$-th element of the sequence in time $\poly(n)$. We similarly use the term \emph{$\QP$-uniform} when the construction runs in quasipolynomial time. More generally, we may say \emph{$\TIME[t(n)]$-uniform} for any time-constructible function $t : \bbN \to \bbN$.

A proof system $S$ \emph{simulates} a system $Q$, written $S \geq Q$, if there is a constant~$c \in \bbN$ such that for every $\varphi \in \TAUT$ we have $\size{S}{\varphi} \leq \size{Q}{\varphi}^c$. We say that $S$ \emph{p-simulates} $Q$ and write $S \psim Q$ if there exists a polynomial-time computable function $f$ such that for every $\varphi\in \TAUT$ and every $\pi : Q\vdash \varphi$, we have $f(\pi) : S \vdash \varphi$. Two proof systems $S$ and $Q$ are \emph{polynomially equivalent} if $S \psim Q$ and $Q \psim S$. 

A \emph{restriction} for a propositional formula $\varphi(x_1, \dots, x_n)$ is a partial assignment $\rho : \{ x_1, \dots, x_n\} \to \{x_1, \dots, x_n, 0, 1\}$ mapping every variable to either a constant or to itself. We write $\varphi_{\restriction \rho}$ for the formula $\varphi(\rho(x_1), \dots, \rho(x_n))$, and we say that a proof system~$S$ is \emph{closed under restrictions} if there exists a constant $d\in \bbN$ such that for every tautology $\varphi$ and every restriction $\rho$, it holds that $\size{S}{\varphi_{\restriction \rho}} \leq \size{S}{\varphi}^d$. All the explicit proof systems dealt with in this work are closed under restrictions, and for a proof $\pi$ of $\varphi$, we denote by $\pi_{\restriction\rho}$ the proof of the restricted formula $\varphi_{\restriction \rho}$ obtained by applying the restriction line by line and performing the necessary clean-up. The notion of restriction is a particular case of a \emph{ formula substitution}: a partial function mapping $\{ x_1, \dots, x_n\}$ to propositional formulas. For a substitution $\sigma$ sending the variables in $\varphi(x_1, \dots, x_n)$ to formulas $\psi_1$, \dots, $\psi_n$, we write $\varphi(\psi_1, \dotsm \psi_n)$ for the formula obtained by substituting $\psi_i$ for~$x_i$.

\subsubsection{Resolution}
\label{subsec:prelim-resolution}
Resolution ($\Res$) is a refutation system for CNF formulas. Accordingly, we often write $\pi : \Res \vdash \neg\varphi$ to express that $\pi$ is a Resolution refutation of $\varphi$.

A \emph{literal} is a propositional atom or its negation, a \emph{clause} is a disjunction of literals, and we see clauses as sets of literals. We write $C \subseteq D$ to express that $C$ is a subclause of $D$.  A \emph{Resolution refutation} of an unsatisfiable CNF formula $\varphi = C_1 \land \dots \land C_m$ over variables~$x_1, \dots, x_n$ is a sequence $D_1, \dots, D_s$ of clauses over $x_1, \dots, x_n$ such that $D_s = \bot$, denoting the empty clause, and for every $i \in [s]$, the clause $D_i$ is either (a) one of the clauses $C_1, \dots, C_m$ of $\varphi$, or (b) a \emph{weakening} of a previous clause, meaning that $D_i \supseteq D_j$ for some $1\leq j < i$, or (c) has been obtained from two previous clauses $D_j = A \lor x$ and $D_k = B \lor \neg x$, for~$j, k < i$, by an application of the \emph{Resolution rule}:
    \begin{prooftree}
        \AxiomC{$A \lor x$}
        \AxiomC{$B \lor \neg x$}
        \RightLabel{\footnotesize{\;\;(Res)}}
        \BinaryInfC{$A \lor B$}
    \end{prooftree}
    \newcommand{\w}{\operatorname{w}}
where we say that $A \lor B$ is obtained by \emph{resolving} over $x$.

The \emph{length} of $\pi$, denoted $\prooflength{\pi}$, is precisely $s$, while we denote by $|\pi|$ the length of $\pi$ as binary string, where $\prooflength{\pi} \leq |\pi|$. We write $\varphi \vdash_\Res^s \bot$ to indicate that there exists a Resolution refutation of $\varphi$ in length $s$. Resolution is \emph{implicationally complete}, meaning that for every pair of CNF formulas $\varphi$ and $\psi$, if $\varphi \models \psi$, there exists a derivation of each clause of $\psi$ from the clauses of $\varphi$. We write $\varphi \vdash_\Res \psi$ to express that such a collection of derivations exists. The \emph{width} of a clause $\w(C)$ is the number of literals in $C$, and the width of a refutation $\pi$ is naturally defined as $\w(\pi) \coloneq \max_{C\in\pi} \w(C)$. To every Resolution refutation $\pi$ we can associate a directed acyclic graph in a natural way, with the root labeled by the empty clause $\bot$. The number of vertices in this graph is exactly $\prooflength{\pi}$.

We often see a Resolution refutation as a \emph{Prover--Delayer game}
\autocite{Pudlak00, AD08}. The game is played between two players, Prover
and Delayer. For a fixed formula $\varphi$ over $n$ variables, in each
round, the Prover either (a) asks for the value of a variable of~$\varphi$ and stores the answer in memory, or (b) chooses to forget
the value of a previously queried variable. When a query is issued,
the Delayer answers by choosing a truth-value for the variable, which
is stored in memory by the Prover. The Prover wins when the partial
assignment kept in memory falsifies a clause of $\varphi$. The key
measure for us is the amount of memory of the Prover. Indeed, if a
formula $\varphi$ requires width $w$ to be refuted in Resolution,
there is a strategy for the Delayer such that by the time Prover wins,
he must have $\Omega(w)$ variables in memory. The notion is readily
generalized to other notions of width, like the idea of block-width we
use later.

\subsubsection{Frege systems}
\label{subsec:prelim-Frege}
A \emph{Frege system} \autocite{CR79} consists of a finite set of axiom schemas and inference rules that are sound and implicationally complete for the language of propositional tautologies built from the Boolean connectives negation ($\neg$), conjunction ($\land$), and disjunction ($\lor$). A \emph{Frege proof} is a sequence of formulas where each formula is obtained by either substitution of an axiom schema or by application of an inference rule on previously derived formulas. The specific choice of rules does not affect proof size up to polynomial factors, as long as there are only finitely many rules and these are sound and implicationally complete \autocite{CR79}.

A generalization of such systems are \emph{Extended Frege} ($\EF$) systems \autocite{CR79}, in which proof lines can be succinctly written as Boolean circuits rather than formulas \autocite{jerabek:phd-thesis}. We often consider extensions of Extended Frege by sets of additional axioms. For a set $A \subseteq \TAUT$ of tautologies such that $A \in \P$, the system $\EF + A$ refers to Extended Frege together with the axiom schemas that allow substitution instances of any formula in $A$. For a circuit class $\ComplexityFont{C}$, one can consider the proof system $\ComplexityFont{C}$-Frege, in which lines are restricted to be Boolean circuits of that type. We are particularly interested in the $\ACZF{d}$ systems, known as \emph{constant-depth Frege systems}, in which lines are restricted to be Boolean circuits of unbounded fan-in and depth at most $d$.

For the sake of precision, most of our formal statements about constant-depth Frege systems are stated for a specific Tait-style calculus defined below, although they could be phrased equally well in a Gentzen-style sequent calculus formalism like $\LK$ (see \autocite[§3.4]{Krajicek19}). We follow here the presentation of \citeauthor{BPT14} \autocite[§1.1]{BPT14}, but see also other sources \autocite{Buss04, BB05, BB19}.

Formulas in the $\PK$ refutation system are formed from literals together with conjunctions and disjunctions of unbounded fan-in (denoted $\bigwedge$ and $\bigvee$). In particular, negations only appear at the atomic level, and we use $\neg \varphi$ for a formula $\varphi$ as an abbreviation for the corresponding formula where negations are pushed to the atoms, interchanging connectives where needed.

For such a formula $\varphi$, its \emph{depth} is defined recursively to be $\depth(\ell) \coloneq 0$ for every literal $\ell$, and $\depth(\bigwedge_{i=1}^k\varphi_i) =\depth(\bigvee_{i=1}^k\varphi_i) \coloneq 1+ \max_{i \in [k]} \depth(\varphi_i) $. A \emph{cedent} is a sequence of formulas, separated by commas, interpreted as their disjunction.
\ifcameraready
The system $\PK$ consists of two introduction rules, an excluded middle axiom for atoms, and a weakening rule,
\[
\hbox{
  \AxiomC{$\Gamma, \varphi$}
  \AxiomC{$\Gamma, \psi$}
   \RightLabel{\footnotesize{\;\;($\land$-Intro)}}
  \BinaryInfC{$\Gamma, \varphi \land \psi$}
  \DisplayProof
  \qquad
  \AxiomC{$\Gamma, \varphi, \psi$}
   \RightLabel{\footnotesize{\;\;($\lor$-Intro)}}
  \UnaryInfC{$\Gamma, \varphi \lor \psi$}
  \DisplayProof
}
\]
\[
\hbox{
  \AxiomC{}
  \RightLabel{\footnotesize{\;\;(Atomic LEM)}}
  \UnaryInfC{$p, \neg p$}
  \DisplayProof
  \quad
  \AxiomC{$\Gamma$}
   \RightLabel{\footnotesize{\;\;(Weak)}}
  \UnaryInfC{$\Gamma, \Delta$}
  \DisplayProof
}
\]
together with a cut rule that generalizes the Resolution rule,
\[
\hbox{
    \AxiomC{$\Gamma, \varphi_1 \land \cdots \land \varphi_m$}
  \AxiomC{$\Gamma, \neg\varphi_1, \dots, \neg\varphi_m$}
   \RightLabel{\footnotesize{\;\;(Cut)}}
  \BinaryInfC{$\Gamma$}
  \DisplayProof
}   
\]
\else
The system $\PK$ consists of two introduction rules and an excluded middle axiom for atoms,
\[
\hbox{
  \AxiomC{$\Gamma, \varphi$}
  \AxiomC{$\Gamma, \psi$}
   \RightLabel{\footnotesize{\;\;($\land$-Intro)}}
  \BinaryInfC{$\Gamma, \varphi \land \psi$}
  \DisplayProof
  \qquad
  \AxiomC{$\Gamma, \varphi, \psi$}
   \RightLabel{\footnotesize{\;\;($\lor$-Intro)}}
  \UnaryInfC{$\Gamma, \varphi \lor \psi$}
  \DisplayProof
  \quad
  \AxiomC{}
  \RightLabel{\footnotesize{\;\;(Atomic LEM)}}
  \UnaryInfC{$p, \neg p$}
  \DisplayProof
}
\]
together with weakening and a cut rule that generalizes the Resolution rule,
\[
\hbox{
  \AxiomC{$\Gamma$}
   \RightLabel{\footnotesize{\;\;(Weak)}}
  \UnaryInfC{$\Gamma, \Delta$}
  \DisplayProof
  \qquad
  \AxiomC{$\Gamma, \varphi_1 \land \cdots \land \varphi_m$}
  \AxiomC{$\Gamma, \neg\varphi_1, \dots, \neg\varphi_m$}
   \RightLabel{\footnotesize{\;\;(Cut)}}
  \BinaryInfC{$\Gamma$}
  \DisplayProof
}
\]
\fi
where $\Gamma$ and $\Delta$ are possibly empty cedents. A \emph{PK-derivation} of a formula $\varphi$ from a set $\Gamma$ of disjunctions (interpreted as their conjunction), is a sequence of disjunctions ending in $\varphi$, such that each disjunction in the sequence is either in $\Gamma$, or follows from earlier disjunctions in the sequence by one of the rules above. A \emph{$\PK$-refutation} of a set of disjunctions $\Gamma$ is a $\PK$-derivation of the empty disjunction $\bot$ from $\Gamma$. We write $\PK \vdash \neg \varphi$ to indicate that $\PK$ refutes $\varphi$, when interpreted as a set of disjunctions.

For every $d \in \bbN$, we write $\PK_d$ for the subsystem of $\PK$ in which every formula in a cedent has depth at most $d$. We also define the systems $\PK_{d}^k$ for $k\in\bbN$ to be the systems in which the formulas have depth $d + 1$, but where all gates at depth $d$ have fan-in at most $k$. The system $\PK_0$ is readily seen to be precisely Resolution ($\Res$). For $k \geq 2$, the system $\PK_{0}^k$ is denoted by $\Res(k)$ and known as \emph{$k$-DNF Resolution} \autocite{Krajicek01}.


Finally, for every $d\in \bbN$, we denote by $\PK_{d+\nicefrac{1}{2}}$ the subsystem of $\PK$ in which a refutation $\pi$ has size $S$ if and only if every cedent has size at most $S$ and the refutation has depth $d+1$ but every formula at depth $d+1$ has fan-in at most $\log S$. This follows the notation $\LK_{d+\nicefrac{1}{2}}$ of Krajíček \autocite{Krajicek01}. The system $\PK_{0+\nicefrac{1}{2}}$ is denoted $\Res(\log)$. We sometimes use the notation $\ACZF{d}$ to refer to $\PK_d$.

\subsubsection{Automatability}
\label{subsec:prelim-automatability}
Following \citeauthor{BPR00} \autocite{BPR00}, for a time-constructible
function $T :\bbN \to \bbN$, we say that a proof system $S$ is
\emph{automatable in time $T$} if there exists an algorithm that given
a propositional tautology $\varphi$, outputs an $S$-proof of $\varphi$
in time $T(|\varphi| + \size{S}{\varphi})$. When $T$ is a polynomial,
we simply say $S$ is \emph{automatable}. We say that $S$ is
\emph{weakly automatable} \autocite{AB04} if there exists a proof system
$Q$ that is automatable and p-simulates $S$. This is readily seen to
be equivalent to saying that there is an algorithm automating $S$
except that proofs are output in a possibly different system $S'$. In
this case we say that $S$ is \emph{weakly automated by $S'$}.

\subsection{Bounded arithmetic}
\label{subsec:prelim-BA}
Theories of bounded arithmetic act as a uniform counterpart to propositional proof systems. Below we present the main theories used in this work.

\subsubsection{Cook's theory $\PVO$}
\label{subsec:prelim-PVO}
Cook's theory $\PVO$ \autocite{Cook75, KPT91} is a universal theory whose vocabulary $\calL_\PV$ consists of a function symbol for each polynomial-time function, and the axioms are precisely the recursive definitions of these functions via composition and limited recursion on notation, in the style of Cobham's functional definition of $\FP$ \autocite{Cobham65}. The theory further admits induction on quantifier-free formulas, which define precisely polynomial-time predicates. The formal definition of $\PVO$ is rather technical and the details are not particularly relevant to our proofs, so we refer the reader to Krajíček's textbook \autocite[Definition 5.3]{Krajicek95} for the details.

\subsubsection{Buss's theories: $\mathsf{S}^i_2$ and $\mathsf{T}^i_2$}
Let $\calL_\PA$ denote the language of {Peano Arithmetic}, $\calL_\PA \coloneq \{ 0, 1, +, \cdot, <\} $. The axioms of $\PA$ consist first of the axioms of Robinson's arithmetic $\mathsf{Q}$, which define the basic behavior of the symbols of $\calL_{\PA}$ (see, for example, \autocite[§7.4.3]{Krajicek19} for a definition), together with the \emph{Induction Schema}
\begin{equation}
    \label{axiom:ind}
    \varphi(0) \land \forall x  \ (\varphi(x) \to \varphi(x+1)) \to \forall x \ \varphi(x),
    \tag{$\textsc{Ind}_\varphi$}
\end{equation}
available for every formula $\varphi$.

Buss's theories operate over the \emph{first-order language of bounded arithmetic}, defined by the symbols $\mathcal{L}_{\text{BA}} \coloneqq \{ 0, 1, {+}, {\cdot}, {<}, {|\cdot|}, {\lfloor \cdot/2 \rfloor}, {\#} \}$. This extends the language of Peano Arithmetic $\calL_\PA$ above by the symbols $|x|$, $\lfloor x/2 \rfloor$ and $x \# y$. The notation $|x|$ denotes the length of the binary encoding of the number $x$, defined as $\lceil \log (x + 1) \rceil$, while the \emph{smash symbol} $x\#y$ stands for $2^{|x|\cdot|y|}$.

For a term $t$ in the language of bounded arithmetic and a variable $x$ that does not appear in $t$, a formula of the form  $\forall x \ (x < t \to \varphi(x))$ or $\exists x \ (x < t \land \varphi(x))$ is called a \emph{bounded formula}. The quantifiers guarded by the bounds on $x$ are called \emph{bounded quantifiers} and we simply write $\forall x < t\ \varphi(x)$ and $\exists x < t \ \varphi(x)$. If the bounded quantifiers are of the form $\forall x < |s|$ or $\exists x < |s|$ for some term $s$, then they are called \emph{sharply bounded} quantifiers. The \emph{hierarchy of bounded formulas} consists of the classes $\Sigma_i^b$ (and $\Pi_i^b$), for $i \in \bbN$, which are defined by counting the alternations of bounded quantifiers ignoring the sharply bounded ones, starting with an existential (respectively, universal) one. The class $\Delta_i^b$ consists of all formulas that admit an equivalent definition in both $\Sigma_i^b$ and $\Pi_i^b$. In particular, the class $\Delta_0^b = \formSigmab^b_0 = \formPib^b_0$ stands for all formulas with sharply bounded quantifiers only.

The theories $\BussSitwo$ and $\BussTitwo$ of Buss \autocite{Buss86} extend Robinson's arithmetic $\mathsf{Q}$ by a set $\operatorname{BASIC}$ of simple axioms for the new function symbols (see, e.g., \autocite[Definition 5.2.1]{Krajicek95} for the complete list). On top of this, the $\mathsf{S}^i_2$ theories have the \emph{Polynomial Induction} schema (\textsc{PInd}) for $\Sigma_i^b$-formulas: for every $\varphi \in \Sigma_i^b$, the theory contains the axiom
\begin{equation}
    \label{axiom:poly-ind}
    \varphi(0) \land \forall x \ (\varphi (\lfloor x/2 \rfloor) \to \varphi(x)) \to \forall  x\ \varphi(x).
    \tag{$\textsc{PInd}_\varphi$}
\end{equation}

The theory $\BussTitwo$ is defined analogously, except we allow instead the standard induction schema (\ref{axiom:ind}) for $\formSigmab^b_i$ formulas. The theory $\BussTtwo$ is defined as $\BussTtwo \coloneq \bigcup_{i \in \bbN} \BussTitwo$, and similarly $\BussStwo \coloneq \bigcup_{i \in \bbN} \BussSitwo$. It holds that $\BussTtwo = \BussStwo$, and $\PVO \subseteq \SOT$. The theories $\mathsf{S}^i_1$ and $\mathsf{T}^i_1$ are defined analogously, except for the absence of smash~($\#$).

Unlike $\calL_{\PV}$, the language $\calL_{\text{BA}}$ of bounded arithmetic does not contain a function symbol for every function in $\FP$. However, every such $f \in \FP$ is $\formSigmab^b_1$-definable in $\SOT$, meaning that there exists a $\formSigmab^b_1$ formula whose interpretation over the standard model $\bbN$ defines $f$ and such that $\SOT$ proves the totality of this definition. In the rest of the paper we abuse notation and work instead over $\BussTitwo(\calL_{\PV})$ and $\BussSitwo(\calL_{\PV})$, which are the theories in the language of bounded arithmetic extended by all $\PV$ function symbols, meaning that we have a fresh symbol for each function in $\FP$, and the corresponding induction is now available for all $\Sigma_i^b(\PV)$ formulas.

The \emph{Log-notation} $n \in \Log$ is a short-hand for the formula $\exists x \ |x|=n$. A set $X$ is called a \emph{bounded definable set with parameter $y$} if there exists an arithmetic formula $\varphi(x,y)$ and some term $t(y)$ such that $X=\{ x<t(y) \mid \varphi(x,y) \}$. We sometimes also adopt the standard set-theoretic notation denoting the interval $[0, a)$ directly by $a$. Then, a Boolean circuit $C : 2^k \to 2$ naturally defines a bounded definable set $X_C = \{ x <2^k \mid C(x) = 1\}$ with parameter $C$, and there exists a $\PV$ function $\operatorname{Count}$ which counts $\Log$-sized initial segments of circuit-definable sets; i.e., in the standard model, $\operatorname{Count}(C,a)$ is the cardinality of the set $X_C \cap [0,|a|)$. Importantly, if $2^k \in \Log$, then we can \emph{exactly count} the size of $X_C$ when working in Buss's hierarchy. This proves crucial when carrying out different combinatorial arguments.

Following standard notation (see, for example, \autocite[324-325]{HP93}), we denote by $(\cdot)_k$ the $\PV$-function that given a number $X$ in the theory, interpreted as a sequence, outputs the $k$-th element in the sequence. In this way, for $\ell \in \Log$ and $n$ not necessarily in $\Log$, we can quantify over sequences of length $\ell$ with elements from $[n]$: the sequence $X \in [n]^\ell$ is encoded by a number and we refer to the $k$-th element in the sequence by $(X)_k$. We denote by $\len(X)$ the $\PV$ function that outputs the number $\ell$ of elements in the sequence.

\subsubsection{\citeauthor{jerabek:phd-thesis}'s theories $\APC$ and $\APCtwo$}
\label{subsubsec:prelim-APC}
When a definable set is too large to count exactly, we can often still approximately count it using  \citeauthor{jerabek:phd-thesis}'s framework for approximate counting \autocite{jerabek:phd-thesis, Jerabek2007}. For a class $\ComplexityFont{C}$ of functions and $f \in \ComplexityFont{C}$, the \emph{dual weak pigeonhole principle for $f$}, denoted by $\dWPHP(f)$, is the formula stating that if $f$ is a function from $[2^n]$ to $[(1+1/n)\cdot 2^n]$, then it is not surjective. More formally, 
\ifcameraready
\begin{equation}
\tag{$\dWPHP$}
\begin{multlined}[c][0.78\displaywidth]
    \dWPHP(f) \coloneq \forall n \in\Log \ \forall z  \\ \exists y<(1+1/n)2^n \ \forall x <2^n\ f(x, z) \neq y.
\end{multlined}
\end{equation}
\else
\begin{equation}
    \tag{$\dWPHP$}
    \dWPHP(f) \coloneq \forall n \in\Log \ \forall z \ \exists y<(1+1/n)2^n \ \forall x <2^n\ f(x, z) \neq y.
\end{equation}
\fi
We write $\dWPHP(\ComplexityFont{C})$ for the set of all such sentences for $f \in \ComplexityFont{C}$.

We will overload notation and write $\PVO$ for the class $\FP$ of polynomial-time functions, and $\PV_2$ for $\FP^\NP$. We can then define the theories $\APC \coloneq \PVO + \dWPHP(\PVO)$, and $\APCtwo \coloneq \BussTonetwo + \dWPHP(\PV_2)$. It is known that $\APC \subseteq \BussTtwotwo$, and $\APCtwo \subseteq \BussTthreetwo$ \autocite{PWW88, MPW02, krajicek_mpw, JerabekHashing}.

In this framework, a map of the form $f : A \EmilInj B$ indicates that $f$ is an injection, and $f : A \EmilSurj B$ indicates that it is a surjection. For definable sets $X$ and $Y$ and $\varepsilon \leq 1$, we say that \emph{the size of $X$ is approximately less than the size of $Y$ with error $\varepsilon$}, written as $X \preceq_\varepsilon Y$, if there exists a circuit $G$ and $v \neq 0$ such that
\begin{equation}
    G : v \times (Y \EmilU \varepsilon 2^n) \EmilSurj v \times X,
\end{equation}
where $\EmilU$ denotes disjoint union.

Let us state a few useful properties of the $\preceq_\varepsilon$ relation, which we invoke later.

\begin{lemma}[\cite{Jerabek2007}, Lemmas 2.10.(ii)--(iii) and 2.11.(ii)]
\label{lemma:trivial-Emil}
Let $X,Y,Z\subseteq 2^n$ be sets definable by Boolean
circuits, let $\varepsilon,\delta\leq 1$, and let
$s,t\leq 2^n$. The following hold:
\begin{enumerate}[label=(\roman*)]
    \item The theory $\PVO$ proves that if
    $X\subseteq Y$, then $X\preceq_0 Y$;

    \item The theory $\PVO$ proves that if
    $X\preceq_\varepsilon Y$ and
    $Y\preceq_\delta Z$, then
    $X\preceq_{\varepsilon+\delta} Z$;

    \item The theory $\APC$ proves that if
    $s\preceq_\varepsilon X\preceq_\delta t$,
    then for every $\xi>0$ with
    $\xi\leq 1$ and $\xi^{-1}\in\Log$,
    \[
        s\leq t+(\varepsilon+\delta+\xi)2^n.
    \]
\end{enumerate}
\end{lemma}

Using the $\preceq_\varepsilon$ relation we can simulate union bound arguments as follows.

\begin{lemma}[{\autocite[Proposition 2.15]{Jerabek2007}}]
\label{lemma:Emil-union-bound}
    Let $b\in \{1,2\}$ and let $\{ X_i \mid i <m\}$ be a $\PV_b$-definable set of subsets of $2^n$, each of them defined by a sequence of circuits. Provably over $\mathsf{APC}_b$, for every $\varepsilon, \xi \leq 1$ with $\xi^{-1} \in \Log$ and $\{ s_i \mid i<m\}$ a sequence of numbers such that $X_i \preceq_{\varepsilon} s_i$ for every $i < m$, then
    \[ \sum_{i<m} X_i \preceq_{\varepsilon+\xi} \sum_{i <m} s_i,\]
    where the disjoint sum $\sum_{i <m} X_i \coloneq \bigcup_{i <m} (X_i \times \{ i\})\subseteq 2^n \times m$ is considered a subset of $2^{n + |m|}$. Furthermore, the same holds for $\succeq$ in place of $\preceq$. 
\end{lemma}

We also invoke several useful basic algebraic facts formalized by \citeauthor{jerabek:phd-thesis} in $\PVO$. In particular, $\PVO$ can perform basic field arithmetic over $\operatorname{GF}(q)$ as long as $q \in \Log$ (see \autocite[§4.3.1]{jerabek:phd-thesis}). Similarly, $\PVO$ can carry out basic linear algebra efficiently over such fields. In particular, for matrices represented as numbers, the rank of such matrices is computed by a $\PV$ function. We refer the reader to \autocite[§4.3.2]{jerabek:phd-thesis} for more details. Below we state a couple of useful results.

\begin{lemma}[Root upper bound in $\PVO$ {\autocite[Lemma 4.3.6]{jerabek:phd-thesis}}]
    \label{lemma:FTA-easy}
    Provably in $\PVO$, for every finite field $\bbF$ and $P$ a non-zero polynomial of degree $d$ over $\bbF$, it holds that $P$ has at most $d$ roots.
\end{lemma}

\begin{lemma}[Gaussian elimination in $\PVO$ {\autocite[Theorem 4.3.19]{jerabek:phd-thesis}}]
\label{thm:Gaussian-elim-PV}
There is a $\PV$-function $\operatorname{Sol}$ that provably in $\PVO$ computes a solution to a system of linear equations over $\bbF(x)$ if one exists, and a $\PV$-function $\operatorname{Basis}$ that provably in $\PVO$ computes a basis for the space of all solutions of a homogeneous linear system over $\bbF(x)$, where $\bbF$ is a finite field.
\end{lemma}

\subsubsection{Relativized theories and the 
Paris--Wilkie translation}
\label{subsubsec:prelim-Paris-Wilkie}
We can extend the language $\mathcal{L}_\text{BA}$ of bounded arithmetic by any countable number of relation symbols of any finite arity. We denote such a language by $\mathcal{L}_\text{BA}(\alpha)$ where the symbols added are unspecified, and write, for instance, $\mathcal{L}_\text{BA}(R, S)$, when two explicit symbols $R$ and $S$ are being added. The theories $\BussSitwo(\alpha)$ and $\BussTitwo(\alpha)$ are the natural extensions of $\BussSitwo$ and $\BussTitwo$ where the corresponding induction schema is allowed for relativized formulas.

By relativizing the theories we obtain a crucial connection to constant-depth Frege systems. We follow here the presentation of \citeauthor{Krajicek19} \autocite[§8.2]{Krajicek19} of the \emph{Paris--Wilkie translation} \autocite{PW85}. Let $\phi(x_1, \dots, x_k)$ be a $\formSigmab^b_i(R)$-formula in the language $\mathcal{L}_{\text{BA}}(R)$, where $R$ is a new relation symbol. (For simplicity, we assume here that $R$ is a binary relation symbol, but it is straightforward to generalize the definition to any number of relation symbols of finite arity.) For every such $\formPib^b_i(R) \cup\formSigmab^b_i(R)$-formula $\phi(x_1, \dots, x_k)$ and natural numbers $\bar{n} = n_1, \dots, n_k \in \bbN$, the translation produces the formula $\PW{\phi}_{\bar{n}}$ in negation normal form over the variables $p_{i, j}$ for $i, j \in [\max\{n_1, \dots, n_k\}]$, defined recursively as follows.

\begin{itemize}
    \item If $\phi$ is an atomic formula like $t(\bar{x}) = s(\bar{x})$ or $t(\bar{x}) \leq s(\bar{x})$, where $s$ and $t$ are terms, then $\PW{\phi}_{\bar{n}} \coloneq \top$ if $\bbN \models \phi(\bar{n})$, and $\bot$ otherwise.
    \item If $\phi$ is the atomic formula $R(t(\bar{x}), s(\bar{x}))$ for terms $s$ and $t$ and $i$ and $j$ are, respectively, the values of $t(\bar{n})$ and $s(\bar{n})$ over $\bbN$, then we define  $\PW{\phi}_{\bar{n}} \coloneq p_{i, j}$.
    \item If $\phi = \neg\psi$, then $\PW{\phi}_{\bar{n}} \coloneq \neg\PW{\psi}_{\bar{n}}$, with negations propagated to the atoms.
    \item If $\phi = \psi \lor \chi$, then $\PW{\phi}_{\bar{n}} \coloneq \PW{\psi}_{\bar{n}} \lor \PW{\chi}_{\bar{n}}$, and if $\phi = \psi \land \chi$, then $\PW{\phi}_{\bar{n}} \coloneq \PW{\psi}_{\bar{n}} \land \PW{\chi}_{\bar{n}}$.
    \item If $\phi(\bar{x}) = \forall y\leq t(\bar{x}) \ \psi(\bar{x}, y)$, then $ \PW{\phi}_{\bar{n}} \coloneq \bigwedge_{m\leq t(\bar{n})} \PW{\psi(\bar{x}, y)}_{\bar{n},m}$.
    \item If $\phi(\bar{x}) = \exists y\leq t(\bar{x}) \  \psi(\bar{x}, y)$, then $\PW{\phi}_{\bar{n}} \coloneq \bigvee_{m\leq t(\bar{n})} \PW{\psi(\bar{x}, y)}_{\bar{n},m}$.
\end{itemize}

The connection we later use can be stated as follows, and goes back to the seminal work of Paris and Wilkie \autocite{PW85}. The version we use for translating from Buss's hierarchy can be traced back to \citeauthor{Krajicek01} \autocites[§5]{Krajicek94}[Corollary 9.1.4]{Krajicek95}. See also \autocite[Lemma 10.5.2 and §10.6.1]{Krajicek19}.


\begin{theorem}[The Paris--Wilkie translation for $\BussTitwo(\alpha)$]
    \label{thm:Paris-Wilkie-translation}
    Let $d \geq 1$, let $\phi(x) = \exists y \leq x \ \forall z\leq x \ \psi(x, y, z)$ be a $\formSigmab^b_{d+1}(\alpha)$-formula, where $\psi$ is $\formSigmab^b_{d-1}(\alpha)$. If $\mathsf{T}_2^{d}(\alpha) \vdash \forall x \ \phi(x)$, then there are $\QP$-uniform $\PK_{(d-1) + \nicefrac{1}{2}}$-refutations of the sets  
    \ifcameraready
    \[ \begin{split}
        \{ [     \neg \PW{\psi(x,y, 1)}_{n, 1}, \dots, \neg \PW{\psi(x,y, 1)}_{n, n}      ],  \dots, \\ \quad [     \neg \PW{\psi(x,y, n)}_{n, 1}, \dots,  \neg \PW{\psi(x,y, n)}_{n, n}      ] \}_{n \in \bbN} ,
    \end{split}\]
    \else
    \[
\left\{
\left[
\neg\langle\psi\rangle_{n,a,0},
\ldots,
\neg\langle\psi\rangle_{n,a,n}
\right]
\;\middle|\;
0\leq a\leq n
\right\}_{n\in\mathbb N},
\]
    \fi
    in size $\quasipoly(n)$, where each $[\cdots]$ denotes a cedent of propositional
translations of $\neg\psi(x,y,z)$, with $x=n$ and
$y=a$ fixed, and $z$ ranging from $0$ to $n$. Furthermore, if the theory is $\BussToneone(\alpha)$ and the formula is $\forall \formSigmab^b_1(\alpha)$, the translation produces polynomial-size Resolution refutations, and if the theory is $\BussTtwotwo(\alpha)$ and the formula is $\forall \formSigmab^b_1(\alpha)$, the translation gives quasipolynomial-size $\Res(\log)$ refutations.
\end{theorem}

\subsection{The  \texorpdfstring{$\Reff$}{Ref} formulas and the Proof Analysis Problem}
\label{subsec:prelim-Reff}
Given a propositional CNF formula $\varphi$ and a size parameter $s \in \bbN$, we define a CNF formula $\Reff_s(\varphi)$ stating that $\varphi$ has a Resolution refutation consisting of at most $s+1$ clauses. Several alternative encodings of these formulas have appeared in the literature (see \autocite[§2.4]{PAP} for a discussion on the different variants), but we stick here to the natural and robust unary encoding used in \autocite{AM20} and \autocite{PAP}. The only change we introduce is that we index the blocks up to $s+1$, with $s+1$ corresponding to the root block, while in \autocite{PAP} the blocks were numbered up to $s$. This change simplifies bookkeeping in some later proofs.

The formula $\Reff_s(\varphi)$ consists of $s+1$ so-called \emph{blocks} of variables, each representing a clause in the purported Resolution refutation. More precisely, blocks indexed from 1 to $s$ represent clauses in the refutation, and block $s+1$ represents the empty clause at the end of the refutation. Each block has variables to represent the literals that appear in this block, how it was obtained (either resolved or weakened from an axiom), and it contains \emph{pointer variables} to indicate from which blocks it was derived. Crucially, the formula includes \emph{enabling variables} used to deactivate blocks.

\myparagraph{The variables of $\Reff_s(\varphi)$}
\label{subsec:vars-of-reff}
We assume $\varphi$ is a CNF formula over $n$ variables $x_1, \dots, x_n$ and $m$ clauses, and let $\Lit_n \coloneq \{x_1, \dots, x_n, \neg x_1, \dots, \neg x_n\}$. We define the following variables with their corresponding meaning,
\ifcameraready
\begin{itemize}
    \item $\alit^A_\ell$: literal $\ell \in \Lit_n$ is present in the clause $A \in [m]$ of $\varphi$;
    \item $\enable^B$: block $B \in [s+1]$ is enabled;
    \item $\derived^B$: block $B\in [s+1]$ is obtained by a Resolution step; 
    \item $\weak^B_A$: block $B\in[s+1]$ is obtained by weakening from clause $A \in [m]$ of $\varphi$;
    \item $\lit^B_\ell$: literal $\ell\in\Lit_n$ is present in the block $B\in[s+1]$;
    \item $\res^B_{x_i}$: block $B\in[s]$ is obtained by resolving over the variable $x_i$; 
    \item $\lpoint^B_{B'}$: block $B \in [s+1]$ is resolved on the left from block $B' \in [s+1]$, $B'<B$;
    \item $\rpoint^B_{B'}$: block $B\in[s+1]$ is resolved on the right from block $B' \in [s+1]$, $B'<B$.
\end{itemize}
\else

\bigskip
\begin{tabular}{lll}
    $\alit^A_\ell$ & : & literal $\ell \in \Lit_n$ is present in the clause $A \in [m]$ of $\varphi$; \\
    $\enable^B$ & : & block $B \in [s+1]$ is enabled; \\
    $\derived^B$ & : & block $B\in [s+1]$ is obtained by a Resolution step;\\ 
    $\weak^B_A$ & : & block $B\in[s+1]$ is obtained by weakening from clause $A \in [m]$ of $\varphi$; \\
    $\lit^B_\ell$ & : & literal $\ell\in\Lit_n$ is present in the block $B\in[s+1]$; \\
    $\res^B_{x_i}$ & : & block $B\in[s+1]$ is obtained by resolving over the variable $x_i$; \\
    $\lpoint^B_{B'}$ & : & block $B \in [s+1]$ is resolved on the left from block $B' \in [s+1]$, $B'<B$; \\
    $\rpoint^B_{B'}$ & : & block $B\in[s+1]$ is resolved on the right from block $B' \in [s+1]$, $B'<B$.
\end{tabular}
\bigskip
\fi

Building on these variables, the $\Reff_s(\varphi)$ formula is defined as follows. We write the clauses as implications for the sake of readability.

\begin{definition}[The $\Reff$ formulas]
\label{def:Reff-formulas}
Let $n, m, s \in \bbN$, and let $\Lit_n \coloneq \{x_1, \dots, x_n, \neg x_1, \dots, \neg x_n\}$ denote the set of possible literals over $n$ variables. The $\Reff_s(\varphi)$ formula is built from the variables defined above, together with the conjunction of the following clauses:

\ifcameraready
{\allowdisplaybreaks
    \begin{axiomdef}{1}{Ref}
        &\text{ For every }B, B'\in[s+1], B' < B, i\in[n], \ell \in \Lit_n, \ell \neq x_i, \notag \\ 
        &\quad \left( \enable^B \land \res^B_{x_i} \land \lpoint^B_{B'} \land \lit^{B'}_\ell \right) \to \lit^{B}_\ell;\label{axiom:must-appear-after-res-left} \\
        &\text{ for every }B, B'\in[s+1], B' < B, i\in[n], \ell \in \Lit_n \ell \neq \neg x_i, \notag \\
        &\quad\left( \enable^B \land \res^B_{x_i} \land \rpoint^B_{B'} \land \lit^{B'}_\ell \right) \to \lit^{B}_\ell;\label{axiom:must-appear-after-res-right}\\
        &\text{ for every } B\in[s+1], A\in[m], \ell\in \Lit_n, \notag \\
        &\quad \left( \enable^B \land \weak^B_A \land \alit^A_\ell\right) \to \lit^B_\ell; \label{axiom:must-appear-after-weak}\\
        &\text{ for every }B \in [s+1], \notag \\
        &\quad \left( \enable^B \land \derived^B \right ) \to \bigvee_{i\in[n]} \res^B_{x_i};\label{axiom:must-resolve} \\
        &\text{ for every }B \in [s+1], \notag \\
        &\quad \left( \enable^B \land \derived^B \right ) \to \bigvee_{\substack{B' \in [s+1] \\ B' < B}} \lpoint^B_{B'}; \label{axiom:must-point-left} \\
        &\text{ for }B \in [s+1],\notag \\
        &\quad \left( \enable^B \land \derived^B \right ) \to \bigvee_{\substack{B' \in [s+1] \\ B' < B}} \rpoint^B_{B'}; \label{axiom:must-point-right} \\
        &\text{ for every }B \in [s+1], \notag \\
        &\quad \left( \enable^B \land  \neg \derived^B \right) \to \bigvee_{A\in[m]} \weak^B_A;\label{axiom:must-weaken} \\
        &\text{ for every}B, B' \in [s+1], B' < B, \notag \\
        &\quad \left( \enable^B \land  \lpoint^B_{B'} \right) \to \enable^{B'}; \label{axiom:must-enable-left} \\
        &\text{ for every }B, B' \in [s+1], B' < B, \notag \\
        &\quad \left( \enable^B \land  \rpoint^B_{B'} \right) \to \enable^{B'}; \label{axiom:must-enable-right} \\
        &\text{ for every }\ell \in \Lit_n, \notag \\
        &\quad \neg \lit^{s+1}_\ell; \label{axion:root-empty}\\
        &\quad\enable^{s+1}. \label{axiom:root-enabled}
    \end{axiomdef}
}
\else
{\allowdisplaybreaks
    \begin{axiomdef}{2}{Ref}
        & \left( \enable^B \land \res^B_{x_i} \land \lpoint^B_{B'} \land \lit^{B'}_\ell \right) \to \lit^{B}_\ell  &&\quad\text{ for }B, B'\in[s+1], B' < B, i\in[n], \ell \in \Lit_n \setminus \{ x_i\},\label{axiom:must-appear-after-res-left} \\
        & \left( \enable^B \land \res^B_{x_i} \land \rpoint^B_{B'} \land \lit^{B'}_\ell \right) \to \lit^{B}_\ell  &&\quad\text{ for }B, B'\in[s+1], B' < B, i\in[n], \ell \in \Lit_n \setminus \{\neg x_i\},\label{axiom:must-appear-after-res-right}\\
        & \left( \enable^B \land \weak^B_A \land \alit^A_\ell\right) \to \lit^B_\ell  &&\quad\text{ for }B\in[s+1], A\in[m], \ell\in \Lit_n,\label{axiom:must-appear-after-weak}\\
        & \left( \enable^B \land \derived^B \right ) \to \bigvee_{i\in[n]} \res^B_{x_i}  &&\quad\text{ for }B \in [s+1],\label{axiom:must-resolve} \\
        & \left( \enable^B \land \derived^B \right ) \to \bigvee_{\substack{B' \in [s+1] \\ B' < B}} \lpoint^B_{B'}  &&\quad\text{ for 
        }B \in [s+1],\label{axiom:must-point-left} \\
        & \left( \enable^B \land \derived^B \right ) \to \bigvee_{\substack{B' \in [s+1] \\ B' < B}} \rpoint^B_{B'}  &&\quad\text{ for }B \in [s+1],\label{axiom:must-point-right} \\
        & \left( \enable^B \land  \neg \derived^B \right) \to \bigvee_{A\in[m]} \weak^B_A  &&\quad\text{ for }B \in [s+1],\label{axiom:must-weaken} \\
        & \left( \enable^B \land  \lpoint^B_{B'} \right) \to \enable^{B'}  &&\quad\text{ for }B, B' \in [s+1], B' < B\label{axiom:must-enable-left} \\
        & \left( \enable^B \land  \rpoint^B_{B'} \right) \to \enable^{B'}  &&\quad\text{ for }B, B' \in [s+1], B' < B\label{axiom:must-enable-right} \\
        & \neg \lit^{s+1}_\ell &&\quad\text{ for }\ell \in \Lit_n ,\label{axion:root-empty}\\
        & \enable^{s+1}. && \label{axiom:root-enabled}
    \end{axiomdef}
}
\fi
We also consider a variant of $\Reff_s(\varphi)$ where the $\enable^B$ variables do not appear. This will be denoted by $\ReffZero_s(\varphi)$, and is simply $\Reff_s(\varphi)$ after setting $\enable^B \mapsto 1$ for every $B \in [s+1]$.
\end{definition}

\begin{remark}[Number of variables]
    \label{rem:num_of_vars}
    The formula $\Reff_s(\varphi)$ is defined over $N = \Theta(s^2 + sm + sn + mn)$ variables and $M=\Theta(s^2n^2 + smn)$ clauses. For the case when the $\alit$ variables are restricted to encode a $k$-CNF formula over $n$ variables and $s = n^c$ for some constant $c\geq 1$, we have $m = O(n^k)$ and $N = O(n^{\max\{ {2c}, {c+k} \}})$.
\end{remark}

The following remark will be useful for the new extraction algorithm.

\begin{remark}[Simulating weakenings]
    \label{rem:simulate-weakening}
    The $\weak$ variables only allow weakening from clauses of $\varphi$. To encode that a block $B$ is a weakening of some other block $B'$, one can simply set $\lpoint^B_{B'} \mapsto 1$,  $\rpoint^B_{B'} \mapsto 1$, and $\lpoint^B_{C} \mapsto 0$ and $\rpoint^B_{C} \mapsto 0$ for every $C \neq B'$. Note that the clauses of $\Reff$ do not impose that the left and right pointers need to be different, so this is a valid assignment. One can then fix $\res^B_{x_i} \mapsto 1$ for some $i \in [n]$ and $\res^B_{x_j} \mapsto 0$ for every $j \in [n] \setminus \{i\}$, and note that even when we say the block was obtained by resolving over $x_i$, both literals $x_i$ and $\neg x_i$ may still appear in $B$.
\end{remark}

\myparagraph{Blocks and block-width}
We say that a variable of $\Reff_s(\varphi)$ \emph{belongs to block $B$} if it is indexed by $B$ as a superscript. We also say that the variable \emph{mentions $B$}. An important measure for us will be the \emph{block-width} of a given clause $C$ over the variables of $\Reff_s(\varphi)$. This is defined as the number of different blocks, without counting the root block, mentioned by the variables of the literals in $C$. We denote this measure by $\bw(C)$, and note that for every clause $C$, $\bw(C) \leq s+1$. We generalize the measure to refutations by taking $\bw(\pi)$ to be the maximum block-width over all the clauses in $\pi$. We also generalize this measure to partial assignments over the variables of $\Reff_s(\varphi)$ in the natural way.

\myparagraph{Encoding as first-order formulas}
We will later work in the relativized theories $\BussTitwo(\Phi,\Pi)$, where the language is the language of bounded arithmetic expanded by two new relation symbols $\Phi$ and $\Pi$. These relation symbols are intended to encode, respectively, a CNF formula $\varphi$ over $n$ variables and $m$ clauses and a Resolution refutation $\pi$ of the formula $\Reff_s(\varphi)$. The proof $\pi$ is encoded with the variables of the formula $\Reff_t(\Reff_s(\varphi))$. When accessing the bits of $\Pi$, we fix an encoding as follows.

There are eight types of variables in $\Reff$, and each of them is indexed by up to two values. Let us fix an encoding by mapping $1, 2, \dots, 8$ to $\alit, \enable, \derived, \dots, \rpoint$. We also fix that the literals $x_1, \dots, x_n, \neg x_1, \dots, \neg x_n$ are represented by the numbers $1, \dots, n, n+1, \dots, 2n$.  Then, the value of $\Pi(1, A, \ell)$ for $A \in \{ 1, \dots, m\}$ and $\ell \in \{ 1, \dots, 2n \}$ indicates the value of the variable $\alit^A_\ell$. Similarly, $\Pi(3, B)$ for $B\in \{1, \dots, s+1\}$ indicates the value of $\derived^B$, and so on. Hence, in what follows, we abuse notation and write directly the names of the propositional variables instead of the exact indexing of $\Pi$. That is, we write $\Pi(\derived^B)$ instead of $\Pi(3, B)$, and so on.

The formula $\varphi$ is similarly encoded by a new relation symbol $\Phi$
\ifcameraready
\else
(see \Cref{subsec:prelim-sat-reflection} below for details),
\fi
such that $\Phi(\ell, k)$ indicates whether literal $\ell$ is present in clause $k$. From these we can define first a formula $\ResRef(\Pi, \Phi, n,m, t)$ to be the $\formPib^b_2(\Phi,\Pi)$-formula verifying that $\Pi$ encodes a Resolution refutation of the formula $\Phi$ in at most $t$ clauses. It consists simply of the conjunction of the first-order version of the clauses in \Cref{def:Reff-formulas}.

The additional layer of complication comes from considering Resolution refutations of $\Reff_s(\varphi)$. To encode that $\Pi$ is a Resolution refutation of $\Reff_s(\varphi)$ rather than of $\Phi$, we first define a formula $\foReff(\Phi, n, m,  s; \ell, k)$, where $\Phi, n ,m, s$ are the parameters of $\Reff_s(\varphi)$, and for a clause $k \in [M(n,m,s)]$ and a literal $\ell \in [2N(n,m,s)]$, the formula indicates whether the literal is present in that given clause. Observe how this formula does not involve any new quantifiers. We can now consider the composed formula
\ifcameraready
\begin{equation}
    \!\!\!\ResRef(\Pi, \foReff(\Phi, n, m, s), N(n,m,s), M(n,m,s), t)
\end{equation}
\else
\begin{equation}
    \ResRef(\Pi, \foReff(\Phi, n, m, s), N(n,m,s), M(n,m,s), t)
\end{equation}
\fi
stating that $\Pi$ is a Resolution refutation of $\Reff_s(\varphi)$ in at most $t$ clauses. The latter is still $\formPib^b_2(\Phi, \Pi)$, since $\foReff$ did not increase the complexity. We note that for every fixed $\varphi$, the formula $\PW{\ResRef}_{n,m,s,t}$ obtained from the Paris--Wilkie translation on this formula is precisely $\Reff_t(\Reff_s(\varphi))$.

\begin{remark}[Complexity in the exponential-size regime]
    \label{rem:ref-in-exp-regime}
    If we assume $n, m, s\in \Log$, then the $\ResRef$ formula in the definition above becomes $\formPib^b_1(\Phi,\Pi)$. It is easy to see that the big disjunctions in the clauses of $\Reff$, which now correspond to existential quantifiers, will be sharply bounded if $n$, $m$ and $s$ are in $\Log$.
\end{remark}

\ifcameraready
\else
\subsubsection{The $\Satf$ formula}
\label{subsec:prelim-sat-reflection}
We also have the $\Satf(\varphi, \alpha)$ formula, encoding that a CNF formula $\varphi$ is satisfied by an assignment $\alpha$. The variables we consider are

\bigskip

\begin{tabular}{lll}
    $\alpha_i$ & : & value assigned by $\alpha$ to variable $x_i$; \\
    $\alit^A_\ell$ & : & literal $\ell \in \Lit_n$ is present in clause $A \in [m]$; \\
    $\sat^A_\ell$ & : & clause $A \in [m]$ is satisfied because literal $\ell \in \Lit_n$ evaluates to $1$ under $\alpha$.
\end{tabular}
\medskip

\begin{definition}[The $\Satf$ formulas]
\label{def:satf}
Let $n, m \in \bbN$, let $\varphi$ denote the set of variables of the form $\alit^A_\ell$ as above and let $\alpha$  denote the set of variables $\alpha_i$ as above. The formula $\Satf(\varphi, \alpha)$ is the CNF formula over the variables in $\varphi$, $\alpha$ and additionally all the variables $\sat^A_\ell$ above consisting of the conjunction of the following clauses,
{\allowdisplaybreaks
\begin{axiomdef}{2}{Sat}
        & \neg\sat^A_\ell \lor \alit_\ell^A &&\quad\text{ for all }A\in[m]\text{ and }\ell\in \{x_1, \dots, x_n, \neg x_1, \dots, \neg x_n\},\label{sat-1}\\
        &\neg \sat^A_{x_i} \lor \alpha_i &&\quad\text{ for all }A \in [m]\text{ and } i \in [n],\label{sat-2}\\
        &\neg \sat^A_{\neg x_i} \lor \neg \alpha_i &&\quad\text{ for all }A \in [m]\text{ and } i \in [n],\label{sat-3}\\
        & \bigvee_{i \in [n]} \sat^A_{x_i} \lor \sat^A_{\neg x_i} &&\quad\text{ for all }A\in[m].\label{sat-4}
    \end{axiomdef}}
\end{definition}

Resolution can convert between the $\Satf$ encoding and the \say{native} encoding of a formula, in the sense that for every CNF formula $\varphi$, if $\varphi$ has a Resolution refutation of length $s$, then $\Satf_{\restriction\varphi}$ has a Resolution refutation of length $O(s+nm)$, and if $\Satf_{\restriction\varphi}$ has a Resolution refutation of length $s$, then $\varphi$ has a Resolution refutation of length $O(s)$ \autocite[Proposition 2.9]{PAP}.

\myparagraph{Encoding as first-order formulas}
Just like with the encoding of Resolution refutations into a predicate $\Pi$, in the context of first-order languages, we usually assume an additional relation symbol $\Phi$ intended to encode a formula $\varphi$, together with a predicate $\alpha$ encoding the assignment. If the intended variables are $x_1, \dots, x_n$ and the formula has $m$ clauses, then we fix again the literals $x_1, \dots, x_n, \neg x_1, \dots, \neg x_n$ to be encoded as $1, \dots, n, n+1, \dots, 2n$, we interpret $\Phi(\ell,k)$ to be true if and only if literal $\ell \in [2n]$ appears in clause $k \in [m]$.

The natural encoding of the satisfiability formula is then the $\formPib^b_2(\Phi,\alpha)$-formula defined as
\newcommand{\Satfo}{\operatorname{Sat}}
\newcommand{\SatfoThree}{\operatorname{Sat}_3}
\begin{equation}
    \label{form:Sat-FO}
    \Satfo(\Phi, n, m) \coloneq \forall k \in [m] \ \exists \ell \in [2n] \ \left( \Phi(\ell, k) \land ( (\ell \in[n] \land \alpha(\ell)) \lor (\ell > n \land \neg \alpha(\ell - n))) \right).
    \tag{$\Satfo$}
\end{equation}
Although the Paris--Wilkie translation of $\Satfo$ is not
syntactically identical to the propositional formula $\Satf$ defined earlier, the two encodings are
polynomially equivalent for constant-depth Frege refutations,
up to a fixed additive increase in depth. Indeed, $\Satf$
implies the translation of $\Satfo$ by polynomial-size
constant-depth Frege proofs, while substituting
$\sat^A_\ell$ by the conjunction expressing that $\ell$
occurs in $A$ and is satisfied by $\alpha$ reduces
$\Satf$ to the translation of $\Satfo$. This difference is inessential for our applications.

\fi

\subsubsection{The Proof Analysis Problem}
\label{subsec:prelim-PAP}
We recall the definition of the \emph{Proof Analysis Problem} \autocite[Definition 3.1]{PAP}. Let $Q$ be a propositional proof system. The \emph{Proof Analysis Problem for $Q$} is the language
    \[ \PAP_Q \coloneq \{ (\varphi, \pi, 1^s) \mid \varphi \in \SAT \text{ and } \pi : {Q \vdash \neg \Reff_s(\varphi)} \},\]
where $\Reff_s(\varphi)$ is the formula defined in \Cref{subsec:prelim-Reff}. We denote by $\PAP_Q[s(n)]$ the problem where the size parameter $s$ is restricted to be at least $s(n)$ for $n$ the number of variables of $\varphi$. The problem asks, given the proof of a Resolution lower bound in a fixed proof system $Q$, to decide whether the underlying formula is satisfiable or not. We say that a propositional proof system $Q$ is \emph{analyzable} if there exists some constant $c > 0$ such that $\PAP_Q[n^c] \in \P$. If a propositional proof system $S \geq \Res$ is both automatable and analyzable, then $\P = \NP$ \autocite[Proposition 3.6]{PAP}.

We will denote by $\Extraction_Q$ the \emph{search version of the Proof Analysis Problem for $Q$} \autocite[Definition 3.5]{PAP}, defined as follows: given a CNF formula $\varphi$, a size parameter $s$ in unary and a proof $\pi$ such that $\pi : Q \vdash \neg \Reff_s(\varphi)$, output either a satisfying assignment for $\varphi$, if $\varphi \in \SAT$, or $0$ otherwise. Similarly to $\PAP_Q[s(n)]$, we define $\Extraction_Q[s(n)]$ to be the search problem where the size parameter is at least $s(n)$. (Note that in \autocite{PAP}, this search problem was denoted $\FPAP_Q$.)

\section{The new assignment extraction algorithm}
\label{sec:new-algorithm}
In this section we introduce the new assignment extraction algorithm
and show how to implement the procedure by polynomial-size
$\TC^0$-circuits and quasipolynomial-size $\ACZ$-circuits. All of the
proofs in this section are carried out in natural language, so for now
we are not concerned with the feasibility of the arguments.

In the introduction (\Cref{subsubsec:intro-techniques-new-algo}) we
discussed the existence of two algorithms: one in quasipolynomial-size
$\ACZ$ and another one in polynomial-size $\TC^0$. These are actually
two distinct algorithms, with somewhat different approaches. However,
from the $\TC^0$ algorithm alone, one can also recover a
quasipolynomial-size $\ACZ$ extraction circuit just by carefully
removing the threshold gates with a quasipolynomial increase in
size. This is the approach we take in this section.

\Cref{subsec:old-algorithm} revisits the Prover--Delayer game used for
the original assignment extraction algorithm of
\citeauthor{PAP}~\autocite{PAP} together with the setup and terminology
needed to reason about it. Crucially, this section introduces a small
modification to the algorithm that proves useful to reason about the
new algorithm. The latter is presented in \Cref{subsec:new-algo}.

Throughout the rest of the section,  we fix $\varphi$ to be a CNF formula over $n$ variables and $m$ clauses, and let $s \in \bbN$ be an integer parameter used in the formulas $\EReff_s(\varphi)$ and $\Reff_s(\varphi)$. Recall that $\EReff$, as defined in \Cref{def:Reff-formulas}, is just the version of $\Reff$ in which all $\enable$ variables are assigned to $1$.

\subsection{The Prover--Delayer game, revisited}
\label{subsec:old-algorithm}
We start by revisiting the original algorithm by \citeauthor{PAP}
\autocite{PAP}. The idea there was to look at the block-width lower bound
of Atserias and Müller~\autocite{AM20} as a Prover--Delayer game and
consider the strategy of the Delayer that leads to the lower
bound. Then, traversing the refutation following this strategy yields
an algorithm that finds the satisfying assignment. See
\Cref{subsec:prelim-resolution} for more details on Prover--Delayer
games, and \Cref{subsubsec:intro-techniques-new-algo} for more details
on the way the original algorithm worked. Note that throughout this
section we consider the formula $\ReffZero_s(\varphi)$, where the
$\enable$ variables are not present (see \Cref{def:Reff-formulas}).

Like in the original algorithm, the Delayer keeps a \emph{reservation}
in memory, which embeds partial information about the purported
refutation into the canonical exponential-size tree-like refutation
(see \Cref{fig:prove-delayer-example}). In \autocite{AM20} and \autocite{PAP},
this was achieved by taking the non-root blocks of
$\ReffZero_s(\varphi)$ 
and arranging them on a grid with $n$ layers and $s/n$ blocks per
layer, the idea being that the $i$-th layer corresponds to the clauses
appearing at layer $i$ in the exponential-size tree-like refutation.

We now introduce a generalization of this. We arrange the~$s$ non-root
blocks of $\ReffZero_s(\varphi)$ into $\ell \coloneq n + w_0$ layers,
for a parameter $w_0 \geq 0$ to be fixed later, each containing
$\lfloor s/\ell \rfloor$ blocks (with the remainder blocks left from
the flooring operations collected all in the last layer), and one
additional layer on top with the root block. We think of the root
block as positioned at layer $0$. Here $w_0$ is a parameter to be
fixed later, and, in particular, for $w_0 = 0$, we recover the grid
that the Delayer used in the original strategy.  For two blocks
$b, c \in [s]$, we say that a partial assignment $R$ to the variables
of $\ReffZero_s(\varphi)$
\begin{itemize}
    \item \emph{fixes the derived variable of $b$} if it sets a value for $\derived^b$;
    \item \emph{fixes the resolved variable of $b$ to a variable $x_i$} if $R(\res^b_i) = 1$ and $R(\res^b_{i'}) = 0$ for all $i' \neq i$;
    \item \emph{fixes the weak pointer of $b$ to an axiom $A \in [m]$} if $R(\weak^b_A) = 1$ and $R(\weak^b_{A'}) = 0$ for all $A' \neq A$;
    \item \emph{fixes the left (respectively, right) pointer of $b$ to $c$} if it sets $R(\lpoint^b_{c}) = 1$ (respectively, $R(\rpoint^b_{c}) = 1$) and $R(\lpoint^b_{c'}) = 0$ (respectively, $R(\rpoint^b_{c'})
    = 0$) for all $c \neq c'$, and there is no $b' \neq b$ such that $R(\lpoint^{b'}_c) = 1$ or $R(\rpoint^{b'}_c) = 1$.
\end{itemize}

The Delayer strategy manipulates very structured partial assignments
fixing such pointers called \emph{reservations}. The formal definition
follows.


\begin{definition}[Reservations for $\EReff_s(\varphi)$]
  \label{def:reservation}
  Fix a set $B \subseteq [s+1]$ of blocks and a number $\ell \in \bbN$
  of layers.  A \emph{basic reservation} $R$ for blocks $B$ and $\ell$
  layers is a partial assignment for $\EReff_s(\varphi)$ fixing the
  variables of type $\lpoint, \rpoint, \weak, \derived$ and $\res$ for
  all blocks in $B$ such that the following holds. For all $b\in B$ at
  layer $i$ it holds that
  \begin{enumerate}[label=(\roman*)]
  \item if $i < \ell$, then $R$ fixes the derived variable of $b$ to
    $1$ and the resolved variable to $x_{\min\{n, i+1\}}$;
  \item if $i < n$, then the reservation $R$ fixes the left and right
    pointers of $b$ to blocks $c$ and $c'$ with $c \neq c'$ in layer
    $i + 1$ and no other block points at blocks $c, c'$;
  \item if $n \leq i < \ell$, then the reservation $R$ fixes both the
    left and right pointers to the same block $c$ at layer $i+1$, and
    no other block points at $c$;
  \item if $i=\ell$, then the reservation $R$ fixes the weak pointer
    of $b$ to some $A \in [m]$.
  \end{enumerate}
  The block-width of $R$ is 
  $\bwhat(R) = |B\setminus \{s+1\}|$. A reservation $R$
  \emph{reserves} a block $b \in [s+1]$ if either $b\in B$, or there
  is some $b_0 \in B$ such that $R$ fixes the left or right pointer of
  $b_0$ to $b$.
  
  A \emph{full reservation} (or simply \emph{reservation}, when it is
  clear from context) for blocks $B$ and $\ell$ layers is the
  extension of a basic reservation that additionally contains
  information about the literals present in the blocks: for every
  block $b\in B$ at layer $i \leq \ell$, a full reservation fixes the
  variables $\lit^b_{x_k}$ to determine that $b$ contains exactly
  $j \coloneq \min\{n, i\}$ literals over the variables
  $x_1, \dots, x_j$ and no two literals for the same variable.
\end{definition}

Observe that a reservation $R$ for $B$ reserves at most
$2\bwhat(R) = 2|B\setminus\{s+1\}|$ blocks at any layer (except perhaps at layer~$1$ where the bound is $2 + \bwhat(R)$).
The high-level strategy of the Delayer is to maintain a (full)
reservation $R$ that maps patches of the trivial exponential-size
tree-like refutation into $s$ blocks.
The block-width of $R$ is always bounded by the block-width of the
assignment kept in memory by the Prover and $R$ thus reserves only few
blocks.
This enables the Delayer to extend the reservation~$R$ upon any query
by the Prover in a consistent manner as explained next.

%

When playing over a grid with $\ell > n$ layers, the strategy for the
Delayer when queried about a block at layer
$i \in [n-1] \cup \{\ell\}$ is identical to the original strategy.
However, if queried about a block at layers $n$ through $\ell -1$, the
Delayer simply answers that the block is a \say{copy} of a block at
the layer above, that is, a trivial weakening without altering the
clause. We call this the \emph{extended strategy}.

Recall that our ultimate goal is to extract a satisfying assignment to
the CNF formula $\varphi$ provided a Resolution refutation $\pi$ of
$\EReff_s(\varphi)$.
This can be achieved by traversing~$\pi$ according to the Delayer
strategy.
In the following we thus present the Delayer strategy in the form of
an algorithm $\Delayer_\ell(\varphi, \pi, 1^s)$ attempting to extract
a satisfying assignment to $\varphi$ from a refutation $\pi$ of
$\EReff_s(\varphi)$.


\begin{boxAlgo}{Extended strategy for the delayer ($\Delayer_\ell(\varphi, \pi, 1^s)$)}{extended-delayer}
    Let $R$ denote the empty reservation. Traverse the proof dag $\pi$ starting at the root $C \coloneq \bot$ according to the following instructions.
    \begin{enumerate}

  \item If $C$ is a leaf of $\pi$, then halt and output failure.
    
  \item If $C$ is obtained by weakening a clause $C' \subseteq C$,
    then set $C \coloneq C'$ and go to Step 4.

  \item If $C$ is derived from clauses $A \lor v$ and $B \lor \neg v$
    by resolving over $v$, then attempt the following reservations.
    \begin{enumerate}
      
    \item If $v$ belongs to a block $b$ on layer $0 \leq i < n$, we have two cases.
      
      \begin{enumerate}
      \item If $b$ is not reserved in $R$, update $R$ to encode that $b$ contains the clause $\bigvee_{j =1}^i x_j$ and is derived by resolving two unreserved blocks on layer $i+1$.

      \item If $b$ is reserved in $R$ but has no children attached, then it is pointed to by a block $b_0$ in layer $i-1$. Let~$\calC$ denote the clause that $b$ should encode: if $b_0$ points at $b$ on the left and $b_0$ encodes $\calC_0$, then $\calC \coloneq \calC_0 \lor x_{i}$ and otherwise, if $b_0$ points at $b$ on the right, then $\calC \coloneq \calC_0 \lor \neg x_{i}$. Update $R$ to encode that $b$ contains the clause $\calC$ and is derived by resolving two unreserved blocks on layer $i+1$.
      \end{enumerate}
      \textbf{If this update fails} because there are not enough
      unreserved blocks on layer $i+1$, halt and output the clause $C$.
      %
      \item If $v$ belongs to a block $b$ on layer $n \leq i < \ell$, we
        distinguish two cases:
        \begin{enumerate}
        \item If $b$ is not reserved in $R$, update $R$ to encode that $b$ contains the clause $\bigvee_{j =1}^n x_j$ and fix both of its pointers to the same unreserved block on layer $i+1$.

        \item If $b$ is reserved in $R$ but has no children attached, then it is pointed to by a block $b_0$ in layer $i-1$ encoding $\calC_0$. If $i > n$, then set $\calC \coloneq \calC_0$ to be the clause that $b$ encodes, and point both of its pointers to the same unreserved block on layer $i+1$. If $i = n$, then $\calC$ is obtained by checking whether $b_0$ points at $b$ on the left or on the right, as in Step 3.(a).ii.
        \end{enumerate}
      %
        \textbf{If this update fails} because there are not enough
      unreserved blocks on layer $i+1$, halt and output the clause $C$.
      
    \item If $v$ belongs to a block $b$ on layer $\ell$, we distinguish two cases.
      
      \begin{enumerate}
      \item If $b$ is not reserved in $R$, then update $R$ so that $b$ encodes the clause $\calA \coloneq \bigvee_{j=1}^n x_j$ and it points to an axiom of $\varphi$ this clause is a weakening of.

      \item If $b$ is reserved in $R$ but has no children attached, then it is pointed to by a block $b_0$ in layer $\ell-1$ encoding $\calC_0$. If $\ell > n$, set $\calA \coloneq \calC_0$ to be the clause that $b$ encodes, find the first clause in $\varphi$
        such that $\calA$ is a weakening of it 
        and let $R$ point to this clause. If $\ell = n$, then $\calA$ is obtained by checking whether $b_0$ points to $b$ on the left or on the right, as in Step 3.(a).ii.
        
      \end{enumerate}

      \textbf{If this update fails} because no axiom could be
      found, then halt and output the assignment to the variables
      $x_1, \dots, x_n$ given by $\neg \calA$, the negation of the
      clause encoded by~ $\calA$.
    \end{enumerate}

    \textbf{If the reservation succeeds}, if $R(v) = 1$, then update
    $C \coloneq B \lor \neg v$, and otherwise update
    $C \coloneq A \lor v$.

  \item Clean up: remove any information from $R$ about blocks not mentioned in $C$. Go to Step
    1.
  \end{enumerate}
\end{boxAlgo}

The following invariant is key for the analysis of the algorithm. 

\begin{definition}[Reservation invariant and Delayer-consistency]
  \label{def:reservation-invariant}
  A full reservation $R$ satisfies the \emph{reservation
    invariant} with respect to a clause $C$ over the variables of
  $\ReffZero_s(\varphi)$ if the following conditions hold:
  \begin{enumerate}[label=(\roman*)]
  \item the reservation $R$ falsifies $C$; \label{item:inv-C}
  \item \label{item:inv-axiom} the reservation $R$ does not falsify any axiom of
    $\ReffZero_s(\varphi)$; 
  \item the reservation $R$ is over blocks mentioned in $C$, and in particular $\bwhat(R)=\bwhat(C)$. \label{item:inv-bw}
  \end{enumerate}
  A clause $C$ is \emph{Delayer consistent} if there is a full reservation that satisfies the reservation invariant with respect to $C$.
\end{definition}

The key property of the Delayer strategy is that if the reservation invariant holds at the beginning of an iteration and the algorithm successfully completes one more iteration, then the reservation invariant holds at the end of the iteration.

\begin{lemma}
  \label{lemma:invariant-one-more-step}
  Let $\pi$ be a Resolution refutation of $\ReffZero_s(\varphi)$, let
  $\ell \in \mathbb{N}$, and consider a run of
  $\Delayer_\ell(\varphi, \pi, 1^s)$.
  Consider the state of the algorithm at the beginning of
  iteration $i$: denote by $C_{i}$ the clause and by $R_{i}$ the
  reservation considered.
  If $R_{i}$ witnesses that $C_i$ is Delayer consistent, then $R_{i+1}$ witnesses that $C_{i+1}$ is Delayer consistent, provided the algorithm does not fail in
  iteration $i$.
\end{lemma}

\begin{proof}[Proof sketch]
  We proceed by induction on the number $i$ of iterations of the
  algorithm. At the first iteration $C_1 = \bot$ and $R_1$ is the
  empty reservation, which readily satisfies all the conditions. It is
  straightforward to verify that if the invariant holds at the
  beginning of some later iteration and the algorithm does not fail,
  then the reservation update was successful and the invariant is thus
  satisfied again.
\end{proof}

Recall also the following connection between failed weakenings and satisfying assignments.

\begin{fact}
    \label{fact:weakening}
    Let $\varphi$ be a CNF formula over $n$ variables. If $C$ is a non-tautological width-$n$ clause over the variables of $\varphi$ that is not the weakening of any clause of $\varphi$, then the unique assignment that falsifies $C$ is a satisfying assignment for $\varphi$.
\end{fact}

We have the following correctness result.

\begin{theorem}[Assignment extraction]
\label{thm:extraction}
Let $\varphi$ be a CNF formula over $n$ variables and $m$ clauses, let $s \in \bbN$, and consider a Resolution refutation $\pi$ of the formula $\ReffZero_s(\varphi)$. For every $\ell \geq n$, it holds that the procedure $\Delayer_\ell(\varphi, \pi, 1^s)$ in Algorithm~\ref{algorithm:extended-delayer} terminates in time $\poly(|\pi|, s, n, m, \ell)$ and outputs either
    \begin{enumerate}[label=(\alph*)]
        \item \label{case:high-bw} 
        a clause $C \in \pi$ of block-width at least $1/2( \lfloor s/\ell \rfloor -1)$, or
        \item \label{case:assign} 
        a satisfying assignment for $\varphi$.
    \end{enumerate}
\end{theorem}

\begin{proof}
    At the beginning of the algorithm, the empty reservation satisfies the reservation invariant with respect to the empty clause. Given a Resolution refutation $\pi$ we claim that the algorithm always halts before reaching a leaf. Indeed, suppose for the sake of contradiction that the algorithm terminates at a leaf $C$ of $\pi$ and denote by $R$ the reservation satisfying the reservation invariant with respect to $C$, obtained by executing the algorithm until reaching $C$. By item \ref{item:inv-axiom} of the invariant, $R$ does not falsify any axiom of $\ReffZero_s(\varphi)$. However, according to item \ref{item:inv-C} of the invariant, $R$ falsifies $C$, which is a leaf of $\pi$ and hence a weakening of an axiom of $\ReffZero_s(\varphi)$. This is a contradiction and establishes the claim.
    
    Thus the algorithm halts before reaching a leaf of $\pi$ for one of the following two reasons.
    \begin{enumerate}
        \item[(a)] The algorithm attempted to reserve the children of a block at level $1 \leq i < \ell$ by finding unreserved blocks on layer $i + 1$, but there were no free blocks left on layer $i+1$. This means that $R$ already reserved at least $\lfloor s/\ell \rfloor - 1$ blocks on layer $i+1$, so $\bwhat(R) \geq 1/2(\lfloor s/\ell \rfloor - 1)$. By \cref{item:inv-bw} of the invariant we have that $\bwhat(R) = \bwhat(C)$, so putting this together we have that when outputting $C$ we are outputting a clause of block-width at least $1/2(\lfloor s/\ell \rfloor - 1)$, as desired.
    
        \item[(b)] The reservation $R$ had a clause $\calA$ encoded in a block at layer $\ell$, but it failed to find an axiom of $\varphi$ that $\calA$ was a weakening of. Since the block is at layer $\ell$, it encodes a width-$n$ clause which is not the weakening of any clause of $\varphi$. By \Cref{fact:weakening}, $\neg \calA$ encodes a satisfying assignment of $\varphi$, as desired. \qedhere
    \end{enumerate}
\end{proof}

\subsection{The new algorithm}
\label{subsec:new-algo}
We have two parameters $s_0$ and $w_0$ to be fixed later. We arrange a
numbering of the $s$ blocks of $\Reff_s(\varphi)$ into
$\ell \coloneq n + w_0$ layers, each containing
$\lfloor s/\ell \rfloor$ blocks (with the remainder blocks left from
the flooring operations collected all in the last layer), plus one
additional layer on top with a single block corresponding to the
root. The root block lies at layer $0$. Note that the following algorithm gets as input a refutation $\pi$ of $\Reff_s(\varphi)$ and not of $\EReff_s(\varphi)$.

\begin{boxAlgo}{Efficient assignment extraction ($\Extractor_{s_0, w_0}(\varphi, \pi, 1^s)$)}{logspace}
    If the all-zeroes assignment satisfies $\varphi$,
output it and halt. If $|\pi| \geq s_0$, then perform a brute-force search to find the lexicographically-first satisfying assignment~$\alpha$ of~$\varphi$.
    Otherwise, if $|\pi| < s_0$, let $\pi = (C_1, \ldots, C_t)$ and proceed for $i=1, \ldots, t$ as follows.
    
    \begin{enumerate}
        \item Let $w \coloneq \bwhat(C_i)$, and denote by $b_1, \dots, b_w \in [s]$ the non-root blocks mentioned in $C_i$. If $w > w_0$, move to $C_{i+1}$; otherwise, continue.
        \item Consider every subset
    $B \subseteq \{ b_1, \dots, b_w\}$. For each such set $B$, consider, for every $b\in B$, the clauses~$\calC_{b} \coloneq \{ z  \in \Lit_n \mid \neg \lit^b_{z} \in C_i \text{ or }\lit^b_{\neg z} \in C_i \}$ and let~$\calA_B \coloneq \bigcup_{b \in B} \calC_b$. If $\calA_B$ is falsifiable and the unique falsifying assignment~$\alpha_B$ of $\calA_B$ satisfies $\varphi$, then extend $\alpha_B$ to an arbitrary full assignment $\alpha \supseteq \alpha_B$ and output $\alpha$.
    \end{enumerate}
\end{boxAlgo}

\subsubsection{Correctness}
\label{subsubsec:corretness-new-algorithm}

Before we can argue correctness of Algorithm~\ref{algorithm:logspace},
recall the restriction argument from \autocite{AM20}. This is the random
restriction argument that reduces the block-width of a Resolution
refutation $\pi$ to $O(\log |\pi|)$. We restate it here highlighting
the fact that the restriction $\rho$ does not only \emph{reduce} the
block-width, but actually \emph{kills} all clauses of high
block-width.
Let us say that a restriction $\rho\in\{0,1, *\}^N$ to the $N$
variables of $\Reff_s(\varphi)$ is \emph{$d$-disabling} if it
satisfies that (i) exactly $d$ blocks are disabled and the rest are
enabled, (ii) every variable belonging to a disabled block is assigned
a value, and (iii) apart from the enabling variables, no variables
belonging to enabled blocks are assigned.

\begin{lemma}[Random restriction argument {\autocites{AM20}[Lemma 4.2]{PAP}}]
  \label{lemma:restriction}
  There exists $\delta \geq 1$ such that for every Resolution
  refutation $\pi$ of the formula $\Reff_s(\varphi)$, there exists an
  $\lceil{s/2}\rceil$-disabling restriction
  $\rho \in \{0, 1, * \}^{N}$ to the $N$ variables of
  $\Reff_s(\varphi)$ such that
  \begin{enumerate}[label=(\roman*)]
  \item the restriction $\rho$ does not falsify $\Reff_s(\varphi)$,
    and
  \item for every clause $C \in \pi$, if
    $\bwhat(C) > \delta\log|\pi|$, then $C_{\restriction \rho} = 1$.
  \end{enumerate}
\end{lemma} 

\begin{proof}
  Without loss of generality, let $s$ be even, since if $s$ is odd,
  then $\pi$ can be turned into a refutation of $\Reff_{s-1}(\varphi)$
  by the restriction that disables and fully restricts one block. This
  reduces the block-width of $\pi$ by at most $1$.

  Consider the following distribution of random restrictions: pair all
  $s$ non-root blocks into $s/2$ pairs, and for each pair, with
  probability $1/2$, decide which block in the pair is going to be
  disabled. The root block indexed by $s+1$ is always enabled. Now, if
  a block is disabled, choose a value uniformly and set all of the
  remaining variables in the block to take this value. We denote by
  $\rho$ a restriction obtained in this way.

  It is clear that a restriction $\rho$ sampled in this way is exactly
  $\lceil s/2\rceil$-disabling and does not falsify
  $\Reff_s(\varphi)$.  We claim that for every Resolution derivation
  $\pi$ of $\Reff_s(\varphi)$, there exists a restriction that
  satisfies all clauses of block-width at least $\delta \log |\pi|$. Indeed, if $\ell$ is a literal corresponding to the
  variable $\enable_i$ determining whether a certain block is
  disabled, then $\Pr_\rho[\ell_{\restriction \rho} = 1] = 1/2$. For
  every other literal $\ell$ in a block~$B_i$,
   \ifcameraready
   \begin{align}
        &\Pr_\rho [ B_i \text{ is disabled and }\ell_{\restriction \rho} = 1 \ ] \\
        &\quad = \Pr_\rho [ B_i \text{ is disabled}] 
        \cdot \Pr_\rho[\ell_{\restriction \rho} = 1 \mid  B_i \text{ is disabled}] \\ 
        &\quad = \frac{1}{2} \cdot \frac{1}{2} = \frac{1}{4}\,.
    \end{align}
   \else
    \begin{align}
        \Pr_\rho [ B_i \text{ is disabled and }\ell_{\restriction
          \rho} = 1 \ ]
        &=
        \Pr_\rho [ B_i \text{ is disabled}] \cdot
        \Pr_\rho[\ell_{\restriction \rho} = 1 \mid  B_i \text{ is
          disabled}] \\
        &= \frac{1}{2} \cdot \frac{1}{2} = \frac{1}{4}\,.
    \end{align}
    \fi
    Hence, for every literal $\ell$ not from the root block, $\Pr_\rho[\ell_{\restriction \rho} = 1] \geq 1/4$.

    Now, if $C$ is a clause such that $\bwhat(C)\geq w$, hence mentioning at least $w$ non-root blocks, we have that
        \begin{equation}
        \Pr_\rho [C_{\restriction \rho} \neq 1] \leq (3/4)^{w/2}\,,    
        \end{equation}
    where the $1/2$ in the exponent comes from the fact that if two consecutive blocks are present, meaning that they were paired together and only one of them was enabled, their values depend on each other.

    Then, if $\pi$ was indeed a Resolution derivation of $\Reff_s(\varphi)$, by a union bound,
    \ifcameraready
    \begin{align}
      \Pr_{\rho}[\bwhat(\pi_{\restriction \rho}) \geq w] &\leq \prooflength{\pi} \cdot (3/4)^{w/2}
        \\ &\leq |\pi| \cdot  (3/4)^{w/2} ,
    \end{align}
    \else
    \begin{equation}
        \Pr_{\rho}[\text{there is }C\in\pi \text{ where }\bwhat(C)\geq w \text{ and } C_{\restriction \rho} \neq 1] \leq \prooflength{\pi} \cdot (3/4)^{w/2}
        \leq |\pi| \cdot (3/4)^{w/2} .
    \end{equation}
    \fi
    We want to choose $w$ such that $|\pi| \cdot (3/4)^{w/2} < 1$. This bound is met by choosing $w > 2\log |\pi|/(\log 4/3)$, which yields the value of the constant $\delta \coloneq 2/\log(4/3) \geq 1$ in the statement. 
\end{proof}

With the restriction argument at hand we are now ready to argue
correctness of Algorithm~\ref{algorithm:logspace}. We prove its
correctness by showing that if the new algorithm fails, then the
Delayer is able to play forever following the strategy outlined in
Algorithm~\ref{algorithm:extended-delayer}. This contradicts the fact
that the algorithm always terminates.

\begin{theorem}[Correctness of Algorithm~\ref{algorithm:logspace}]
  \label{thm:correctness-logspace}
  There exist $s_0 = s_0(s)$, $w_0= w_0(\pi)$, and $n_0 \in \bbN$ such
  that for every satisfiable CNF formula $\varphi$ over $n \geq n_0$
  variables, $s\geq n^2$, and every Resolution refutation $\pi$ of
  $\Reff_s(\varphi)$, the procedure
  $\Extractor_{s_0, w_0}(\varphi, \pi, 1^s)$ in
  Algorithm~\ref{algorithm:logspace} outputs a satisfying assignment
  of $\varphi$.
\end{theorem}

\begin{proof}
  Fix $w_0 = w_0(\pi) \coloneq \lceil \delta \log |\pi|\rceil$, where
  $\delta$ is the constant given by Lemma~\ref{lemma:restriction}. Fix
  $\varepsilon \coloneq 1/8\delta$ and
  $s_0 = s_0(s) \coloneq 2^{\varepsilon\sqrt{s}}\ge 2^{\varepsilon n}$.  In the case of the
  brute-force search at the beginning of
  Algorithm~\ref{algorithm:logspace}, correctness is immediate, so we
  focus on the case $|\pi|<s_0$.

  Suppose for contradiction that there exist $\pi$ and $\varphi$ such
  that Algorithm~\ref{algorithm:logspace} iterates through all the
  clauses of $\pi$ without ever finding a satisfying assignment. In
  the following we show that this implies that the original algorithm
  (Algorithm~\ref{algorithm:extended-delayer}) does not always
  terminate thereby contradicting \Cref{thm:extraction}.

  Let $\rho$ denote the restriction as guaranteed to exist by
  Lemma~\ref{lemma:restriction} and consider
  $\pi_{\restriction\rho}$. By Lemma~\ref{lemma:restriction} it holds
  that $\bwhat(\pi_{\restriction\rho}) \leq \delta\log|\pi| \leq w_0$.
  Fix $\ell \coloneq n+4w_0$. Since $|\pi|<s_0$ and
  $\varepsilon=1/8\delta$ it holds that
  $\delta\log|\pi| < \delta\log s_0 = \delta\varepsilon\sqrt{s} = \frac{\sqrt{s}}8$,
  and therefore $w_0<\frac{\sqrt{s}}8+1$.
  Since $s\geq n^2$, we have $n\leq\sqrt{s}$, and hence
  \begin{equation}
    \ell
    <
    \sqrt{s}
    +4\left(\frac{\sqrt{s}}8+1\right)
    =
    \frac32\sqrt{s}+4
    \leq
    \frac53\sqrt{s}\,,
  \end{equation}
  where the last inequality holds for sufficiently large $s$.

  Up to a renaming of blocks, $\pi_{\restriction\rho}$ is a Resolution
  refutation of $\EReff_{\lfloor s/2\rfloor}(\varphi)$.
  We claim that the block\nobreakdash-width
  $\bwhat(\pi_{\restriction\rho}) < 1/2(\lfloor s/\ell \rfloor - 1)$,
  that is, the block-width of $\pi_{\restriction\rho}$ is less than
  the block-width guaranteed in
  case \ref{case:high-bw} of Theorem~\ref{thm:extraction}. Indeed,
  \begin{align}
    \label{eq:in-bound}
    \frac12\left(
      \left\lfloor
        \frac{\lfloor s/2\rfloor}{\ell}
      \right\rfloor-1
    \right) 
    &\geq
    \frac12\left(
      \frac{s/2-1}{(5/3)\sqrt{s}}-2
    \right)\\
    &=
    \frac{3\sqrt{s}}{20}
    -1-\frac{3}{10\sqrt{s}}\\
    &> \label{eq:strictineq}
    \frac{\sqrt{s}}8\\
    &>
    \delta\log|\pi|
    \geq
    \bwhat(\pi_{\restriction\rho})\,,
  \end{align}
  where \cref{eq:strictineq} holds for sufficiently large $s$.
    
  
  Suppose we run the original algorithm
  (Algorithm~\ref{algorithm:extended-delayer}) on
  $\pi_{\restriction \rho}$.
  By \cref{eq:in-bound}, we are guaranteed to end up
  in case \ref{case:assign} of \Cref{thm:extraction}; the algorithm
  outputs a satisfying assignment.

  According to the proof of \Cref{thm:extraction} it is even possible
  to characterize the state in which
  Algorithm~\ref{algorithm:extended-delayer} finds the satisfying
  assignment.

    First, suppose that the Delayer fails in Step 3(c).i
of Algorithm~\ref{algorithm:extended-delayer}, when attempting to reserve a
previously unreserved block in the final layer.
In this case, the clause assigned to that block is
$A=x_1\lor\cdots\lor x_n$. Since the update fails,
$A$ is not a weakening of any axiom of $\varphi$.
By \Cref{fact:weakening}, the all-zero assignment satisfies
$\varphi$. But Algorithm~\ref{algorithm:logspace} explicitly checks
this assignment, contradicting our assumption. Hence we may assume that the Delayer fails in
Step 3(c).ii.

  Denote by $C$ the clause and by $R$ the reservation
  Algorithm~\ref{algorithm:extended-delayer} considers right before
  outputting the satisfying assignment.
  Note that $R$ satisfies the reservation invariant.
  The clause $C$ mentions some block $b$ at layer $\ell - 1$ and $R$
  encodes a width-$n$ clause $\calB$ appearing in this block.
  Furthermore, $R$ has a block $b'$ reserved at layer $\ell$ weakened
  from $b$ but not mentioned in $C$.
  The algorithm attempts to extend the reservation to block $b'$ but
  fails in the process.
  In a bit more detail, since $b'$ is a trivial weakening of $b$ it
  must also encode the width-$n$ clause $\calB$.
  In the process of extending the reservation $R$ the algorithm
  attempts to find a clause $A$ of~$\varphi$ that $\calB$ is a
  weakening of, but fails.

  Let us use the above observation of how the Delayer conceded to
  argue correctness of the new extraction algorithm. 
  Let $B$ be the set of blocks in the connected component of~$b$ in the reservation~$R$.
  We claim that $B$ consists of blocks on distinct levels ending on level $\ell-1$
  with the block $b$ (see \Cref{fig:prove-delayer-example}).
  Indeed, since by \cref{item:inv-bw} of the reservation invariant it
  holds that $\bwhat(R) = \bwhat(C)$, and we have that $\bwhat(C) \leq w_0$ and
  $\ell = n + 4w_0$, it follows that the chain starting at $b$ is of length
  at most $w_0$.
  This implies in particular that $B$ is not connected to level $n$;
  the claim follows.



  Consider the clause $C' \in \pi$ such that
  $C = C'_{\restriction \rho}$.
  Since $C$ survived the restriction it holds that $\bwhat(C') \leq w_0$.
  %
  This implies that the clause $C'$ is evaluated by the \say{new}
  Algorithm~\ref{algorithm:logspace} and the set of blocks $B$ is
  considered.
  As we assumed that Algorithm~\ref{algorithm:logspace} does not
  output a satisfying assignment it holds that the partial assignment
  $\alpha_B$ does not satisfy $\varphi$.
  Hence there exists a clause $A$ of $\varphi$ that is not satisfied
  by $\alpha_B$.
  This clause $A$ is of the form $A = A_0 \lor A_1$ with
  $A_0 \subseteq \calA_B$ and $\calA_B \cap A_1 = \emptyset$.
  
  Consider the reservation $R'$ that modifies $R$ on $B$ as follows:
  the topology between the blocks of $B$ is left intact but every
  block in $B$ now contains the literals $\calA_B \cup A_1$ along with
  positive literals $x_j$ for every variable
  $x_j$ not mentioned in $\calA_B \cup A_1$.
  By construction the reservation $R'$ satisfies the reservation
  invariant with respect to $C$, but the Delayer no longer concedes in
  this round:
  when queried about block $b'$ the Delayer answers that it is a
  weakening of $A$. Algorithm~\ref{algorithm:extended-delayer} thus
  does not fail in this iteration.
  %
  This shows that Algorithm~\ref{algorithm:extended-delayer} never
  terminates thereby contradicting \Cref{thm:extraction}. The
  statement follows.
\end{proof}

\subsubsection{Complexity of the algorithm}
\label{subsec:complexity-new-algo}

\Cref{thm:correctness-logspace} shows that the new algorithm is
correct, but it says nothing about its complexity. It is not hard to
see that the procedure can be implemented in logarithmic space
($\FL$), but we can do better. In the following we show that
$\Extraction_\Res[n^2]$ lies in $\P$-uniform $\FTC^0$.
This puts $\Extraction_\Res[n^2]$ in $\FP$, whereas before it was only
known to be in $\FZPP$ ($\Extraction_\Res[n^3]$ was shown to be in
$\FP$.)

Recall that $\P$-uniform $\FTC^0$ is the class of relations $R \subseteq \{ 0,1\}^* \times \{ 0,1\}^*$ for which there exists a Turing machine $D$ and a constant $d \in \bbN$ such that on input $1^n$, the machine $D(1^n)$ runs in time $\poly(n)$ and outputs the description of a multi-output circuit $C_n$ of depth $d$, unbounded fan-in and threshold gates, such that for every $x \in \{0, 1\}^n$, it holds that $(x, y) \in R$ for some $y \in \{0,1\}^*$ if and only if $(x, C_n(x)) \in R$.

\begin{theorem}[$\Extraction_\Res$ is $\FTC^0$]
    \label{thm:FTC0}
    The search version of the Proof Analysis Problem for Resolution is solvable by $\P$-uniform multi-output circuits of polynomial size and constant depth with threshold gates whenever $s \geq n^2$. That is, it holds that $\Extraction_\Res[n^2] \in \P\text{-uniform } \FTC^0$.
\end{theorem}
    
\begin{proof}
    For the sake of precision, let us fix the following natural binary encoding for $\Extraction_\Res$. An input $(\varphi, \pi, 1^s)$ will be of the form $(1^n, 1^m, C_1, \dots, C_m, 1^t, \pi, 1^s)$. Here, the first part of the tuple corresponds to the encoding of $\varphi$, a CNF formula over $n$ variables and $m$ clauses $C_1, \dots, C_m$, and we assume that these clauses are initially represented as strings of length $2n$ with indicators for every possible literal. The Resolution refutation of $\Reff_s(\varphi)$ is encoded by $1^t$ and $\pi$, where $\pi$ is an assignment to the $N \coloneq N(n,m,t, s) = \poly(n,m,t,s)$ variables of $\Reff_t(\Reff_s(\varphi))$, and the number of variables $N$ can be easily computed in polynomial time. For the purpose of unique decoding, we assume that the tuple $(1^n, 1^m, C_1, \dots, C_m, 1^t, \pi, 1^s)$ is encoded by bit-doubling: each bit is duplicated and $01$ is used as separators. We assume that there are no separators between the clauses $C_1, \dots, C_m$, so a correct input contains only five separators.

    Now, when dealing with binary strings of even length $\ell$, there
    are at most $O(\ell^4)$ possible ways of interpreting such strings
    as a tuple of the form
    $(1^n, 1^m, C_1, \dots, C_m, 1^t, \pi, 1^s)$. This is because we
    can choose values for $n$, $m$, $t$ and $s$ in the interval
    $[\ell/2 - 5]$, where $\ell/2 - 5$ comes from the fact that we
    duplicated every bit and introduced 10 bits for the five
    separators, not counting separators between $C_1$ and $C_m$. One
    can then check that this choice of $n$, $m$, $t$ and $s$ conforms
    to the desired pattern: the segment for the clauses $C_1$ to $C_m$
    has length exactly $2nm$, and the segment for $\pi$ has length
    exactly $N = N(n,m, t, s)$. That is, it must hold that
    $\ell = 2( n + m + 2nm + t + N(n,m,t,s) + s)+ 10$ and
    $s \geq n^2$. There are at most $O(\ell^4)$ such choices for
    $(n,m,t, s) \in [\ell/2 - 5]^4$, hence the upper
    bound. Furthermore, it is easy to see that, due to the
    bit-doubling, every string can only encode correctly one
    input of the form $(1^n, 1^m, C_1, \dots, C_m, 1^t, \pi, 1^s)$, so
    the decoding is unique.

    The $\P$-uniform descriptor machine for inputs of even length $\ell$ now works as follows. On input $1^\ell$, for $\ell$ even, it tries all possible $O(\ell^4)$ ways of separating the lengths, and for each interpretation $(n,m,t,s)$ of the lengths it constructs a different constant-depth multi-output Boolean circuit $D_{n,m,t,s}$, where the $i$-th bit $D^i_{n,m,t,s}$ is defined as follows.
    
    \begin{enumerate}[label=(\alph*)]
        \item If the interpretations of the lengths are inconsistent, in the sense that the string cannot correspond to something of the form $(1^n, 1^m, C_1, \dots, C_m, 1^t, \pi, 1^s)$, then it outputs the constant-$0$ circuit. \label{it:outputzero}
    
        \item If the interpretation of the lengths is valid and $|\pi| \geq 2^{\varepsilon \cdot n}$, then it simply outputs the conjunction of the following two circuits. First, the CNF that checks that $\pi$ is a correct Resolution refutation of $\Reff_s(\varphi)$ in at most $t$ clauses. Second, the standard constant-depth circuit that outputs the lexicographically-first satisfying assignment for $\varphi$. This circuit is built as follows. For every assignment $\alpha \in \{ 0, 1\}^n$, take the CNF formula $\Satf(\varphi, \alpha)$ and let $S_\alpha$ be the output of this circuit. Then, define for every $\alpha \in \{0,1\}^n$ the circuit
            \begin{equation}
                \label{eq:lexicographic}
                L_\alpha \coloneq S_\alpha \land \bigwedge_{\beta \prec \alpha} \neg S_\beta ,
            \end{equation}
            where $\prec$ denotes the lexicographic order on $\{ 0, 1\}^n$. The bit $L_\alpha$ is $1$ if, and only if, $\alpha$ happens to be the lexicographically-first satisfying assignment of $\varphi$. Now, to output the $i$-th bit of $\alpha$, it suffices to output the bit
            \begin{equation}
                A_i \coloneq \bigvee_{\substack{\alpha \in \{ 0,1\}^n \\ \alpha_i = 1}} L_\alpha.
            \end{equation}
            Here $D^i_{n,m,t,s} \coloneq A_i$. This entire circuit has depth $4$ and size $\poly(n, m) \cdot 2^{O(n)}$.
        
        \label{it:bruteforce}

        \newcommand{\cktPresent}{\operatorname{Present}}
        \newcommand{\Th}{\mathsf{Th}}
        
        \item \label{it:nontrivial} If the interpretation of lengths is consistent and $|\pi| < 2^{\varepsilon \cdot n}$, we construct the following circuit, which essentially implements Algorithm~\ref{algorithm:logspace}. Let $w \coloneq \lceil \delta \log |\pi|\rceil$ and fix a clause $C \in \pi$. For every $b\in [s]$, define
            \begin{equation}
                \cktPresent^C_b \coloneq \bigvee_{z \in \Lit(b)} \lit^C_{z},
            \end{equation}
        which checks whether block $b$ is mentioned in $C$.

        Next, suppose $\beta \in \{ 0,1\}^w$ and $b\in [s]$, and define
        \ifcameraready
            \begin{equation}
                \label{eq:HW}
                \!\!\!\! B^C_b\! (\beta)   \coloneq  \cktPresent^C_b  \! \land \bigvee_{\mathclap{k\in[w]}}  \left(\#_k(\cktPresent_{\leq b})  \land \beta_k \right),\!
            \end{equation}
        \else
            \begin{equation}
                \label{eq:HW}
                B^C_b(\beta) \coloneq  \cktPresent^C_b \land \bigvee_{k\in[w]} \left(\#_k(\cktPresent^C_{1}, \dots, \cktPresent^C_{b})  \land \beta_k \right),
            \end{equation}
        \fi
        where $\#_k(p_1, \dots, p_n) \coloneq \Th_k(p_1, \dots, p_n) \land \neg \Th_{k+1}(p_1, \dots, p_n)$ tests if the Hamming weight of the string consisting of bits $p_1, \dots, p_n$ is equal to $k$. The formula $B^C_b(\beta)$ checks whether $b$ belongs to the subset of blocks selected by $\beta$. More precisely, if $b_1 < \cdots < b_r$ are the blocks mentioned in $C$, where $r \leq w$, then $B^C_{b_k}(\beta) = \beta_k$ for every $k \in [r]$, and $B^C_b(\beta) = 0$ for all other blocks. Thus, by ranging over all $\beta \in \{0,1\}^w$, we enumerate all subsets of the blocks mentioned in $C$.

        We can then obtain the $i$-th bit of the candidate assignment the algorithm obtains when inspecting clause $C$ as
            \begin{equation}
                \alpha^C_{\beta, i} \coloneq \bigvee_{b \in [s]} \left( B^C_{b}(\beta) \land \left ( \lit^C_{\neg \lit^b_{\neg x_i}} \lor \lit^C_{\lit^b_{x_i}}\right) \right).
            \end{equation}

        We now want to obtain the first $C\in\pi$ and the first $\beta \in \{0,1\}^w$ such that the assignment $\alpha^C_{\beta} \coloneq (\alpha^C_{\beta, 1},\dots,\alpha^C_{\beta, n})$ satisfies $\varphi$. For this, take
        \ifcameraready
         \begin{equation}
                L_{\alpha^C_{\beta}} \coloneq \Satf(\varphi, \alpha^C_{\beta}) \land \, \bigwedge_{\mathclap{(C', \beta')\prec(C, \beta)}} \, \neg \Satf({\varphi,\alpha^{C'}_{\beta'} }).
        \end{equation}
        \else
            \begin{equation}
                L_{\alpha^C_{\beta}} \coloneq \Satf(\varphi, \alpha^C_{\beta}) \land \bigwedge_{(C', \beta')\prec(C, \beta)} \neg \Satf({\varphi,\alpha^{C'}_{\beta'} }).
            \end{equation}
        \fi
        Analogously to (\ref{eq:lexicographic}), $L_{\alpha^C_{\beta}}$ is 1 if and only if $\alpha^C_{\beta}$ is the first candidate assignment to be satisfying. Now, to recover every bit of the assignment, we simply output
            \begin{equation}
                A_i \coloneq \bigvee_{C \in [\pi]} \bigvee_{\beta \in \{0,1\}^w} L_{\alpha^C_{\beta}} \land {\alpha^C_{\beta, i}}. 
            \end{equation}
    \end{enumerate}
    It is easy to verify that if Algorithm~\ref{algorithm:logspace} found a satisfying assignment when looking at clause $C$ and subset $B$ of the blocks mentioned in $C$, then that same assignment will be considered by the construction above.

    Finally, for each possible interpretation $(n, m, t, s)$ of the lengths of the input, we obtain a circuit $D_{n, m, t, s}$. For each interpretation of the lengths there is also a circuit $\operatorname{Correct}_{n, m, t, s}$ that verifies that the input correctly encodes an $\Extraction_\Res$ instance of the right size. This amounts to checking that the separators are in the right place and the double-bit encoding is correctly implemented, which can all be verified in depth $3$. We can once again repeat the construction in~\ref{it:bruteforce} to output the lexicographically-first assignment that passes the correctness check.

    The circuit above behaves like Algorithm~\ref{algorithm:logspace}. Choosing the same parameters $s_0$ and $w_0$ as in \Cref{thm:correctness-logspace}, we are guaranteed to output a correct satisfying assignment when one exists. It is also clear by construction that the circuit has constant depth. It is only left to argue that the size is polynomial in the size of the input. Indeed, choosing the $s_0$ and $w_0$ as in \Cref{thm:correctness-logspace} the construction in \ref{it:bruteforce} has size $\poly(n, m) \cdot 2^{O(n)}$, but since $|\pi| \geq 2^{\varepsilon \cdot n}$, we have $|\pi| \geq 2^{O(n)}$ and the overall size is $\poly(n, m, |\pi|, s)$, as desired.

    The construction in part \ref{it:nontrivial} also has constant depth. The iteration over all possible subsets is over $2^{w_0} = 2^{\lceil \delta \log |\pi|\rceil} = \poly(|\pi|)$ subsets, so it is polynomial in size.
    \end{proof}

    In the absence of threshold gates, we can still give a non-trivial circuit, at the cost of a quasipolynomial increase in size.

    \begin{theorem}[$\Extraction_\Res$ in quasipolynomial $\ComplexityFont{FAC}^0$]
        \label{thm:FAC0}
        The search version of the Proof Analysis Problem for Resolution is solvable by $\QP$-uniform multi-output circuits of quasipolynomial size and constant depth whenever $s \geq n^2$.
    \end{theorem}

    \begin{proof}
        The construction is almost identical to the one above in \Cref{thm:FTC0}. The main obstacle is that in the absence of threshold gates, we cannot compute Hamming weight as in \cref{eq:HW}. We can side-step this by considering instead all possible subsets $B \subseteq[s]$ containing exactly $w \coloneq \lceil \delta \log |\pi|\rceil$ elements. The descriptor machine writes each such $B$ as an indicator vector of size $s$, and for each $b \in [s]$, we define $B_{b}^C \coloneq B_b \land \operatorname{Present}_{b}^C$, which gives us the exact blocks that are present in both $B$ and $C$. The rest of the construction proceeds as before.

        This change means the big disjunctions over all $\beta\in\{0,1\}^w$ will be replaced by big disjunctions over all possible subsets $B \subseteq[s]$ containing exactly $w$ elements. This gives us a blow-up of 
        \begin{equation}
            |\pi| \cdot {{s}\choose{w}}  = |\pi| \cdot {{s}\choose{\lceil \delta \log |\pi| \rceil}}  \leq |\pi|^{O(1)} \cdot s^{O(\log |\pi|)},
        \end{equation}
        which is quasipolynomial in the size of the input.
    \end{proof}

\section{The restriction argument in  \texorpdfstring{$\APCtwo(\alpha)$}{APC2(alpha)}}
\label{sec:restriction-formalized-APC}

In this section we show that the random restriction argument for $\Reff$ formulas
(\Cref{lemma:restriction}) can be formalized in the relativized theory
$\APCtwo(\alpha)$. In \Cref{subsec:designs} we show that $\PVO$
(unrelativized) can prove the existence of suitable combinatorial designs giving
us $k$-wise independent distributions with the right parameters. In
\Cref{subsec:bad-seeds} we employ \citeauthor{jerabek:phd-thesis}'s $\APC$
framework (again, unrelativized) to show that for clauses of block-width at
least $k$, we can bound the number of \say{bad seeds}---those for which the
$k$-wise independent distribution produces a restriction that does not kill
the clause. \Cref{subsec:formulas-BW} presents suitable relativized formulas
to check that a certain clause has block-width at least $k$. Finally,
\Cref{subsec:good-seeds} shows how we can formalize a union bound argument
implying that there exists at least one \say{good seed} producing a restriction
that kills all wide clauses in a refutation. In this latter proof we encode
Resolution refutations by a new relation symbol. This final proof can be carried out in $\APCtwo(\alpha)$.

\subsection{Construction of designs}
\label{subsec:designs}

The goal is to argue in $\PVO$ that we can construct a combinatorial design
yielding a $k$-wise independent distribution for suitable values of $k$.
Our construction of designs will be the first construction of combinatorial
designs given by Nisan and Wigderson \autocite[Lemma 2.5]{NW94}.
(\citeauthor{Jerabek2007} \autocite[Lemma 2.4]{Jerabek2007} formalized the second,
greedy construction \autocite[Lemma 2.6]{NW94}, used to get the parameters needed
for derandomizing $\BPP$, but the greedy algorithm is sequential and not feasible in these theories.)

\begin{definition}[Combinatorial designs]
    \label{def:combinatorial-design}
    Let $n, m, p, w \in \bbN$ with $w \leq p \leq m$. An
    $(n, m, p, w)$-\emph{design} is a Boolean matrix of size $n \times m$
    such that every row has exactly $p$ ones and every two distinct rows
    intersect in at most $w$ positions.
\end{definition}

Let us construct such designs with the right parameters for our applications.

\begin{lemma}[Existence of designs, in $\PVO$]
    \label{lemma:design}
    There is a $\PV$-function $\Design$ such that for every $n$ and every
    $2^k \in \Log$ satisfying $|n|\leq 2k(2^k+1)$, and every $i\in[n]$,
    the function $\Design(n,2^k,i)$ outputs the $i$-th row of a fixed matrix
    $M$, and $\PVO$ proves that the rows given by this function form an
    $(n,2^{4k},2^{2k},2^k)$-design. Furthermore, there is a $\PV$-function
    that, for every set $I\subseteq[n]$ of indices of size $|I|\leq 2^k$,
    outputs the entire submatrix $M_I$ of $M$ restricted to the rows indexed
    by $I$, and $\PVO$ proves that $\rank_{\bbF_2}(M_I)=|I|$.
\end{lemma}

\begin{proof}
    We fix the field $\bbF\coloneq\operatorname{GF}(2^{2k})$, where elements
    are binary strings of length $2k$. Since $2^k\in\Log$, we can operate
    over $\bbF$ within $\PVO$ (see \Cref{subsubsec:prelim-APC} for details
    about field arithmetic over $\PVO$). We describe the matrix $M$ globally;
    it will be clear from the construction that, given the index $i$ of a row,
    its $i$-th row can be constructed by a $\PV$-function.

    Given $i\in[n]$, consider its binary representation. Since
    $|i|\leq |n|\leq 2k(2^k+1)$, we may pad it with leading zeroes until it
    has length exactly $(2^k+1)2k$. Split the resulting string into $2^k+1$
    segments of $2k$ bits and interpret these segments as the coefficients
    of a polynomial $P_i\in\bbF[X]$ of degree at most~$2^k$. Notice that
    if $i\neq i'$, then $P_i\neq P_{i'}$, since their coefficient vectors
    are distinct.

    A position in the $i$-th row of $M$ is indexed by a pair
    $(a,b)\in\bbF^2$, and we define $M[i,(a,b)]=1$ if and only if
    $P_i(a)=b$. Thus $M$ has dimension $n\times 2^{4k}$. Every row contains
    exactly $2^{2k}$ ones, since for every $a\in\bbF$ there is exactly one
    $b\in\bbF$ such that $P_i(a)=b$.

    Consider now two distinct rows $i\neq i'$. A coordinate $(a,b)$ is
    contained in both rows only if $P_i(a)=P_{i'}(a)=b$. Hence $a$ is a root
    of the non-zero polynomial $P_i-P_{i'}$, whose degree is at most $2^k$.
    By \Cref{lemma:FTA-easy}, available in $\PVO$, there are at most~$2^k$
    such values of $a$. Thus every two distinct rows intersect in at most
    $2^k$ positions, and $M$ is an $(n,2^{4k},2^{2k},2^k)$-design.

    Finally, let $I\subseteq[n]$ have size $r\coloneq |I|\leq 2^k$. Since
    $2^k\in\Log$, the submatrix $M_I$ can be encoded as a number in $\PVO$.
    Suppose for contradiction that $\rank_{\bbF_2}(M_I)<r$. Then some row
    $i\in I$ is an $\bbF_2$-linear combination of the remaining rows in
    $I\setminus\{i\}$. Consequently, every coordinate at which row $i$ is
    $1$ must also be $1$ in at least one of those remaining rows. The support
    of row $i$ is therefore contained in the union, over
    $i'\in I\setminus\{i\}$, of the intersections of the supports of rows
    $i$ and $i'$. Every such intersection has size at most $2^k$, and there
    are at most $r-1\leq 2^k-1$ of them. Hence this union has size at most
    $2^k(r-1)\leq 2^k(2^k-1)<2^{2k}$, contradicting the fact that row~$i$
    has exactly $2^{2k}$ ones. Thus $\rank_{\bbF_2}(M_I)=r$. All the sets counted above are $\Log$-sized, so these counting arguments
    and the construction of $M_I$ can be carried out in $\PVO$.
\end{proof}

\subsection{Bad restrictions from seeds}
\label{subsec:bad-seeds}

Our goal is to sample a restriction that kills clauses of high block-width.
We use the rows of a combinatorial design as a linear map from short seeds
to the bits describing such restrictions.
Recall that we are working with the $\Reff_s(\varphi)$ formula and the
restrictions we are interested in are the so-called \emph{disabling
restrictions} from \Cref{subsubsec:corretness-new-algorithm}. Moreover, we
want to replicate the pairing argument in \Cref{lemma:restriction}: we pair
the $s$ blocks into $s/2$ pairs, and we enable and disable exactly one block
in each pair. Then, the variables of the disabled blocks are all set to the
same value.

Let us first make precise the way of pairing blocks.

\newcommand{\pair}{\operatorname{pair}}
\newcommand{\pw}{\operatorname{pw}}
\newcommand{\pairpos}{\operatorname{pairpos}}

\begin{definition}[Pair-index and pair-width]
    Assume $s$ is even. For every $j\in[s/2]$, we pair blocks $2j-1$ and
    $2j$ and define its \emph{pair-index} to be $\pair(2j-1)=\pair(2j)\coloneq j$. We say that the block $2j-1$ is in position $0$ and the block $2j$ in position $1$, and we define
    $\pairpos(2j-1)\coloneq 0$, and $\pairpos(2j)\coloneq 1$.
    For a variable of $\Reff_s(\varphi)$, its pair-index is the pair-index
    of the block to which the variable belongs. We denote by $\pw(C)$ the
    \emph{pair-width} of a clause $C$, namely, the number of distinct
    pair-indices mentioned in $C$.
\end{definition}

The restrictions we deal with are defined by strings of length~$s$, where
the first $s/2$ bits select the blocks to be disabled in each pair, and the
last $s/2$ bits set the value of variables in the disabled blocks.

More precisely, for a pair-index $j\in[s/2]$, a choice bit $\alpha_j=0$ means
that the block in position $0$ is disabled, while $\alpha_j=1$ means that
the block in position $1$ is disabled. The bit $\beta_j$ gives the common
value assigned to all variables belonging to the disabled block.

\begin{definition}[Restrictions from seeds]
\label{def:restrictions-from-seeds}
    Let $s,n,m\in\bbN$ be the parameters of the formula
    $\Reff_s(\varphi)$, with $s$ even. Let $M$ be an $s\times\ell$
    Boolean matrix. Suppose that a clause $C$ has width~$w$ and
    pair-width $w$, and let its pair-indices be
    $\{i_1,\ldots,i_w\}\subseteq[s/2]$. Define
    $I(C)\coloneq\{i_1,\ldots,i_w,i_1+s/2,\ldots,i_w+s/2\}\subseteq[s]$.

    We interpret a string
    $\mu=\alpha_1\cdots\alpha_w\beta_1\cdots\beta_w\in\{0,1\}^{2w}$
    as the restriction on these $w$ pairs in which $\alpha_j$ determines
    which block of pair-index $i_j$ is disabled and $\beta_j$ gives the value
    assigned to all variables in that disabled block. Given a seed
    $x\in\{0,1\}^{\ell}$, we write
    $\rho_x^{I(C)}\coloneq M_{I(C)}x\in\{0,1\}^{2w}$ for the restriction
    on the pairs occurring in $C$ generated by $x$.
\end{definition}

For every clause $C$, we are interested in bounding the number of \say{bad seeds}, those
leading to restrictions that do not kill the clause.

\begin{definition}[Bad seeds]
\label{def:bad-set}
Let $M$ and $C$ be as in \Cref{def:restrictions-from-seeds}. We denote by
$\BadSet_C$ the set
    \begin{equation*}
        \BadSet_C
        \coloneq
        \set*{x\in\{0,1\}^{\ell}}
        {C{\restriction_{\rho_x^{I(C)}}}\neq 1}
    \end{equation*}
    of \emph{bad seeds} for the clause $C$.
\end{definition}

Observe that for clauses $C$ of both width $w$ and pair-width~$w$, and
assuming $w\in\Log$, the clause can be represented as a string of length
$O(w(\log s + \log n + \log m))$ and hence encoded as a number. Then, given such a clause $C$
and a seed $x\in\{0,1\}^{\ell}$, there is a $\PV$-symbol that checks whether
$x\in\BadSet_C$.

We are now ready to bound the number of bad seeds. The only property of
the design needed in the following lemma is that the $2w$ rows indexed by
$I(C)$ are linearly independent.

\begin{lemma}[Restrictions at clause-level, in $\APC$]
    \label{lemma:restrictions-clause-level}
    The theory $\APC$ proves the following statement. Let $s,n,m\in\bbN$,
    let $w,\ell\in\Log$, and let $C$ encode a clause containing exactly
    $w$ literals over $w$ distinct pair-indices of the variables of
    $\Reff_s(\varphi)$. Let $M$ be an $s\times\ell$ Boolean matrix such
    that $\rank_{\bbF_2}M_{I(C)}=2w$. Provably in $\APC$,
    \[
        \BadSet_C
        \preceq_0
        3^w2^{\ell-2w}
        =
        \left(\frac34\right)^w2^\ell.
    \]
\end{lemma}

\begin{proof}
    Write the literals of $C$ as $\lambda_1,\ldots,\lambda_w$, with
    $\lambda_j$ belonging to the pair of pair-index $i_j$. For every
    $j\in[w]$, let $P_j\subseteq\{0,1\}^2$ be the set of pairs
    $(\alpha,\beta)$ for which the restriction of pair-index~$i_j$ determined
    by the choice bit $\alpha$ and value bit $\beta$ does not satisfy~$\lambda_j$.

    If $\lambda_j$ is an $\enable$-literal, precisely two of the four
    patterns are bad: the value bit $\beta$ is irrelevant, and the choice
    bit must make the $\enable$-literal false. If $\lambda_j$ is any other
    literal, precisely three patterns are bad: two patterns leave its block
    enabled and hence the literal unset, while among the two patterns
    disabling its block exactly one assigns the literal the falsifying value.
    Thus $|P_j|\leq 3$ for every $j\in[w]$.

    For each $j$, fix a $\PV$-computable surjection
    $h_j:\{0,1,2\}\twoheadrightarrow P_j$, repeating one of the two
    patterns when $|P_j|=2$. We construct a surjection
    \begin{equation}
        \label{eq:bad-surjection}
        G:
        \{0,1,2\}^w\times\{0,1\}^{\ell-2w}
        \EmilSurj
        \BadSet_C.
    \end{equation}
    Given $(z_1,\ldots,z_w,\kappa)$ in the domain, write
    $h_j(z_j)=(\alpha_j,\beta_j)$ and let
    $\mu\coloneq\alpha_1\cdots\alpha_w\beta_1\cdots\beta_w
    \in\{0,1\}^{2w}$.

    Since $\rank M_{I(C)}=2w$, the linear map
    $M_{I(C)}:\bbF_2^\ell\to\bbF_2^{2w}$ is surjective. Using Gaussian
    elimination in $\PVO$ (\Cref{thm:Gaussian-elim-PV}), let
    $\nu^\star=\operatorname{Sol}(M_{I(C)},\mu)$ be a fixed solution of
    $M_{I(C)}\nu=\mu$, and let $K$ be a matrix whose columns form a basis
    of $\ker M_{I(C)}$. Since the rank is $2w$, this kernel has dimension
    $\ell-2w$. Define
    $G(z_1,\ldots,z_w,\kappa)\coloneq\nu^\star+K\kappa$.

    By construction, $M_{I(C)}G(z_1,\ldots,z_w,\kappa)=\mu$, and the
    local pattern $(\alpha_j,\beta_j)$ lies in $P_j$ for every $j$.
    Hence none of the literals $\lambda_j$ is satisfied, so the output
    seed belongs to $\BadSet_C$.

    It remains to prove surjectivity. Let $x\in\BadSet_C$, and write
    $\mu=M_{I(C)}x$. Since $x$ is bad, the two bits of $\mu$ corresponding
    to each pair $i_j$ form a pattern in $P_j$. Choose
    $z_j\in\{0,1,2\}$ mapping to that pattern under $h_j$. Let
    $\nu^\star$ be the canonical solution associated with this $\mu$.
    Then $x-\nu^\star\in\ker M_{I(C)}$, so there is a coefficient vector
    $\kappa\in\{0,1\}^{\ell-2w}$ such that
    $x-\nu^\star=K\kappa$. Hence
    $G(z_1,\ldots,z_w,\kappa)=x$.

    The domain has cardinality
    $3^w2^{\ell-2w}=(3/4)^w2^\ell$, which proves the claimed
    $\preceq_0$ bound.
\end{proof}

\subsection{The block-width formulas}
\label{subsec:formulas-BW}

Until now, all arguments were in the unrelativized theories. The clauses
we worked with so far had pair-width $w$ as well as width $w$, for
$w \in \Log$, and hence they could be encoded as numbers, since these are
strings of length $O(w (\log s + \log n + \log m))$. In general, however, we will deal with
clauses encoded as strings given by an external predicate $\Pi$ (see
\Cref{subsec:prelim-Reff} for details on how refutations are encoded in
such a predicate). We will generally assume that if the refutation has
$t$ clauses, we can talk about any particular clause $D_i$ for
$i \in [t]$; this is just a figure of speech to refer to the substring
defined by $\Pi$ containing the information about clause $D_i$. In this
section we define formulas checking that such a clause $D_i$ has high
enough block-width.

Observe first that given numbers $s$, $n$ and $m$, we can determine the
number of variables of the formula $\Reff_s(\varphi)$, and there is a
$\PV$-function $\paramLit$ that given a block $b \in [\splus]$ indicates
which are the numbers encoding the literals of $\Reff_s(\varphi)$ belonging
to block $b$. Note that the specifics of $\varphi$ are not needed here,
only the parameters $s$, $n$ and $m$. We will abuse notation and use $\Lit$
also to refer to the set of numbers itself, although this $\PV$ function
only outputs an upper and a lower bound on the numbers we have to check.

We first define a formula to check whether a certain block $b \in [\splus]$ is
present in the clause $D_i$ of $\Pi$. This is the
$\formSigmab^b_1(\Pi)$-formula
\ifcameraready
\begin{equation}
    \label{form:present}
    \tag{$\presentname$}
    \!\!\!\parampresent \!
    \coloneq \!
    \exists \ell \in \paramLit \,
    \Pi(\lit^i_\ell). \!
\end{equation}
\else
\begin{equation}
    \label{form:present}
    \tag{$\presentname$}
    \parampresent
    \coloneq
    \exists \ell \in \paramLit \
    \Pi(\lit^i_\ell).
\end{equation}
\fi
Next, for $w\in\Log$, we define the
$\formSigmab^b_1(\Pi)$-formula checking that the clause $D_i$ has
block-width at least $w$:
\ifcameraready
\begin{equation}
\tag{$\bwformulaname$}
\begin{split}
    \label{form:bw}
    \parambwformula 
    &\coloneq
    \exists \beta\in[s]^w\, 
    (
        \forall j\in[w]\, \\
        &\present(i,(\beta)_j,s,n,m)
        \, \land \\
        &\forall j<j'\in[w]\,
        (\beta)_j\neq(\beta)_{j'}
    ).
\end{split}
\end{equation}
\else 
\begin{equation}
    \label{form:bw}
    \tag{$\bwformulaname$}
    \parambwformula 
    \coloneq
    \exists \beta\in[s]^w\, 
    \left(
        \forall j\in[w]\,
        \parampresent[(\beta)_j]
        \land
        \forall j<j'\in[w]\,
        (\beta)_j\neq(\beta)_{j'}
    \right).
\end{equation}
\fi
Here $\beta$ represents a sequence of $w$ distinct blocks present in
$D_i$, and $(\beta)_j$ denotes its $j$-th element. Such a sequence is
encoded using $O(w\log s)$ bits. 
Note that we deliberately do not count the root block
towards the block width.
Since $w\in\Log$, the quantifiers over
$j,j'\in[w]$ are sharply bounded, and hence the entire formula is
$\formSigmab^b_1(\Pi)$.

\begin{remark}
    \label{rem:impossibility-below-Sigmab1}
    The formula~(\ref{form:bw}) above is $\formSigmab^b_1(\Pi)$, and this
    existential quantifier will force us to state the final restriction
    argument in $\APCtwo(\alpha)$ rather than $\APC(\alpha)$. In general,
    as long as we use the existing encoding of Resolution refutations in
    $\Pi$, we cannot go below $\formSigmab^b_1(\Pi)$. This is because a
    $\formDeltab^b_0(\Pi)$ formula would amount to a $\PV(\Pi)$-function,
    which corresponds to a decision tree of quasipolynomial size and
    polylogarithmic depth. However, it is easy to see that any decision
    tree deciding whether a clause of $\Pi$ has block-width at least $w$
    must require linear depth.
\end{remark}

\subsection{Existence of good seeds}
\label{subsec:good-seeds}

It now remains to conclude that there exists a \say{good} seed that
generates a restriction satisfying all wide clauses. The main technical
obstacle is that the sets $\BadSet_C$ above were defined only for clauses
$C$ that can themselves be encoded as numbers. In the general setting,
however, a Resolution refutation is encoded by the auxiliary predicate
$\Pi$, and a clause $D_i$ in the refutation may be too large to be
encoded as a number.

If $D_i$ has block-width at least $w$, then it mentions at least $w/2$
distinct pair-indices. We can therefore choose a subclause
$C_i\subseteq D_i$ containing exactly one literal from each of $w/2$
distinct pair-indices. Thus $C_i$ has both standard width and pair-width
exactly $w/2$ and, since $w\in\Log$, it can be encoded as a number.
Killing $C_i$ also kills $D_i$. The choice of $C_i$ will be implemented by a
$\PV_2(\Pi)$-function. This is the point at which we need to work in
$\APCtwo(\alpha)$ rather than $\APC(\alpha)$.

\newcommand{\Good}{\operatorname{Good}}
Before we proceed, let us define what it means for a seed to be \say{good}
for a clause encoded by a string.
For $s\in \bbN$ and $w\in\Log$, let
$q=q(s,w)\coloneq
2^{\lceil\log(\max\{2,w,|s|\})\rceil}$.
Thus $q$ is a power of two, $q\geq w$, $q\geq |s|$, and
$q<2\cdot \max\{2,w,|s|\}$. In particular, $q\in\Log$ whenever $w\in\Log$.
Writing $q=2^k$, we have $|s|\leq q\leq 2k(q+1)$, so the hypothesis of
\Cref{lemma:design} holds when the number of rows is $s$.

\begin{definition}[Good seeds]
    \label{def:good-seeds}
    Assume $s$ is even and let $q=q(s,w)$ be as above. For a seed
    $\sigma\in\{0,1\}^{q^4}$, write
    $\rho^\sigma_r\coloneq\Design(s,q,r)\cdot\sigma^\top$ for
    $r\in[s]$. Thus $\rho^\sigma_1,\ldots,\rho^\sigma_{s/2}$ are the
    choice bits and
    $\rho^\sigma_{s/2+1},\ldots,\rho^\sigma_s$ are the corresponding
    value bits.

    We denote by $\Good(s,n,m,w,\sigma,i)$ the
    $\formDeltab^b_2(\Pi)$-formula expressing the following: if the
    $i$-th clause $D_i$ encoded by $\Pi$ has block-width at least $w$,
    then there is a literal $\ell$ present in $D_i$, belonging to some
    block $j\in[s]$, such that one of the following holds:
    \begin{enumerate}[label=(\alph*)]
        \item $\ell=\enable^j$, $\pairpos(j)=0$, and
        $\rho^\sigma_{\pair(j)}=1$;
        \item $\ell=\neg\enable^j$, $\pairpos(j)=0$, and
        $\rho^\sigma_{\pair(j)}=0$;
        \item $\ell=\enable^j$, $\pairpos(j)=1$, and
        $\rho^\sigma_{\pair(j)}=0$;
        \item $\ell=\neg\enable^j$, $\pairpos(j)=1$, and
        $\rho^\sigma_{\pair(j)}=1$;
        \item $\ell$ is a literal of a variable of block $j$ different
        from $\enable^j$, $\pairpos(j)=0$,
        $\rho^\sigma_{\pair(j)}=0$, and the polarity of $\ell$ agrees
        with the value $\rho^\sigma_{\pair(j)+s/2}$;
        \item $\ell$ is a literal of a variable of block $j$ different
        from $\enable^j$, $\pairpos(j)=1$,
        $\rho^\sigma_{\pair(j)}=1$, and the polarity of $\ell$ agrees
        with the value $\rho^\sigma_{\pair(j)+s/2}$.
    \end{enumerate}
    We say that a seed $\sigma\in\{0,1\}^{\ell}$ is
    \emph{$(\ell,w)$-good} for $\Pi$ if
    $\Good(s,n,m,w,\sigma,i)$ holds for every clause $D_i$ encoded by~$\Pi$.
\end{definition}

We are ready to prove that good seeds exist.

\begin{theorem}[Existence of good seeds, in $\APCtwo(\alpha)$]
\label{thm:good-seeds}
The theory $\APCtwo(\Phi,\Pi)$ proves the following statement. Let
$s,n,m\in\bbN$, let $t>1$, and let
$w\coloneq8\lceil\log t\rceil$, $q\coloneq q(s,w)$, and
$\ell\coloneq q^4$. If $\Phi$ encodes a CNF formula
over $n$ variables and $m$ clauses and $\Pi$ encodes a Resolution
refutation of $\Reff_s(\varphi)$ of length at most $t$, then there exists an $(\ell,w)$-good seed
$\sigma\in\{0,1\}^{\ell}$ for $\Pi$.
\end{theorem}

\begin{proof}
    The case of odd $s$ reduces to the even case by first applying the fixed
one-block restriction used in the odd-$s$ case of
\Cref{lemma:restriction}, so we may assume without loss of generality that $s$ is even.
    Let $w=8\lceil\log t\rceil$, $q=q(s,w)=2^k$, and $\ell=q^4$.
    By the definition of $q$, we have $q\geq w$,
    $|s|\leq2k(q+1)$, and $q\in\Log$. Hence we may invoke
    \Cref{lemma:design} to obtain an
    $(s,q^4,q^2,q)$-design $M$. Moreover, every submatrix formed by at
    most $q$ rows has full row rank over $\bbF_2$. We refer to this
    matrix as $M$, although formally its rows are accessed through the
    function $\Design$.

    Put $r\coloneq w/2$. Let $i\in[t]$ be such that $D_i$ has block-width
    at least $w$. Since every pair contains only two blocks, $D_i$ mentions
    at least $r$ distinct pair-indices. Let $N$ denote the number of variables of $\Reff_s(\varphi)$, and let
    $\operatorname{block}(\lambda)$ denote the $\PV$-function returning
    the block containing the literal encoded by $\lambda$. For a
    number-coded sequence $C\in[2N]^r$, let
    $\operatorname{Valid}(i,C)$ be the sharply bounded formula
    asserting that every literal occurring in $C$ occurs in $D_i$, and
    that the $r$ literals of $C$ belong to pairwise distinct pair-indices.

    Since $D_i$ has block-width at least $w$, some such valid~$C$ exists.
    Define $C_i$ to be the lexicographically first one. Equivalently,
    its graph is defined by
    \ifcameraready
    \begin{equation}
        \begin{split}
        \label{form:Ci-form-Di}
        \operatorname{Choice}&(i,C)
        \coloneq
        \operatorname{Valid}(i,C)
        \land \phantom{a} \\
        &\forall C'\in[2N]^r\,
        \bigl(
            C'<_{\mathrm{lex}}C
            \to
            \neg\operatorname{Valid}(i,C')
        \bigr).
        \end{split}
    \end{equation}
    \else
    \begin{equation}
        \label{form:Ci-form-Di}
        \operatorname{Choice}(i,C)
        \coloneq
        \operatorname{Valid}(i,C)
        \land
        \forall C'\in[2N]^r\,
        \bigl(
            C'<_{\mathrm{lex}}C
            \to
            \neg\operatorname{Valid}(i,C')
        \bigr).
    \end{equation}
    \fi
    Because $r\in\Log$, $\operatorname{Valid}$ is sharply bounded
    relative to $\Pi$, while \cref{form:Ci-form-Di} is
    $\formPib^b_1(\Pi)$. Testing whether $D_i$ has block-width at least
    $w$ is $\formSigmab^b_1(\Pi)$ by \cref{form:bw}. Thus we can define
    in $\PV_2(\Pi)$ a function $f$ which outputs this $C_i$ whenever
    $D_i$ is wide, and outputs the empty sequence otherwise.

    Let $W\coloneq\{i\in[t]:\parambwformula\}$ be the set collecting the clauses of block-width at least $w$. For every
    $i\in W$, the clause $C_i$ has width and pair-width exactly $r$, so
    $|I(C_i)|=2r=w\leq q$. By \Cref{lemma:design},
    $\rank_{\bbF_2}M_{I(C_i)}=w=2r$. Therefore
    \Cref{lemma:restrictions-clause-level} gives
    \begin{equation}
        \label{eq:bad-Ci-bound}
        \BadSet_{C_i}
        \preceq_0
        \left(\frac34\right)^{w/2}2^\ell.
    \end{equation}

    Now, if $W$ is empty then every seed is
    already good. Otherwise, let $i_0$ be the least element of $W$; since \cref{form:bw} is
$\formSigmab^b_1(\Pi)$, bounded minimization is
available in $\PV_2(\Pi)$. Fix
    $B\coloneq3^r2^{\ell-2r}$, and let
    $G_i:[B]\twoheadrightarrow\BadSet_{C_i}$ denote the surjection from
    \Cref{lemma:restrictions-clause-level}. Define $H(i,z)=G_i(z)$ when
    $i\in W$, and $H(i,z)=G_{i_0}(z)$ otherwise. Then $H$ is a
    $\PV_2(\Pi)$-definable surjection
    \begin{equation}
        \label{eq:exact-union-bound}
        [t]\times[B]
        \twoheadrightarrow
        \bigcup_{i\in W}\BadSet_{C_i},
    \end{equation}
    and hence
    $\bigcup_{i\in W}\BadSet_{C_i}
      \preceq_0
      t(3/4)^{w/2}2^\ell$.

    We claim that some seed lies outside this union. Suppose otherwise.
    Then $2^\ell\subseteq\bigcup_{i\in W}\BadSet_{C_i}$, and by
    \Cref{lemma:trivial-Emil}.(i), \cref{eq:exact-union-bound}, and
    transitivity, we get
    $2^\ell\preceq_0t(3/4)^{w/2}2^\ell$.
By \Cref{lemma:trivial-Emil}.(iii), taking
$\xi=1/4$, this implies
$1\leq t(3/4)^{w/2}+1/4$.
But since $w=8\lceil\log t\rceil$, writing
$h\coloneq\lceil\log t\rceil\geq 1$, we have
\begin{equation}
    t\left(\frac34\right)^{w/2}
    \leq 2^h\left(\frac34\right)^{4h}
    =\left(\frac{81}{128}\right)^h
    \leq\frac{81}{128}.
\end{equation}
Consequently,
$1\leq 81/128+1/4=113/128<1$,
a contradiction.

    Hence there exists
    $\sigma\notin\bigcup_{i\in W}\BadSet_{C_i}$. Consider the global
    restriction whose $s$ restriction bits are
    $\rho^\sigma_j=\Design(s,q,j)\cdot\sigma^\top$ for $j\in[s]$.
    For every $i\in W$, the projection of this restriction to $I(C_i)$
    is $M_{I(C_i)}\sigma$, and since
    $\sigma\notin\BadSet_{C_i}$ we have
    $C_i{\restriction_{\rho^\sigma}}=1$. As $C_i\subseteq D_i$, it
    follows that $D_i{\restriction_{\rho^\sigma}}=1$. Thus every clause
    of block-width at least $w$ is killed by the restriction generated
    by~$\sigma$, which is precisely to say that $\sigma$ is
    $(\ell,w)$-good.
\end{proof}

\begin{remark}[Good seeds in the exponential-size regime:
formalization in $\APC(\alpha)$]
    \label{rem:good-seeds-in-exp-regime}
    Suppose we want to prove \Cref{thm:good-seeds} in the
    \emph{exponential-size regime}, where $n,m,s\in\Log$. In this
    setting the argument goes through in $\APC(\Phi,\Pi)$. The design
    construction and \Cref{lemma:restrictions-clause-level} are already
    unrelativized, and the only reason the general proof requires
    $\APCtwo$ is the $\PV_2(\Pi)$-selection of the small subclauses
    $C_i$. When $s,n,m\in\Log$, the relevant searches through the
    variables and blocks of $D_i$ are sharply bounded, so the choice of~$C_i$ can be carried out in $\PV(\Pi)$. Consequently, the uniform
    surjection used in \cref{eq:exact-union-bound} is also available in
    the $\APC(\Phi,\Pi)$ setting.
\end{remark}

We finish this section with the following note on more general forms of XOR
lifting.

\begin{remark}[General XOR lifting]
    The restriction argument employed for $\Reff$ formulas follows the
    pattern of \emph{XOR lifting}, attributed by
    \citeauthor{BenSasson09}~\autocite[§4]{BenSasson09} to Alekhnovich and
    Razborov around 2001, and known as \emph{relativization} in the
    context of model-theoretic techniques for Resolution lower bounds
    \autocite{DR03, Kra11}. Roughly speaking, for a CNF formula $\varphi$,
    the XOR-lifted $\varphi[\oplus]$ is obtained by replacing each
    variable $x_i$ of $\varphi$ by $y_i \oplus z_i$, where the latter
    are new variables. The same restriction argument employed for
    $\Reff$ formulas works here, in the sense that if a formula
    $\varphi$ requires Resolution refutations of width $w$, then
    Resolution refutations of $\varphi[\oplus]$ require exponential size
    $2^{\Omega(w)}$. Following the same arguments in this section, one
    can show that if a theory $T(\alpha)\supseteq\APCtwo(\alpha)$ proves
    a width lower bound for some formula $\varphi$, then this same theory
    proves a size lower bound for the XOR-lifted formula. In the
    exponential-size regime an analogous result is implicit in the
    decision tree reductions of \autocite[Theorem 6.4]{LLR24} and is
    formalizable in $\LLR$. The difference in using $w$-wise
    independence is again that we do not need to assume the parameters
    are in $\Log$ and we can thus derive fixed-polynomial size lower
    bounds.
\end{remark}

\section{The block-width lower bound in \texorpdfstring{$\BussTthreetwo(\alpha)$}{T32(alpha)}}
\label{sec:BW-formalized-Ttwo}

In this section, we prove within $\BussTthreetwo(\alpha)$ the block-width lower bound for $\Reff$ formulas. We follow the new inductive proof explicated in \Cref{subsubsec:intro-techniques-new-algo}. 
The crux of the proof is to define suitable formulas to encode the following properties: for a clause $C$ in a Resolution refutation of $\EReff_s(\varphi)$,

\begin{enumerate}[label=(\roman*)]
    \item \label{item-bw} whether $C$ \emph{has block-width at least $w$};
    \item \label{item-DC} whether $C$ is \emph{Delayer consistent}, that is, whether there exists a full reservation $R$ that satisfies the \emph{reservation invariant} from \Cref{def:reservation-invariant} with respect to $C$;
    \item \label{item-encodes} whether $C$ \emph{encodes a satisfying assignment to $\varphi$}: whether there is a reservation $R$ such that $C$ is Delayer consistent with respect to $R$ and this reservation contains information about a satisfying assignment (a formal definition follows in \Cref{subsubsec:assignment-formulas}).
\end{enumerate}


The concept of block-width is clear: we already gave a $\formSigmab^b_1(\Pi)$-formula (\ref{form:bw}) to check block-width in \Cref{subsec:formulas-BW}. The concept of a clause $C$ \emph{encoding a satisfying assignment}, intuitively, corresponds to the quasipolynomial-time algorithm presented in \Cref{subsubsec:intro-techniques-new-algo} that extracts a satisfying assignment from $C$. The proof will be in $\BussTthreetwo(\Phi, \Pi)$, where again we relativize the theory with access to two new relation symbols. 

The outline of this section is as follows. In \Cref{subsec:width-lb-formulas} we define relativized first-order formulas for the concepts listed above. In \Cref{subsec:bw-proof} we prove the block-width lower bound in $\BussTthreetwo(\Phi, \Pi)$.

\subsection{Bounded formulas for Delayer-consistency and assignment extraction}
\label{subsec:width-lb-formulas}

We assume that the relation symbol $\Pi$ encodes a Resolution refutation consisting of $t$ clauses, and that these are listed in $\Pi$ as a topological ordering as $C_1, C_2, \dots, C_t$, where $C_1$ is the empty clause.

\subsubsection{Formula for Delayer consistency}
\label{subsubsec:DC-formulas}

The natural encoding of Delayer consistency has an existential quantifier guessing a reservation, but this is infeasible: full reservations contain information about the literals present in the blocks, so this requires $\poly(n)$ bits to represent. Instead, we observe that to check whether a clause $C$ is Delayer consistent, it essentially suffices to check that a \emph{basic reservation} $R$, as in \Cref{def:reservation}, satisfies some basic properties that allow for it to be extended into a full reservation verifying the reservation invariant. 

Since information about resolved variables is fully determined by the
layer at which blocks appear, a basic reservation can be represented
just by the information about its pointers. As long as a basic
reservation only talks about blocks $B$ mentioned in a clause $C$ of
low block-width, this partial assignment can be encoded
succinctly. In this context, we say that $R$ is a \emph{basic
  reservation for $C$} if $R$ is a basic reservation that fixes the
pointers of all blocks in $B$, and only of such blocks.


Let $w$ be an upper bound on the number of blocks mentioned in $C$.
A basic reservation $R$ for $C$ is encoded as a string over $[\splus]^{3w} \times [m]^w$, which we interpret as follows. For every block $b\in [\splus]$ mentioned in the clause~$C$, there is an entry for it in $R$ with a tuple of the form $\langle b, c, d, j \rangle$, where $c,d\in [\splus]$ are the blocks $b$ points to on the left and right, respectively, in case the block $b$ was derived; otherwise, $j \in [m]$ points to an axiom of $\varphi$ that $b$ was weakened from. Furthermore, if there is a tuple in $R$ starting with $b \in [\splus]$, then $b$ is mentioned in $C$. The string $R$ has then size $O(w \log s + w\log m)$. We use the notation $R(b) = \langle c, d, j \rangle$ to refer to the tuple starting with $b$, and write $b \in \dom(R)$ to indicate that such a tuple exists in $R$. Although $R$ is technically not indexed by values in $[\splus]$, we can always find this entry by the sharply bounded formula $\exists k \in [\len(R)] \ (R)_k = \langle b, c, d, j \rangle$, since $\len(R) \in \Log$ for $w \in \Log$. Whenever $b\in\dom(R)$, we write $R_{\mathrm L}(b)$, $R_{\mathrm R}(b)$, and $R_{\mathrm A}(b)$ for the left-pointer, right-pointer, and axiom coordinates of $R(b)$, respectively. These are just tuple projections: each can be recovered by a sharply bounded search through the entries of $R$. In particular, using this notation below does not introduce new bounded quantifiers.


It is straightforward to see that given $R$, we can write a $\formSigmab^b_2(\Phi, \Pi)$-formula that checks that $R$ is a basic reservation (\Cref{def:reservation}): its pointers only relate blocks between consecutive layers, etc. To be able to talk about layers, we have a $\PV$-function to compute the layer of a given block, defined as
\begin{equation}
    \label{form:layer}
    \tag{$\layer$}
    \layer(b; s, n) \coloneq \min\!\left\{n,\left\lfloor\frac{s-b}{\lfloor s/n\rfloor}\right\rfloor+1\right\}\,,
\end{equation}
although throughout the rest of this section we often omit $s$ and $n$ as inputs, as these are clear from context.

Furthermore, we can also check with a $\formPib^b_2(\Pi)$ formula that $R$ is a basic reservation \emph{for $C$}: a block is mentioned in $C$ if and only if there is a tuple for it in $R$,
\begin{equation}
   \forall b\in[\splus] \, \left( \parampresent \leftrightarrow \exists k\in[\len(R)] \ (R)_{k,1 } = b \right) \,.
   \label{eq:BforC}
\end{equation}
We regard $R$ as a sequence of at most $w$
tuples, so that $\len(R)\leq w$; in particular,
the empty sequence is permitted.


Our goal is now to define a formula that checks whether there exists a
full reservation $R'$ extending~$R$ satisfying the reservation
invariant of \Cref{def:reservation-invariant}.
Note that the pointers of $R$ set up a forest-like structure between
the blocks.
This structure restricts $R'$ in the choice of literals in
  clauses. Since we cannot traverse trees in low depth, in order to
  reason about these constraints, we need an additional existential quantifier to guess some extra
  information regarding which blocks appear in
  what subtrees.
%
Let us specify the required extra information.

For every block $b\in\dom(R)$, we record the \emph{set
of proper descendants of $b$ according to the
pointers fixed by $R$}. These descendants may
include blocks outside $\dom(R)$, namely children
pointed to by $R$ that are not mentioned in $C_i$.
Since $R$ contains at most $w$ tuples, a block can have
at most $2w$ proper descendants (since a binary tree with at most $w$ internal nodes can have at most $w+1$ leaves).

We encode this collection as a sequence $S$ of
at most $w$ tuples of the form
$\langle b,c_1,\ldots,c_k\rangle$, where
$k\leq 2w$. Thus $S$ requires at most
$w(2w+1)$ block entries, or
$O(w^2\log s)$ bits. We write $S(b)$ for the
set of descendants recorded in the tuple
beginning with $b$, and adopt the convention
$S(b)=\varnothing$ whenever $b\notin\dom(R)$.
In order to ensure that this collection $S$ indeed encodes the subtrees of all the nodes in $R$ we impose that $S$ satisfies some basic closure properties. 
To state the precise conditions let us say that $R$ \emph{fixes a pointer of $b$ to $c$} if $R$ fixes the left or right pointer of $b$ to $c$. The conditions we impose are that
\begin{enumerate}[label=(\alph*)]
    \item for every block $b$ mentioned in $R$, there is exactly one tuple beginning with $b$ in $S$; \label{it:firstcondition}
    \item for every $c \in S(b)$, it holds that $\layer(c) > \layer(b)$;
    \item if $c \in S(b)$ and $\layer(c) = \layer(b) + 1$, 
    $R$ fixes a pointer of $b$ to $c$;
    \item if 
    $R$ fixes a pointer of $b$ to $c$, then $c \in S(b)$;
    \item for every $c\in S(b)$, we have that $S(c) \subseteq S(b)$;
    \item if $c \in S(b)$, then either $R$ fixes a pointer of $b$ to
      $c$, or there is some $c' \in S(b)$ such that
      $\layer(c') < \layer(c)$ and $R$ fixes a pointer of $c'$ to $c$. \label{it:lastcondition}
\end{enumerate}

Here, $S$ is a sequence of exactly one tuple
for each block in $\dom(R)$, with no additional
tuples. Whenever $c\notin\dom(R)$, expressions
involving $S(c)$ are interpreted using the
convention $S(c)=\varnothing$.  It is not hard to verify that $S$
  satisfies the above properties if and only if it contains exactly
  one tuple per block $b$ containing all the blocks present in
  the subtree rooted at $b$.  It is straightforward to check that
all of the above properties are $\formPib^b_1(\Pi)$.

With $R$ and $S$ fixed we want to check whether there exists a full
reservation $R'$ that extends $R$, is defined over the same blocks 
$B$ as $R$ is, and satisfies the reservation invariant of
\Cref{def:reservation-invariant}.
%
%
Since $R'$ only adds information to blocks in $B$ it therefore
satisfies item \ref{item:inv-bw} of \Cref{def:reservation-invariant}.
It thus follows that in order to decide whether there is such a full
reservation $R'$ it suffices to verify \ref{item:inv-C} and
\ref{item:inv-axiom} of \Cref{def:reservation-invariant}. In other
words, we need to check that $R'$ falsifies $C$ and does not falsify
any axiom of $\EReff_s(\varphi)$.
We cannot, however, check this explicitly as we do \emph{not} have access to $R'$.
Instead, we
%
exhibit a small certificate that guarantees the existence of such an $R'$.

Note that by virtue of $R$ being a basic reservation, axioms
(\ref{axiom:must-resolve}) through (\ref{axiom:root-enabled}) are satisfied
for all blocks $B$ (and for other blocks these axioms are
unrestricted). Hence, we need to define formulas checking whether there exists a
full reservation $R'$ extending $R$ that satisfies condition \ref{item:inv-C} and
axioms (\ref{axiom:must-appear-after-res-left}) to
(\ref{axiom:must-appear-after-weak}).

First, recall that a basic reservation fixes, for every block mentioned in
$C$, whether the block is derived or weakened, its left and right pointers
in the former case, the axiom it points to in the latter case, and the
variable which it resolves over. 
%
We therefore have to check that every literal of $C$
over variables of type $\derived$, $\res$, $\lpoint$, $\rpoint$, and $\weak$
is falsified by the corresponding value prescribed by $R$. 
%
It is readily verified that this can be expressed as bounded
formulas directly from the encoding of $R$.

It remains to determine whether there is an extension $R'$ of $R$
  assigning the $\lit$-variables such 
that they simultaneously falsify $C$ and satisfy axioms
(\ref{axiom:must-appear-after-res-left})--(\ref{axiom:must-appear-after-weak}).
  The only additional constraints on the assignment of literals originate from
  the forest-like structure of $R$.
  With the help of $S$ we propagate all the relevant constraints
  forced upon $R'$ by $R$ and the minimal partial assignment falsifying $C$.
Let us formally define the formulas that check whether such an $R'$ exists.

Denote by $i\in [t]$ the index of the clause $C = C_i$ we consider, and fix $R$ and $S$.
We check that a block $b$ on layer $h$ only contains literals over
$x_1,\dots,x_h$. Hence, for every $j>\layer(b)$, $\neg \lit^b_{x_j}$ and
$\neg \lit^b_{\neg x_j}$ cannot be present in $C_i$. Formally, we check
that the formula
\begin{equation}
    \forall b\in\dom(R)\ \forall j\in[n]\,
    \Bigl(j>\layer(b)\to
    \neg\Pi\bigl(\lit^i_{\neg\lit^b_{x_j}}\bigr)
    \land
    \neg\Pi\bigl(\lit^i_{\neg\lit^b_{\neg x_j}}\bigr)\Bigr) \,,
    \label{eq:DC-out-of-range}
\end{equation}
is satisfied.

Let us say that the clause $C_i$ \emph{forces} a variable to true (respectively, false) 
if the variable appears only negatively (respectively, only positively) in $C_i$. That is, the 
unique minimal falsifying assignment to $C_i$ sets this variable to true 
(respectively, false).
We say that $C_i$
\begin{itemize}
\item \emph{forces a literal $\ell$ to appear in a block $b$} if $C_i$ forces $\lit_{\ell}^b$ to true;
\item \emph{prevents a literal $\ell$ from appearing in a block $b$} if $C_i$ forces $\lit_{\ell}^b$ to false or forces $\lit_{\lnot \ell}^b$ to true.
\end{itemize}

Since every reserved block contains exactly one polarity of
each variable up to its layer, preventing a literal from
appearing also forces its opposite polarity to appear.
Moreover, the axioms selected by $R$ impose additional
constraints on the contents of blocks in the last layer.
To account for both sources of information, let
$\operatorname{Pos}_{i,R}(c,j)$ and
$\operatorname{Neg}_{i,R}(c,j)$ abbreviate
\begin{align*}
\operatorname{Pos}_{i,R}(c,j) \coloneq {}&
    \Pi(\lit^i_{\neg\lit^c_{x_j}})
    \lor \Pi(\lit^i_{\lit^c_{\neg x_j}}) \lor
    (c\in\dom(R)\land\layer(c)=n
    \land\Phi(x_j,R_{\mathrm A}(c))),\\
\operatorname{Neg}_{i,R}(c,j) \coloneq {}&
    \Pi(\lit^i_{\neg\lit^c_{\neg x_j}})
    \lor \Pi(\lit^i_{\lit^c_{x_j}}) \lor
    (c\in\dom(R)\land\layer(c)=n
    \land\Phi(\neg x_j,R_{\mathrm A}(c))).
\end{align*}
These formulas express that $x_j$ and $\neg x_j$,
respectively, are forced to appear in block $c$,
for $j\leq\layer(c)$.

For a block $b$ mentioned in $C_i$ and its subtree
$S(b)$, we need to ensure that the information forced
at different blocks is consistent. In particular, for
every variable $x_j$ with $j\leq\layer(b)$, no two
blocks in $S(b)\cup\{b\}$ can be forced to contain
opposite polarities of $x_j$. This is expressed by

\begin{align}
    \label{eq:DC-propagation}
    \forall b\in\dom(R)\
    \forall c,d\in S(b)\cup\{b\}\
    \forall j\in[\layer(b)]\
    \neg\bigl(
        \operatorname{Pos}_{i,R}(c,j)
        \land
        \operatorname{Neg}_{i,R}(d,j)
    \bigr).
\end{align}

On top of the propagation step above, we also check
that the information forced by $C_i$ and by the axioms of \(\varphi\) specified by \(R\) at final-layer blocks is consistent. In particular, the left subtree
of a block resolving over $x_j$ must contain $x_j$,
while its right subtree must contain $\neg x_j$.
Formally, we check that

\begin{align}
  \label{eq:DC-resolution}
  \begin{split}
  \forall b\in\dom(R)\quad
  \biggl(\layer(b)<n \to {}&
  \forall c\in S(R_{\mathrm L}(b))
       \cup\{R_{\mathrm L}(b)\}\
  \neg\operatorname{Neg}_{i,R}
       (c,\layer(b)+1)
  \\
  &{}\land
  \forall d\in S(R_{\mathrm R}(b))
       \cup\{R_{\mathrm R}(b)\}\
  \neg\operatorname{Pos}_{i,R}
       (d,\layer(b)+1)\biggr).
  \end{split}
\end{align}

where we recall that $R_{\mathrm{L}}(b)$ (respectively, $R_\mathrm{R}(b)$) denotes
the block that $R$ fixes the left (respectively, right) pointer of $b$ to.
 We use the convention $S(c)=\emptyset$ whenever
$c\notin\dom(R)$.

Finally, we need the analogous condition for a block weakened from an axiom
of $\varphi$. Suppose that $b$ is a block on the last layer and the basic
reservation fixes the axiom $R_{\mathrm A}(b)$ of $\varphi$ from which $b$ is weakened.
Every literal appearing in that clause is not prevented from appearing in $b$. Thus we
require
\begin{align}
  \label{eq:DC-weakening}
  \begin{split}
 \forall b\in\dom(R)\ \layer(b)=n
  \ \to\
  \forall j\in[n]\,
  &\left(
    \Phi\bigl(x_j,R_{\mathrm A}(b)\bigr)
    \to
    \neg\Pi\Bigl(\lit^i_{\lit^b_{x_j}}\Bigr)
    \land
    \neg\Pi\Bigl(\lit^i_{\neg \lit^b_{\neg x_j}}\Bigr)
  \right) \land \\
  &\left(
    \Phi\bigl(\neg x_j,R_{\mathrm A}(b)\bigr)
    \to
    \neg\Pi\Bigl(\lit^i_{\lit^b_{\neg x_j}}\Bigr)
    \land
    \neg\Pi\Bigl(\lit^i_{\neg \lit^b_{x_j}}\Bigr)
  \right)\,.
  \end{split}
\end{align}

This completes the description of the conditions we need to check. These are sufficient to enforce axioms
(\ref{axiom:must-appear-after-res-left})--(\ref{axiom:must-appear-after-weak}).

To see this, start with the values of the non-$\lit$ variables
prescribed by the basic reservation $R$. For each variable $x_j$,
consider the connected components of the reserved forest
restricted to layers at least $j$. Within each such component,
the polarity of $x_j$ must be constant. The propagation conditions
guarantee that the information forced by $C_i$ and the selected
axioms is consistent within each component, while the Resolution
conditions ensure compatibility with the polarities imposed by
the pointers of $R$. We can therefore choose a consistent polarity
for each component, respecting all these constraints, and choose
one arbitrarily whenever neither polarity is forced.

This determines the contents of all reserved blocks. By
\cref{eq:DC-out-of-range}, no block is forced to contain a
literal involving a variable whose index exceeds the block's layer, and the resulting assignment satisfies
(\ref{axiom:must-appear-after-res-left}) and
(\ref{axiom:must-appear-after-res-right}).
Moreover, \cref{eq:DC-weakening} ensures compatibility with
(\ref{axiom:must-appear-after-weak}).
Hence the resulting full reservation $R'$ falsifies $C_i$
and does not falsify any axiom of $\EReff_s(\varphi)$.
This shows that $R'$ satisfies the reservation invariant
with respect to $C_i$.

Now, let $\DC(i,R,S,n,m,s,w)$ denote the conjunction of all
formulas we have discussed so far. Note that \cref{eq:BforC} is
$\formPib_2^b(\Pi)$ and the rest are
$\formSigmab_2^b(\Phi,\Pi)$. By the preceding discussion,
$\DC(i,R,S,n,m,s,w)$ holds exactly when $R$ is a basic
reservation for $C_i$, $S$ correctly encodes its subtrees
(by virtue of conditions
\ref{it:firstcondition}-\ref{it:lastcondition}),
and $R$ can be extended to a full reservation $R'$
satisfying the reservation invariant with respect to $C_i$.
For $w\in\Log$, we consider the formula
\begin{align}
    \label{form:DC-with-quantifiers}
    \tag{$\operatorname{Delayer-Cons}$}
    \begin{split}
    \paramDC
    \coloneq &\\
    \exists R \in [\splus]^{3w} &\times [m]^w \ 
    \exists S \in [\splus]^{w(2w+1)} \ 
    \DC(i,R,S,n,m,s,w)\,.
    \end{split}
\end{align}

We conclude that $\paramDC$ is a
$\formSigmab_3^b(\Phi,\Pi)$-formula expressing that
the clause $C_i$ is Delayer consistent.
If $n,m,s\in\Log$, then the formula is
$\formSigmab^b_2(\Phi,\Pi)$.


\subsubsection{Formula for assignment extraction}
\label{subsubsec:assignment-formulas}

Consider a clause $C_i$ in $\Pi$. Recall from
\Cref{subsubsec:DC-formulas} that Delayer-consistency is witnessed by
a basic reservation $R$, together with a collection $S$ encoding its
subtrees, such that $R$ can be extended to a full reservation $R'$
satisfying the reservation invariant. We say that $C_i$ \emph{encodes
  a satisfying assignment} 
if there is such a choice of $R$, $S$, and $R'$ such that from $R'$ we
can deduce a clause 
whose unique 
falsifying assignment satisfies $\Phi$, as formally defined below.
This corresponds precisely to the quasipolynomial-time procedure described
in \Cref{subsubsec:intro-techniques-new-algo}: instead of guessing the
contents of a full reservation $R'$, we guess only its pointers, and use the
information appearing in $C_i$, together with the connected components
determined by those pointers, to reconstruct the contents of a block.

The goal of this section is to formally define a  $\formSigmab^b_3(\Phi,\Pi)$-formula 
$\paramassignment$ that is true if and only if
the clause $C_i$ encodes a satisfying assignment 
(of $\Phi$), as explained above. 
%
To do this, suppose we have a $\formSigmab^b_1(\Phi,\Pi)$-formula $\satclause(k,i,R,S,b,n)$ encoding whether 
the assignment given, when inspecting block $b$,  by propagating the information given by clause
$C_i$, together with the reservation $R$ and the extra information
$S$,  satisfies the $k$-th clause of $\Phi$.
We can now define the formula
\begin{align}
    \label{form:assignment}
    \tag{$\assignmentname$}
    \begin{split}
        \paramassignment
        \coloneq
        \exists R\in[\splus]^{3w}\times[m]^w\,
        \exists S\in[\splus]^{w(2w +1)}\,
        \exists b\in[\splus]\,\\
        \Bigl(
            \DC(i,R,S,n,m,s,w)
            {}\land
            \forall k\in[m]\,
            \satclause(k,i,R,S,b,n)
        \Bigr)\,.
    \end{split}
\end{align}
Formally, we say that the clause $C_i$ \emph{encodes a satisfying assignment}
if (\ref{form:assignment}) holds.

In order to define $\satclause(k,i,R,S,b,n)$, we 
need a $\formSigmab^b_0(\Phi,\Pi)$-formula $\alpha(i,R,S,b,j)$ encoding
whether the variable $x_j$ is forced to $1$ on block $b$ by $C_i, R$ and $S$. 
Now, given a literal $\ell\in[2n]$, where $\ell\in[n]$ represents
$x_\ell$ and $\ell\in[2n]\setminus[n]$ represents
$\neg x_{\ell-n}$, we can check whether the assignment defined above
satisfies $\ell$ by the $\formSigmab^b_0(\Pi)$-formula
\begin{equation}
    \label{form:satlit}
    \tag{$\satlit$}
    \satlit(i,R,S,b,\ell,n)
    \coloneq
    \bigl(
        \ell\in[n]
        \land
        \alpha(i,R,S,b,\ell)
    \bigr)
    \lor
    \bigl(
        \ell\in[2n]\setminus[n]
        \land
        \neg\alpha(i,R,S,b,\ell-n)
    \bigr)\,,
\end{equation}
and then use this formula to define
\begin{equation}
    \label{form:satclause}
    \tag{$\satclause$}
    \satclause(k,i,R,S,b,n)
    \coloneq
    \exists\ell\in[2n]\,
    \Bigl(
        \Phi(\ell,k)
        \land
        \satlit(i,R,S,b,\ell,n)
    \Bigr)\,.
\end{equation}
Thus the candidate assignment satisfies the whole formula precisely
when $\forall k\in[m] \satclause(k,i,R,S,b,n)$. As in
(\ref{form:Sat-FO}), this is a $\formPib^b_2(\Phi,\Pi)$ condition.

It remains to define $\alpha(i,R,S,b,j)$. For this, 
use the following two formulas:
\begin{equation}
    \label{form:below}
    \operatorname{Below}_{R,S}(b,c)
    \coloneq
    (b=c)
    \lor
    \bigl(c\in\dom(R)\land b\in S(c)\bigr)\,,
\end{equation}
encoding that block $b$ is either identical to $c$ or it is a descendant of $c$ according to the pointers fixed by $R$; and for $j\in[n]$, 
\begin{equation}
    \label{form:same-component}
    \operatorname{Same}_{R,S}(b,d,j)
    \coloneq
    \exists c\in\dom(R)\,
    \Bigl(
        j\leq \layer(c)
        \land
        \operatorname{Below}_{R,S}(b,c)
        \land
        \operatorname{Below}_{R,S}(d,c)
    \Bigr)\,.
\end{equation}
The point of the condition $j\leq\layer(c)$ is that every Resolution step
strictly below $c$ resolves over a variable with index larger than $j$.
Consequently, all blocks below $c$ contain the same polarity of $x_j$.


We now reconstruct one particular completion of the contents of $b$. For
$j\in[n]$, let $\alpha(i,R,S,b,j)$ express that the assignment obtained by
falsifying this clause sets $x_j$ to $1$.
There are two ways in which this
can be forced.
First, $C_i$ may explicitly determine the polarity of $x_j$ at some block
$d$ whose $x_j$-content must agree with that of $b$. Since the reservation
falsifies $C_i$, either a positive occurrence of $\lit^d_{x_j}$ or a
negative occurrence of $\lit^d_{\neg x_j}$ forces $\neg x_j$ to be present
in $d$, and hence also in $b$. Second, the pointers themselves may force
$b$ to lie below the right child of a block resolving over $x_j$, which
again forces $\neg x_j$ to be present. We therefore define
\begin{align}
    \label{form:alpha}
    \begin{aligned}
\alpha(i,R,S,b,j) \coloneq {}&
\exists d\in\operatorname{dom}(R)\
\Bigl(
    \operatorname{Same}_{R,S}(b,d,j)
    \land
    \operatorname{Neg}_{i,R}(d,j)
\Bigr)
\\
&{}\lor
\exists c\in\operatorname{dom}(R)\
\Bigl(
    j=\layer(c)+1
    \land
    \operatorname{Below}_{R,S}(b,R_R(c))
\Bigr).
\end{aligned}
\end{align}
All the quantifiers ranging over $\dom(R)$ or over the
entries of a subtree $S(c)$ are sharply bounded, and
hence the formula $\alpha$ above is
$\formSigmab^b_0(\Phi,\Pi)$. Let us justify the definition.
The first disjunct accounts for the information forced
by $C_i$ or by an axiom selected in the reservation.
In either case, $\operatorname{Neg}_{i,R}(d,j)$ forces
$\neg x_j$ to appear in $d$, and
$\operatorname{Same}_{R,S}(b,d,j)$ guarantees that the
same polarity must appear in $b$. The second disjunct
accounts for the information imposed by the Resolution
pointers: a block at layer $j-1$ resolves over $x_j$,
and its right child must contain $\neg x_j$. Hence,
if $b$ lies below this right child, then $\neg x_j$
must appear in $b$. If neither condition holds, we
complete the contents of $b$ by taking $x_j$.
By the consistency conditions included in $\DC$,
this completion does not contradict any information
forced by $C_i$, the selected axioms, or the pointers
of $R$.

This completes the definition of (\ref{form:assignment}).

\subsection{The block-width lower bound}
\label{subsec:bw-proof}

With the formal definitions in place, we are ready to prove the block-width lower bound. We start by recalling the following simple fact.

\begin{lemma}
    \label{lemma:fact}
    The theory $\mathsf{T}^1_1(\Phi, C)$ proves the following statement. For every $n$ and $m$, if $\Phi$ encodes a CNF formula $\varphi$ over $n$ variables and $m$ clauses and $C$ encodes a non-tautological width-$n$ clause that is not the weakening of any clause of $\varphi$, then the formula $\varphi$ is satisfied by the unique assignment that falsifies $C$.
\end{lemma}

\begin{proof}[Proof sketch]
    The proof does not use any induction. The definition of the assignment and the satisfiability condition 
    is encoded analogously to (\ref{form:satclause}) (see \Cref{subsec:prelim-sat-reflection} for details). Suppose for contradiction that the assignment does not satisfy $\varphi$, meaning that there is a clause $C_i$ of $\varphi$, with $i \in [m]$, such that~$C_i$ is falsified by our assignment. By definition of the assignment, this means $C_i$ is a subclause of $C$, so in fact~$C$ is a weakening of $C_i$. Contradiction. 
\end{proof}

We now move to the actual inductive argument.

\begin{lemma}
    \label{lemma:bw-auxiliary}
    The theory $\BussTthreetwo(\Phi, \Pi)$ proves the following $\forall\formSigmab^b_3(\Phi, \Pi)$ statement. For every $n$, $m$, $s$ and $t$ and every $w \in \Log$ with $w \le 1/2(\lfloor s/n\rfloor -1)$, 
    suppose~$\Phi$ encodes a CNF formula $\varphi$ over $n$ variables and $m$ clauses, suppose $\Pi$ encodes a length-$t$ Resolution refutation of $\EReff_s(\varphi)$ and suppose that  for every enabled clause $i \in [t]$, 
    \begin{enumerate}[label=(\alph*)]
        \item the clause $C_{i}$ 
        has block-width less than $w$; and \label{item:bw-auxlemma}
        \item $C_i$ does not encode a satisfying assignment. \label{item:sat-auxlemma}
    \end{enumerate}
    Then, for every $i \in [t]$, it holds that there exists $j > i$ such that $j \leq  t$, the clause $C_j$ is enabled and $C_j$ is Delayer consistent.
\end{lemma}

\begin{proof}
     Let us first check that the complexity of the statement is
    $\forall\formSigmab^b_3(\Phi,\Pi)$. 
    Formally, the statement is that for all $n, m, s, t$, for every $w \in \Log$ and for every $i\in [t]$, either $\Phi$ does not
    encode a CNF formula, or $\Pi$ does not encode 
a correct Resolution refutation of $\Reff_s(\varphi)$, or there exists $i' \in [t]$ such that \ref{item:bw-auxlemma} or \ref{item:sat-auxlemma} are false, or there exists $j\in [t]$ such that $j>i$ and $C_j$ is Delayer consistent.
    
    The formulas checking that $\Pi$ encodes
    a correct Resolution refutation of $\Reff_s(\varphi)$ and that $\Phi$
    encodes a CNF formula are both $\formPib^b_2(\Phi,\Pi)$, so their negation is 
    $\formSigmab^b_2(\Phi,\Pi)$.
    Since the formula 
    (\ref{form:bw}) encoding that $C_i$ has block-width at least $w$ is $\formSigmab^b_1(\Phi,\Pi)$,
    and the formula 
    (\ref{form:assignment}) is a $\formSigmab^b_3(\Phi,\Pi)$-formula, then we have that 
    the formula 
    \begin{equation}\label{eq:BWorSAT}
        \exists i' \in [t] \, \left( \parambwformula[i'] \lor  \paramassignment[i']\right)\,, 
    \end{equation}
    is
    $\formSigmab^b_3(\Phi,\Pi)$. Finally, the formula encoding Delayer consistency  (\ref{form:DC-with-quantifiers}) 
    is $\formSigmab^b_3(\Phi,\Pi)$, and therefore the formula 
    \begin{equation}\label{eq:DC}
            \exists j  \in [t] \, (j > i \land \paramDC[j])
    \end{equation}
    is also $\formSigmab^b_3(\Phi,\Pi)$.
    
    We can therefore proceed by
    $\formSigmab^b_3(\Phi,\Pi)$-induction on $i$, available in
    $\BussTthreetwo(\Phi,\Pi)$.
Consider the refutation $\Pi = (C_1, \ldots, C_t)$ in a top-down ordering, 
    that is, where
    $C_1=\bot$ and every clause used to derive $C_i$ has index strictly
    larger than $i$.
    We prove the statement by induction on $i = 1, \ldots, t$. Throughout this proof, all references to clauses
of $\Pi$ are understood to range over
enabled clauses. Since $\Pi$ is a valid refutation,
every enabled derived clause has enabled premises.

    Before setting up the induction, note that it is straightforward to verify that the update of a reservation according to the Delayer strategy $\Delayer_n(\Phi, \Pi, 1^s)$ as in Algorithm~\ref{algorithm:extended-delayer} can be expressed by a bounded formula. More precisely, given the basic reservation $R$ and the tree-structure $S$ as in \eqref{form:DC-with-quantifiers} witnessing that a clause $C_i$ is Delayer consistent, it is simple to define an updated reservation $R'$ and $S'$ following Algorithm~\ref{algorithm:extended-delayer} using only $\formSigmab^b_3(\Phi,\Pi)$ formulas. We implicitly use this observation throughout the inductive argument.

    Now, for the base case $i=1$, consider the clause $C_1=\bot$ and the empty reservation. If this is a Resolution
    step, the required reservation update following Algorithm~\ref{algorithm:extended-delayer} is possible: 
    no block is 
    reserved, and hence the update cannot fail
    because of a shortage of free blocks. The resulting reservation witnesses that the chosen child is Delayer consistent,  establishing
    item~\eqref{eq:DC}.

    Suppose now that the claim holds for $i$ and let us prove it for
    $i+1$. If \cref{eq:BWorSAT} holds, there is nothing to prove. Otherwise, 
    the induction hypothesis yields an index $j$ with $i<j\leq t$ 
    such that $C_j$ is Delayer consistent. If $j>i+1$, then the same
    $C_j$ witnesses the final statement for $C_{i+1}$. Hence it only remains to
    consider the case $j=i+1$.

    Suppose $C_{i+1}$ is Delayer consistent and does not encode a satisfying
    assignment. Fix witnesses $R$ and $S$ to its Delayer-consistency, so that $\DC(i+1,R,S,n,m,s,w)$ holds. In particular, $R$ can be extended to a full reservation $R'$
    satisfying the reservation invariant with respect to $C_{i+1}$. Let us now do a case distinction on how $C_{i+1}$ is derived.
    \begin{itemize}
        \item If $C_{i+1}$ is obtained by weakening a clause
    $C_k\subseteq C_{i+1}$, where $k>i+1$, then the same full reservation
    falsifies $C_k$ and does not falsify any axiom of
    $\EReff_s(\varphi)$. After deleting from the reservation any
    information about blocks not mentioned in $C_k$, we obtain a
    reservation satisfying the reservation invariant with respect to
    $C_k$. Hence $C_k$ is Delayer consistent and since $k > i+1$ we are done.

    \item Suppose that $C_{i+1}$ is derived by a Resolution step. Starting
    from the full reservation $R'$ extending $R$, choose the child $C_k$ to
    which the Delayer moves according to
    Algorithm~\ref{algorithm:extended-delayer}. If the corresponding
    reservation update succeeds, it is easy to show that the updated reservation witnesses that $C_k$ is
    Delayer consistent. We therefore only need to rule out the two ways in
    which this update can fail.

    \begin{itemize}
        \item
        Suppose first that the update fails because the Delayer needs to
        attach fresh children to a block at some layer $h<n$, but there
        are not enough unreserved blocks on layer $h+1$. Exactly as in the
        block-counting argument for the Delayer strategy, this implies that
        at least $\lfloor s/n\rfloor-1$
        blocks on that layer are already reserved. Since each block
        mentioned in the current clause can account for at most two
        reserved blocks in this update, we obtain
        \begin{equation}
            \bw(C_{i+1})
            \geq
            \frac{1}{2}
            \left(
                \left\lfloor\frac{s}{n}\right\rfloor-1
            \right)
            \ge w.
        \end{equation}
        This implies $C_{i+1}$ has large block-width, contradicting our assumption that \cref{eq:BWorSAT} does not hold.

        \item If the update concerns a previously unreserved
block $b'$ in the final layer, then the Delayer
attempts to assign to it the clause
$A=x_1\lor\cdots\lor x_n$. Failure means that
$A$ is not a weakening of any axiom of $\Phi$.
Consequently, the all-zero assignment satisfies
$\Phi$, by \Cref{lemma:fact}.

Moreover, since $b'$ is unreserved, it is
neither in $\dom(R)$ nor pointed to by $R$.
Thus, for every $q\in[n]$, the formula
$\alpha(i+1,R,S,b',q)$ is false. Hence
$R,S,b'$ witness
$\operatorname{Assignment}(i+1,n,m,s,w)$,
contrary to our assumption that (5.9)
does not hold.

We may therefore assume that the block $b'$
was already reserved as a child of another
block.
        The remaining possibility is the final-layer case. That is,
        $C_{i+1}$ mentions a block $b$ at layer $n-1$, the reservation $R$ points from
        $b$ to a reserved block $b'$ at layer $n$, and when moving to $C_k$,
        the block $b'$ becomes relevant. This would imply that extending the reservation
        requires assigning $b'$ to an axiom of $\Phi$. The only way the update of the reservation can fail is if 
        no such extension is possible.

        Consider the width-$n$ clause
        $\mathcal B=\mathcal B_{R,S,b'}$ whose contents are reconstructed
        from $C_{i+1}$, $R$, and $S$ by the definition of
        $\alpha(i+1,R,S,b',\cdot)$ in \cref{form:alpha}. More explicitly,
        for every $q\in[n]$, the clause $\mathcal B$ contains
        $\neg x_q$ precisely when $\alpha(i+1,R,S,b',q)$
        holds, and contains $x_q$ otherwise. By
        (\ref{form:DC-with-quantifiers}) and the consistency conditions
        comprising $\DC$, this is a well-defined non-tautological
        width-$n$ clause. 

        We claim that $\mathcal B$ is not a weakening of any clause of
        $\Phi$. Indeed, if $\mathcal B$ were a weakening of some axiom
        $A$ of $\Phi$, then we could extend $R$ by declaring that $b'$ is
        weakened from $A$ and by assigning to $b'$ the contents
        $\mathcal B$. Because of the definition of $\DC$, this would give a full reservation
        satisfying the reservation invariant after the update, which we assumed failed.

        Hence $\mathcal B$ is a non-tautological width-$n$ clause that is
        not a weakening of any clause of $\Phi$. By
        \Cref{lemma:fact}, its unique falsifying assignment satisfies
        $\Phi$. But this falsifying assignment is exactly the assignment
        represented by $ \alpha(i+1,R,S,b',\cdot)$. 
        Moreover, $R$ and $S$ satisfy $\DC$. Therefore the same witnesses
        $R,S,b'$ satisfy every conjunct in
        \cref{form:assignment}. Consequently $C_{i+1}$ encodes a
        satisfying assignment, contradicting our assumption that \cref{eq:BWorSAT} does not hold.
    \end{itemize}

    Both possible failures of the Delayer update are therefore impossible. The chosen child $C_k$, with $k>i+1$, is Delayer consistent. This completes the induction. \qedhere
    \end{itemize}
\end{proof}

The universal statement above now implies an existential statement: there is either a wide clause or a satisfying assignment.

\begin{theorem}[Assignment or wide clause, in $\BussTthreetwo(\alpha)$]
    \label{thm:assignment-or-wide}
    The theory $\BussTthreetwo(\Phi, \Pi)$ proves the following $\forall \formSigmab^b_3(\Phi,\Pi)$ statement. For every $n$, $m$, $s$ and $t$ and every $w \in \Log$ with $w \le 1/2(\lfloor s/n\rfloor -1)$, suppose $\Phi$ encodes a CNF formula over $n$ variables and $m$ clauses, and suppose $\Pi$ encodes a length-$t$ Resolution refutation of $\EReff_s(\varphi)$. Then, there exists $i\in[t]$ such that either
    \begin{enumerate}[label=(\alph*)]
        \item \label{item:encodes-ass} the clause $C_i$ is enabled and encodes a satisfying assignment to $\varphi$, or else
        \item \label{item:w-plus-one} the clause $C_i$ is enabled and the block-width of $C_i$ is at least $w$.
    \end{enumerate}
\end{theorem}

\begin{proof}
The complexity of the statement is indeed $\forall\formSigmab^b_3(\Phi, \Pi)$, as already discussed in \Cref{lemma:bw-auxiliary}. To complete the proof, working within the theory, suppose for contradiction that for every $i \in [t]$, neither \ref{item:encodes-ass} nor \ref{item:w-plus-one} holds. Then, by \Cref{lemma:bw-auxiliary}, for every $i \in[t]$ there exists $j$ such that $i < j \leq t$ and $C_j$ is Delayer consistent. But then for the case $i = t$ we get that there is a $j$ such that $t <j \leq t$, which is absurd. 
\end{proof}

\begin{remark}[The block-width lower bound in the exponential-size regime: formalization in $\BussTonetwo(\alpha)$]
    \label{rem:bw-in-exp-regime}
    Suppose we wanted to prove \Cref{thm:assignment-or-wide} above in the so-called exponential-size regime, where $n$, $m$ and $s$ are assumed to be in $\Log$. In this setting, the proof goes through in $\BussTonetwo(\Phi, \Pi)$. This is because when assuming $s, n, m\in\Log$, the universal quantifiers in the definition of Delayer-consistency in \Cref{subsubsec:DC-formulas} are now sharply bounded, making that overall formula $\formSigmab^b_1(\Pi)$. The same happens with the assignment extraction formula (\ref{form:assignment}), where the universal quantifiers become sharply bounded. This overall makes \Cref{lemma:bw-auxiliary} available in $\BussTonetwo(\Phi, \Pi)$.
\end{remark}

\section{The Atserias--M\"uller lower bound in  \texorpdfstring{$\BussTtwo(\alpha)$}{T2(alpha)}}
\label{sec:AM-complete-formal}

With both the restriction argument and the block-width lower bound ready, we combine them into the final statement of the lower bound. As previously discussed, depending on the regime we find ourselves in, we can slightly vary the power of the theory. In the exponential-size regime we can work in $\LLR$, while in the polynomial-size regime we are forced to go to $\BussTfourtwo(\alpha)$. \Cref{subsec:tedious-restrictions} shows how to manipulate restrictions within these theories, and \Cref{subsec:poly-regime} and \Cref{subsec:exp-regime} complete the arguments for the final lower bound.

\subsection{Applying the restrictions}
\label{subsec:tedious-restrictions}

In the next proposition, $\rho$ is a new relation symbol encoding a restriction in the natural way and we show that the fact that Resolution is closed under restrictions is easily derivable already in $\BussToneone(\alpha)$. The idea is that from $\Pi$ and $\rho$ we can define a relation $\Pi'$ encoding $\Pi_{\restriction{\rho}}$.

\begin{proposition}
    \label{prop:closure-restrictions}
    The theory $\BussToneone(\Phi, \Pi, \Phi', \Pi', \rho)$ proves that Resolution is closed under restrictions: for every $n\geq 1$, $m$ and $t$, if $\Phi$ encodes a CNF formula over $n$ variables and $m$ clauses and $\Pi$ encodes a Resolution refutation of $\Phi$ of length $t$, $\Phi'$ encodes $\Phi_{\restriction\rho}$ and $\Pi'$ encodes $\Pi_{\restriction\rho}$, then $\Pi'$ is a length-$t$ Resolution refutation of $\Phi'$. Furthermore, the formulas relating $\Phi$ and $\Pi$ to $\Phi'$ and $\Pi'$ are $\formDeltab^b_2(\Phi, \Pi, \Phi', \Pi', \rho)$.
\end{proposition}

\begin{proof}
    It suffices to see that there are $\formDeltab^b_2(\Phi, \Pi, \Phi', \Pi', \rho)$ formulas defining the new formula and refutation after the restriction. For a clause $A$ of $\Phi$, let
    \begin{equation}
        \label{form:killed-axiom}
        \operatorname{Killed}_{\rho}(A)
        \coloneq
        \exists \ell\in[2n]\,
        \bigl(\Phi(\ell,A)\land \rho(\ell)=1\bigr).
    \end{equation}
    
    Since our encoding has a fixed number of clauses, we represent a clause that is satisfied by $\rho$ by the fixed tautology $x_1\vee\neg x_1$. Thus we define $\Phi'$ by
    \begin{align}
        \Phi'(\ell,A)\longleftrightarrow
        &\ \bigl(\neg\operatorname{Killed}_{\rho}(A)
        \land \Phi(\ell,A)\land \rho(\ell)=\star\bigr) \notag\\
        &\ \lor\bigl(\operatorname{Killed}_{\rho}(A)
        \land (\ell=x_1\lor \ell=\neg x_1)\bigr).
        \label{eq:restricted-Phi}
    \end{align}
    This convention is semantically equivalent to deleting the satisfied clauses, while keeping the clause indices unchanged.

    To illustrate the definition of $\Pi'$, let us first note that a proof line $B$ is killed if either it was already disabled or some literal in it is set to true by $\rho$. The remaining proof variables can be defined by
    \begin{align}
        \Pi'(\enable^B) &\longleftrightarrow  \Pi(\enable^B) \land \forall \ell \in [2n] \ \left( \Pi(\lit^B_\ell) \to \rho(\ell) \neq 1 \right) \\
       \Pi'(\alit^A_\ell) &\longleftrightarrow \Phi'(\ell,A) \\
        \Pi'(\derived^B) &\longleftrightarrow \Pi(\derived^B) \\
        \Pi'(\weak^B_A) &\longleftrightarrow \Pi(\weak^B_A)\land \neg\operatorname{Killed}_{\rho}(A) \\
        \Pi'(\lit^B_\ell) &\longleftrightarrow \Pi(\lit^B_\ell) \land \rho(\ell) = \star \\
        \Pi'(\res^B_{x_i}) &\longleftrightarrow \Pi(\res^B_{x_i}) \\
        \Pi'(\lpoint^B_{B'}) &\longleftrightarrow \left( \Pi(\lpoint^{B}_{B'}) \land \forall \ell \in [2n] \ \left( \Pi(\lit^{B'}_\ell) \to \rho(\ell) \neq 1 \right) \right) \\
        &\phantom{\longleftrightarrow} \quad \lor \left( \Pi(\rpoint^B_{B'}) \land \exists  i \in [n] \ (\Pi(\res^B_{x_i})\land \rho(x_i) =1 ) \right) \notag \\
        \Pi'(\rpoint^B_{B'}) &\longleftrightarrow \left( \Pi(\rpoint^{B}_{B'}) \land \forall \ell \in [2n] \ \left( \Pi(\lit^{B'}_\ell) \to \rho(\ell) \neq 1 \right) \right) \\
        &\phantom{\longleftrightarrow} \quad \lor \left( \Pi(\lpoint^B_{B'}) \land \exists  i \in [n] \ (\Pi(\res^B_{x_i})\land \rho(x_i) =0) \right). \notag
    \end{align}
    The right-hand side formulas are all $\formDeltab^b_2(\Phi, \Pi, \Phi', \Pi', \rho)$, so the entire conjunction of double implications has the same complexity.

    Once the definitions are in place, it is easy to verify that the statement follows. If the resolved variable is unset, a surviving Resolution inference remains a Resolution inference. If the resolved variable is assigned, exactly the premise falsifying that value survives, and the two new pointers are both redirected to that premise; as observed in \Cref{rem:simulate-weakening}, this is a valid way to encode a weakening. Finally, if an axiom is killed, every proof line weakened from it is killed as well. Hence every enabled line of $\Pi'$ is a valid Resolution or weakening step from earlier enabled lines, and $\Pi'$ is a Resolution refutation of $\Phi'$.
\end{proof}

The following is the propositional version of the previous proposition, which will be useful later.

\begin{proposition}[Resolution proves that Resolution is closed under restrictions]
    \label{prop:Res-closed-restrictions}
    There are polynomial-size $\P$-uniform Resolution derivations of the form
    \begin{equation*}
        \Reff_t(\varphi, \pi) \land (\pi' = \pi_{\restriction\rho}) \land (\varphi' =\varphi_{\restriction\rho}) \vdash_\Res \Reff_t(\varphi', \pi'),
    \end{equation*}
    where, once $\rho$ is fixed, $(\pi' = \pi_{\restriction\rho}) \land (\varphi' =\varphi_{\restriction\rho})$ are depth-$2$ formulas.
\end{proposition}

\begin{proof}
    Immediate from the Paris--Wilkie translation in \Cref{thm:Paris-Wilkie-translation}. 
\end{proof}

We next record explicitly the form of the disabling restriction that will be used below. Put $s_{\mathrm e}\coloneq 2\lfloor s/2\rfloor$ and $s'\coloneq \lfloor s/2\rfloor$. We pair the first $s_{\mathrm e}$ blocks consecutively. If $s$ is odd, the remaining block is disabled deterministically. For $w\in\Log$, let $q\coloneq q(s_{\mathrm e},w)$ be the power of two from \Cref{thm:good-seeds}, and let $\ell\coloneq q^4$. Given a seed $\sigma\in\{0,1\}^{\ell}$, write
$\eta_r\coloneq \Design(s_{\mathrm e},q,r)\cdot\sigma^\top$ for $r\in[s_{\mathrm e}]$. For every $j\in[s']$, the bit $\eta_j$ chooses the disabled block in pair $j$, and $\eta_{j+s'}$ gives the common value assigned to the remaining variables of that disabled block. The root is always enabled.

Let $\rho_\sigma$ denote this disabling restriction. We further extend it to a restriction $\widehat\rho_\sigma$ by setting to $0$ every pointer variable whose source block survives but whose target block is disabled. This extra clean-up can only satisfy additional literals and never revives a killed clause. If $\operatorname{surv}_{\sigma}(j)\coloneq 2j-\eta_j$ denotes the surviving block from pair $j$,
then, after renaming the surviving block $\operatorname{surv}_{\sigma}(j)$ to $j$ and the old root to $s'+1$, the restricted formula $\Reff_s(\varphi)_{\restriction\widehat\rho_\sigma}$ is precisely the enable-free formula $\EReff_{s'}(\varphi)$, up to clauses already satisfied by the restriction.

\begin{lemma}[Applying a good seed]
    \label{lemma:restriction-applied}
    The theory $\APCtwo(\Phi,\Pi)$ proves the following. Let $w\coloneq 8\lceil\log t\rceil$, let $s_{\mathrm e}=2\lfloor s/2\rfloor$, let $q=q(s_{\mathrm e},w)$, and let $\ell=q^4$. If $\Pi$ encodes a length-$t$ Resolution refutation of $\Reff_s(\varphi)$, then there is a seed $\sigma\in\{0,1\}^{\ell}$ and a $\formDeltab^b_2(\Phi,\Pi)$-definable relation $\widehat\Pi^{\sigma}$ such that:
    \begin{enumerate}[label=(\roman*)]
        \item $\widehat\Pi^{\sigma}$ encodes a length-$t$ Resolution refutation of $\EReff_{s'}(\varphi)$, where $s'=\lfloor s/2\rfloor$;
        \item every enabled clause of $\widehat\Pi^{\sigma}$ has block-width at most $w-1$.
    \end{enumerate}
    If $n,m,s\in\Log$, the same statement is available in $\APC(\Phi,\Pi)$ and the definition of the relation $\widehat\Pi^{\sigma}$ is sharply bounded relative to $\Phi$ and $\Pi$.
\end{lemma}

\begin{proof}
    If $s$ is even, take the good seed given by \Cref{thm:good-seeds}. If $s$ is odd, leave the last block unpaired and disable it deterministically, and repeat the proof of \Cref{thm:good-seeds} on the remaining $s-1$ blocks. This does not change the estimate: since $w$ is even, every clause mentioning at least $w$ blocks mentions at least $w/2$ distinct pair-indices among the paired blocks, so the same width-$w/2$ subclause and the same bad-seed bound can be used.

    For this structured formula we use the same line-by-line restriction clean-up as in \Cref{prop:closure-restrictions}, but we omit clauses of the inner formula that become true rather than representing them by tautological placeholders. After applying $\widehat\rho_\sigma$ and renaming the variables, the unsatisfied clauses are exactly the clauses of $\EReff_{s'}(\varphi)$. The resulting proof relation $\widehat\Pi^\sigma$ is still $\formDeltab^b_2(\Phi,\Pi)$-definable, since this is the same local clean-up and renaming as in \Cref{prop:closure-restrictions}. If an original clause has block-width at least $w$, the good-seed property already kills it under $\rho_\sigma$, and hence also under the stronger restriction $\widehat\rho_\sigma$. Restriction and renaming cannot increase block-width, so every enabled clause in the resulting refutation has block-width at most $w-1$.

    In the exponential-size regime all ranges occurring in the restriction and clean-up definitions are in $\Log$, so the defining relation becomes sharply bounded and \Cref{rem:good-seeds-in-exp-regime} gives the result in $\APC(\Phi,\Pi)$.
\end{proof}

Finally, for a seed $\sigma$, let $\operatorname{Ren}_{\sigma}$ be the $\PV$-function sending a literal over the variables of $\EReff_{s'}(\varphi)$ to the corresponding literal over the surviving blocks of $\Reff_s(\varphi)$. For a clause index $i$, define the restricted-clause predicate by
\begin{equation}
    \label{form:restricted-clause-predicate}
    \Pi^{\sigma}_{\mathrm{cl}}(\lit^i_L)
    \coloneq
    \Pi\bigl(\lit^i_{\operatorname{Ren}_{\sigma}(L)}\bigr)
    \land
    \widehat\rho_\sigma\bigl(\operatorname{var}(\operatorname{Ren}_{\sigma}(L))\bigr)=\star.
\end{equation}
We also let $\operatorname{Survives}(i,\sigma)$ express that no literal of $C_i$ is set to true by $\widehat\rho_\sigma$. Now let
$\paramRassignment$ denote the formula obtained from \cref{form:assignment}, with the block parameter replaced by $s'=\lfloor s/2\rfloor$ and every query to the clause predicate $\Pi$ replaced by the sharply bounded predicate $\Pi^{\sigma}_{\mathrm{cl}}$, together with the conjunct $\operatorname{Survives}(i,\sigma)$. 

Because \cref{form:restricted-clause-predicate} is sharply bounded relative to $\Pi$, while $\operatorname{Survives}(i,\sigma)$ is $\formPib^b_1(\Pi)$, and $\RAssignmentname$ is 
$\formSigmab^b_3(\Phi,\Pi)$.
If $n,m,s\in\Log$, 
$\RAssignmentname$
collapses to $\formSigmab^b_1(\Phi,\Pi)$.


\subsection{Formalization in the polynomial-size regime:  \texorpdfstring{$\BussTfourtwo(\alpha)$}{T42(alpha)}}
\label{subsec:poly-regime}

Our first statement is for the most general setting, where we deal with any CNF formula without any bounds on the width, and we are in the polynomial-size regime. The name \say{polynomial-size regime} might look like a misnomer, since in the following statement we prove instead an exponential lower bound on the length $t$ of the refutations. The crucial point here is that we do not assume that either $s$ or $n$ is in $\Log$, so we cannot feasibly express $2^{n}$. In the following statements, we say that $C_i$ \emph{encodes a satisfying assignment} to mean that $\RAssignmentname$ is true for some seed $\sigma$.

\begin{theorem}[Atserias--M\"uller in $\BussTfourtwo(\alpha)$, polynomial-size regime]
\label{thm:poly-regime-ref-lwb-all-CNF}
There exists a positive constant $\varepsilon\in\bbQ$ and $n_0 \in \bbN$ such that the theory $\BussTfourtwo(\Phi, \Pi)$ proves the following $\forall \formSigmab^b_3(\Phi, \Pi)$ statement. For every $n$ and $m$, for every $s$ and every $t > 1$, suppose $\Phi$ encodes a CNF formula over $n\geq n_0$ variables and $m$ clauses, and suppose $\Pi$ encodes a length-$t$ Resolution refutation of $\Reff_s(\varphi)$. If $|t| < \varepsilon \cdot s/n$, then there exists  $i \in [t]$ such that $C_i$ encodes a satisfying assignment.
\end{theorem}

\begin{proof}
    Fix $\varepsilon=1/64$ and $n_0=1$. Let
    $w=8\lceil\log_2 t\rceil$, $s'=\lfloor s/2\rfloor$,
    $q=q(2s',w)$, and $\ell=q^4$.
    Since $\lceil\log_2 t\rceil\leq |t|$, the hypothesis gives $w\leq 8|t|\le \frac{s}{8n}$.
    Moreover, since $n$ is a positive integer, $\lfloor s'/n \rfloor = \lfloor {s}/{2n}\rfloor \geq s/{2n}-1$.
    Consequently,
    \begin{equation}
        \label{eq:parameter-section6}
        \frac12\left(
            \left\lfloor\frac{s'}{n}\right\rfloor-1
        \right)
        \geq 
        \frac{s}{4n} - 1
        \geq w.
    \end{equation}

    By \Cref{lemma:restriction-applied}, working in
    $\APCtwo(\Phi,\Pi)$, we obtain a seed
    $\sigma\in\{0,1\}^{\ell}$ and a
    $\formDeltab^b_2(\Phi,\Pi)$-definable length-$t$
    Resolution refutation $\widehat{\Pi}^{\sigma}$ of
    $\EReff_{s'}(\varphi)$ in which every enabled clause
    has block-width at most $w-1$.

    By \cref{eq:parameter-section6}, we may apply
    \Cref{thm:assignment-or-wide} to
    $\widehat{\Pi}^{\sigma}$ with width parameter $w$.
    It follows that some enabled clause of
    $\widehat{\Pi}^{\sigma}$ encodes a satisfying
    assignment to $\Phi$. Let $i\in[t]$ be its index in the original refutation.
    By construction, $C_i$ survives the restriction and
    the clause analyzed in $\widehat{\Pi}^{\sigma}$ is
    its renamed restriction. Thus
    $\paramRassignment[w]$
    holds.

    Note that the restriction argument is available in
    $\APCtwo(\Phi,\Pi)\subseteq\BussTthreetwo(\Phi,\Pi)$.
    Applying the block-width lower bound to the
    $\formDeltab^b_2(\Phi,\Pi)$-definable relation
    $\widehat{\Pi}^{\sigma}$ raises the induction level
    from $\BussTthreetwo$ to $\BussTfourtwo$.
    The conclusion remains $\formSigmab^b_3(\Phi,\Pi)$,
    since $\operatorname{RAssignment}$ has this
    complexity and the existential quantifiers over
    the number-coded seed $\sigma$ and the index $i$
    can be merged with its leading existential block.
\end{proof}



\subsection{Formalization in the exponential-size regime:  \texorpdfstring{$\LLR$}{T12 plus dWPHP(PV1(alpha))}}
\label{subsec:exp-regime}

In the so-called exponential-size regime we assume the initial parameters of the $\Reff$ formula are in $\Log$. This means many of the quantifiers in our definitions become sharply bounded, leading to the following result.

\begin{theorem}[Atserias--M\"uller for all formulas in $\LLR$, exponential-size regime]
\label{thm:exp-regime-ref-lwb}
There exists a positive constant $\varepsilon\in\bbQ$ and $n_0 \in \bbN$ such that the theory $\LLR$ proves the following $\forall \formSigmab^b_1(\Phi, \Pi)$ statement. For every $n\in\Log$, $m\in\Log$ and $s\in \Log$, and every $t > 1$, suppose $\Phi$ encodes a CNF formula over $n\geq n_0$ variables and $m$ clauses, and suppose $\Pi$ encodes a length-$t$ Resolution refutation of $\Reff_s(\varphi)$. If $t < 2^{\varepsilon \cdot s/n}$, then there exists $i\in[t]$ such that $C_i$ encodes a satisfying assignment.
\end{theorem}

\begin{proof}
    The parameters are in the proof of \Cref{thm:poly-regime-ref-lwb-all-CNF}. Because $n,m,s\in\Log$, the restriction in \Cref{lemma:restriction-applied} is available already in $\APC(\Phi,\Pi)$ as discussed in  \Cref{rem:good-seeds-in-exp-regime}, and the block-width lower bound is available in $\BussTonetwo(\Phi,\Pi)$ as in \Cref{rem:bw-in-exp-regime}; substituting the sharply bounded definition of $\widehat\Pi^\sigma$ does not increase the complexity of the induction.
\end{proof}


\section{Propositional consequences}
\label{sec:propositional-consequence}

In this section we show how the new formalizations extend some of the propositional corollaries of \citeauthor{PAP} \autocite{PAP} from Extended Frege to constant-depth systems, with a quasi-polynomial blow-up in parameters. \Cref{subsec:fragments} shows how to apply the Paris--Wilkie translations to get Resolution lower bounds formalized in the propositional setting. \Cref{subsec:Reff-complete-automatability} applies these to argue that automating on $\Reff$ formulas suffices to automate on all formulas.

\subsection{Propositional fragments of the Atserias--Müller lower bound}
\label{subsec:fragments}

We start by showing that if a proof system can refute some formula efficiently, then it can also refute a lower bound on its corresponding $\Reff$ formula. Before we state this formally, we prove a simple technical lemma showing how the depth of a propositional proof increases when substituting formulas of bounded depth.
\begin{lemma}
    \label{lemma:size-depth-increase}
Let $d\in\bbN$, and let $\pi$ be a
$\PK_{d+1}$-refutation of $\varphi$ where every
connective in $\pi$ at depth $d+1$ has arity at
most $r$. Let $\psi$ be a formula of depth $k \geq 1$,
whose top connective has arity $\ell$, and whose
bottom connectives have arity at most $b$.
It holds that the substitution of $\psi$ into $\pi$,
denoted by $\pi(\psi)$, is a
$\PK_{d+k}$-refutation of $\varphi(\psi)$
of size $|\pi(\psi)| =
O\left(|\pi|\cdot|\psi|\cdot\ell^r\right)$,
where the connectives at depth $d + k$ have arity at most $b$ if $k \geq 2$ and at most $rb$ if $k = 1$.
\end{lemma}

\begin{proof}
Since every line of $\pi$ has depth at most $d+1$,
substitution of a depth-$k$ formula for atoms
initially increases the depth to at most $d+k+1$.
To reduce the depth, consider the original
bottom-layer gates of $\pi$, each of arity at
most $r$, whose inputs are replaced by copies
of $\psi$. If the bottom gate of $\pi$ matches the top gate
of $\psi$, we merge the two gates, reducing the
depth to at most $d+k$. The bottom fan-in remains
at most $b$ when $k\geq 2$, while for $k=1$
the merged gate has arity at most $r\ell = rb$. 
Otherwise, we distribute the bottom gate of
$\pi$ over the top gates of the substituted
formulas. This produces at most $\ell^r$
terms.
If $k\geq 2$, this reduces the depth to $d+k-1$; we can see this as a depth-$(d+k)$ formula by adding an extra dummy gate at the bottom of fan-in $1$.
If $k=1$, the resulting formula 
has depth $d+k$ and the bottom gates have arity at most $r$.
Combining the cases, the bottom fan-in is
bounded by $b$ for $k\geq 2$, and by $rb$ for $k=1$.
Each line $\gamma$ of $\pi$ increases in size
by at most a factor $O(|\psi|\ell^r)$.
Summing over all lines yields $|\pi(\psi)| =
O\left(|\pi|\cdot|\psi|\cdot\ell^r\right)$.
%
%
\end{proof}

We are ready to apply the propositional translations to our formalizations.

\begin{theorem}[Propositional fragments of Atserias--Müller]
\label{thm:fragment-characterization}
Let $d\in\bbN$, let $\varphi$ be a CNF formula over
a large enough number $n$ of variables and $m$
clauses, let $s\geq n^2$, and let $t>1$ satisfy
$|t|<\varepsilon s/n$, where $\varepsilon>0$
is the constant from \Cref{thm:poly-regime-ref-lwb-all-CNF}.
The following hold:
\begin{enumerate}[label=(\roman*)]
    \item if $\PK_d \vdash \neg \varphi$ in size $\ell$, then $\PK_{\max\{d+\nicefrac{1}{2}, \ 3+\nicefrac{1}{2}\}} \vdash \neg \Reff_{t}(\Reff_{s}(\varphi))$ in size $\quasipoly(n, m, s, t, \ell)$;
    \item if $\PK_{(d+3)+\nicefrac{1}{2}} \vdash \neg \varphi$ in size $\ell$, then $\PK_{(d+3)+\nicefrac{1}{2}} \vdash \neg \Reff_{t}(\Reff_{s}(\varphi))$ in size $\quasipoly(n, m, s, t, \ell)$;
\end{enumerate}
Furthermore, in both cases, given the proof of $\neg \varphi$ in the left-hand side, one can obtain a proof of $\neg\Reff_{t}(\Reff_{s}(\varphi))$ in quasipolynomial time.
\end{theorem}

\begin{proof}
    In \Cref{thm:poly-regime-ref-lwb-all-CNF} we showed that $\BussTfourtwo(\Phi,\Pi)$ proves the Atserias--Müller lower bound for CNF formulas of arbitrary width. Since the complexity of the statement is $\forall \formSigmab^b_3(\Phi, \Pi)$, we can apply the Paris--Wilkie translation (\Cref{thm:Paris-Wilkie-translation}) to get that there are quasipolynomial-size derivations of the form
    \begin{equation}
        \label{eq:AM-in-PK}
        \Reff_{t}(\Reff_s(\varphi), \pi) \vdash_{\PK_{3+\nicefrac{1}{2}}} \Satf_1(\varphi, \alpha_1(\pi)), \dots, \Satf_T(\varphi, \alpha_T(\pi)).
    \end{equation}
    Here, $T = \quasipoly(n,m,s,t)$ is the upper bound of the total number of possible candidate assignments we try. That is, the Paris--Wilkie translation would produce a big disjunction when translating the existential quantifiers in (\ref{form:assignment}), which we have represented here as a cedent. For each candidate assignment we have a small subformula $\alpha_i(\pi)$ which is just the polylogarithmic-width disjunct coming from the translation of (\ref{form:alpha}). Note that we also see $\Reff_{t}(\Reff_s(\varphi), \pi)$ as a set of formulas, rather than as a syntactic conjunction.

    Suppose now there is a $\PK_d$-refutation of $\varphi$ in size $\ell$. Then, with at most a linear blow-up, there is a $\PK_d$ refutation of $\Satf_{\restriction\varphi}(\alpha)$ (see \Cref{subsec:prelim-sat-reflection}). For each $i \in [T]$, substitute for $\alpha$ the clauses given by $\alpha_i(\pi)$ extracting a candidate assignment. This is a clause of width $\polylog(n, m,s)$, so it follows that we now have a derivation of the form $\PK_{d+\nicefrac{1}{2}} \vdash \neg \Satf_{\restriction\varphi}(\alpha_i)$ in size $O(\ell)$ for every $i \in [T]$. By successively cutting $T$ times with (\ref{eq:AM-in-PK}), we get a $\PK_{\max\{d+\nicefrac{1}{2}, \ 3+\nicefrac{1}{2}\}}$-refutation of $\Reff_t(\Reff_s(\varphi),\pi)$ in size $\quasipoly(n,m,s,t,\ell)$. This proves the first item.

    To prove the second item, repeat the same process using \Cref{lemma:size-depth-increase} to argue about the depth increase after substitution. We have that the proof of $\neg \varphi$ has depth $d+4$ and polylogarithmic bottom fan-in, and the $\alpha_i$ circuits have depth $1$ with polylogarithmic bottom fan-in, so we get a depth-$(d+4)$ proof with polylogarithmic fan-in at the bottom layer, which is precisely $\PK_{(d+3)+\nicefrac{1}{2}}$. The size blow-up, as accounted for by \Cref{lemma:size-depth-increase}, is precisely quasipolynomial.
\end{proof}




We also obtain a converse of this statement building on Pudlák's upper bound construction \autocite{Pudlak03}, which was already formalized in Resolution.

\begin{theorem}[Pudlák's upper bound in Resolution {\autocite[Theorem 7.1]{PAP}}]
    \label{thm:ub-in-Res}
    For every $n, m, s \in \bbN$, 
    there is some $\tau=\Theta(s(m+sn^2))$ such that for $t\geq \tau$
    there exist uniform polynomial-size Resolution derivations of the form
    \[ \Satf(\varphi, \alpha) \land \pi =P(\varphi, \alpha, s)  \vdash_{\Res}  \Reff_t(\Reff_s(\varphi), \pi) \,,\]
    where the $\Satf$ formula is for CNF formulas with $n$ variables and $m$ clauses.
\end{theorem}

In the statement above,  $\pi = P(\varphi, \alpha, s)$ stands for the conjunction of clauses describing the upper bound construction of Pudlák. As shown in \autocite{PAP}, this circuit $P$ is actually of very small depth. In fact, once $\varphi$ is fixed, the $\pi$ variables are simply $2$-extensions over the variables of $\alpha$. In other words, for every fixed $\varphi$, the formula $\pi = P(\varphi, \alpha, s)$ amounts to the set of all $O(n^2)$ possible extensions for two literals over $n$ variables.
This leads to the following converse statement to \Cref{thm:fragment-characterization}.

\begin{theorem}
\label{thm:pudlak-going-back}
There is a function $p(n, m, s) =\poly(n,m, s)$ such that for every CNF formula $\varphi$ over $n$ variables and $m$ clauses, every $s \in \bbN$, and $d \in \bbN$,
\begin{enumerate}[label=(\roman*)]
\itemsep=0pt
    \item if $\PK_d \vdash \neg \Reff_{p(n,m,s)}(\Reff_s(\varphi))$ in size $\ell$, then $\PK_{d+1}^2 \vdash \neg \varphi$ in size $\poly(n, m, s, \ell)$;
    \item if $\PK_{d+\nicefrac{1}{2}} \vdash \neg \Reff_{p(n,m,s)}(\Reff_s(\varphi))$ in size $\ell$, then $\PK_{d+\nicefrac{1}{2}} \vdash \neg \varphi$ in size $\poly(n, m, s, \ell)$.
\end{enumerate}
Furthermore, in both cases, given a proof of the left-hand side statement, one can obtain a proof of the right-hand side statement in polynomial time.
\end{theorem}

\begin{proof}
    To prove the first item, we use the fact that from \Cref{thm:ub-in-Res} we have Resolution derivations of the form
    \begin{equation}
       \{ \Satf(\varphi, \alpha) \} \cup\{  \pi =P(\varphi, \alpha, s)  \} \vdash_{\Res}  \{ \Reff_t(\Reff_s(\varphi), \pi)\},
    \end{equation}
    for $t$ the parameter $\tau$ in the statement of \Cref{thm:ub-in-Res}, which will also be our polynomial $p$ in this statement. We write the premises and conclusion in set brackets to emphasize that these are sets of formulas interpreted as their conjunction, but without this explicitly counting towards the depth. 
    
    For $\varphi$ fixed, Resolution can go from the formula $\Satf$ to the native encoding of $\varphi$ with a linear blow-up (see \Cref{subsec:prelim-sat-reflection}). Thus, chaining this derivation with the refutation given by assumption in $\PK_d$, we have
    \begin{equation}
       \{ \varphi \} \cup\{  \pi =P(\varphi, \alpha, s)  \} \vdash_{\Res}  \{ \Reff_t(\Reff_s(\varphi), \pi)\} \vdash_{\PK_d} \bot,
    \end{equation}
    in size $\poly(n,m,s,\ell)$ for $t$ the aforementioned polynomial $p(n,m,s)$. 

    Since once $\varphi$ is fixed the conjunct $\pi = P(\varphi, \alpha,s)$ is simply defining every variable in $\pi$ in terms of a conjunction of two literals, we can simply substitute these small conjunctions in the proof, getting a refutation in $\PK_{d+1}^{2}$.
    
    The proof of the second item is analogous, except in this last step, when substituting the extensions given by $\pi$, we start from a derivation in $\PK_{d+\nicefrac{1}{2}}$ of size $S \coloneq \poly(n,m,s,\ell)$, and \Cref{lemma:size-depth-increase} now guarantees that after substitution we stay in $\PK_{d+\nicefrac{1}{2}}$ with a blow-up in size of $O(2^{\log S}) = O(S)$.
\end{proof}

\subsection{The  \texorpdfstring{$\Reff$}{Ref} formulas are complete for automatability}
\label{subsec:Reff-complete-automatability}

\citeauthor{PAP} \autocite{PAP} proved that Resolution lower bound statements are complete for the automatability and weak automatability of strong propositional proof systems. In particular, this was proven for any so-called \emph{reasonably strong} system: extensions of Extended Frege by some polynomial-time recognizable set of axioms. 

In what follows, we shall say that a proof system $S$ is \emph{automatable in time $T$ on $\Reff$ formulas} if there is an algorithm such that whenever there is a size-$\ell$ $S$-refutation of $\Reff_s(\varphi)$, then the algorithm finds an $S$-refutation in time $T(\ell)$. For $\PK_{d+\nicefrac{1}{2}}$ for large enough $d$, this immediately implies automatability on all CNF formulas. 

\begin{theorem}[Automatability on $\Reff$ formulas]
    \label{theorem:eqivalence-automatability}
    Let $d \in \bbN$ and let $T : \bbN \to \bbN$ be a time-constructible function. If the proof system $\PK_{(d+3)+\nicefrac{1}{2}}$ is automatable on $\Reff$ formulas in time $T$, then $\PK_{(d+3)+\nicefrac{1}{2}}$ is automatable in time $\poly(T(\quasipoly(\cdot)))$ on all CNF formulas.
\end{theorem}

\begin{proof}
    We assume there is an automating algorithm on $\Reff$ formulas and describe the algorithm that extends this to all formulas. If $\varphi$ is a CNF formula and has $\PK_{(d+3)+\nicefrac{1}{2}}$-refutations in size $\ell$, then by \Cref{thm:fragment-characterization}.(ii), we have that there exists a $\PK_{(d+3)+\nicefrac{1}{2}}$-refutation of $\Reff_{p(n,m,s)}(\Reff_{n^3}(\varphi))$ in size $\ell' \coloneq \quasipoly(n,m,s,t,\ell)$, where $p$ is the polynomial in \Cref{thm:pudlak-going-back}. We use the automating algorithm on $\Reff$ formulas to find a refutation of this formula in time $T(\ell')$. Then, by \Cref{thm:pudlak-going-back}.(ii), we can obtain a $\PK_{(d+3) + \nicefrac{1}{2}}$-refutation of $\varphi$ in time $\poly(T(\ell'))$. 
\end{proof}

For $T$ a function of quasipolynomial growth-rate, we can give a cleaner statement.

\begin{corollary}
\label{cor:automatability-equivalence-quasipoly}
Let $d \in \bbN$. The proof system $\PK_{(d+3)+\nicefrac{1}{2}}$ is automatable in quasipolynomial time on all CNF formulas if, and only if, it is automatable on $\Reff$ formulas in quasipolynomial time.
\end{corollary}



If instead of automatability we are interested in \emph{weak automatability}, the situation is a bit more general. 

\begin{theorem}[Weak automatability on $\Reff$ formulas]
    \label{thm:weak-automatability-equivalence}
    Let $d \in \bbN$, let $T : \bbN \to \bbN$ be a time-constructible function and let $S$ be a propositional proof system closed under restrictions, modus ponens and substitutions of polylogarithmic-width clauses. Suppose that $S$ simulates $\PK_{(d+3)+\nicefrac{1}{2}}$. If $S$ is weakly automatable on $\Reff$ formulas in time $T$, then $S$ is weakly automatable on all CNF formulas in time $\poly(T(\quasipoly(\cdot)))$.
\end{theorem}

\begin{proof}
    Suppose $S$ is p-simulated by $Q$ and $Q$ is automatable on $\Reff$ formulas. Now, if $\varphi$ is a CNF formula and has $S$-refutations in size $\ell$, then by \Cref{thm:fragment-characterization}.(ii) and the closure and substitutions properties, we have that there exists an $S$-refutation of $\Reff_{p(n,m,s)}(\Reff_{n^3}(\varphi))$ in size $\ell' \coloneq \quasipoly(n,m,s,t,\ell)$, where $p$ is the polynomial in \Cref{thm:pudlak-going-back}. We use the automating algorithm on $\Reff$ formulas for $Q$ to find a refutation of this formula in time $T(\ell')$ in $Q$. Then, we turn this into a refutation in $Q' \coloneq \EF + \Reflf^Q \psim Q$ \autocite[Theorem 8.4.3]{Krajicek19}, where \Cref{thm:pudlak-going-back} is available and we have all the necessary closure properties. We can now obtain a $Q'$-refutation of $\varphi$ in time $\poly(T(\ell'))$, so $Q'$ weakly automates $S$.
\end{proof}

\section{The Proof Analysis Problem for constant-depth Frege is $\NP$-hard}
\label{sec:PAP-NP-hard}
In this section, we show that the Proof Analysis Problem for constant-depth Frege is $\NP$-hard. Recall that a proof system $S$ is said to be \emph{analyzable} if given an $S$-proof of a Resolution lower bound on $\varphi$, we can decide in polynomial time whether $\varphi$ is satisfiable. Formally, this is the decision problem $\PAP_S$ from \Cref{subsec:prelim-PAP}. If a proof system is both automatable and analyzable, then $\P = \NP$; this makes analyzability a generalization of the Atserias--Müller strategy for proving $\NP$-hardness of automatability. As discussed in \Cref{subsubsec:intro-contributions-PAP-NP-hard}, showing that a proof system is hard to analyze suggests that the Atserias--Müller approach of using $\Reff$ formulas to show $\NP$-hardness of automatability cannot work for $\ACZFnodepth$ (as long as $\Reff$ is the formula for Resolution lower bounds). To show that a proof system is \emph{not} analyzable, we first show that it is able to prove fixed polynomial-size Resolution lower bounds.

In \Cref{subsec:LLR-PHP}, we prove that for the particular case of the pigeonhole principle formulas, lower bounds in the exponential-size regime imply lower bounds in the polynomial-size regime.
In \Cref{subsec:hardness-PAP} we show that the Proof Analysis Problem for constant-depth Frege is $\NP$-hard.

\ifcameraready
Before proceeding any further, we state here the following fact regarding Resolution's ability to show that is itself closed under restrictions.

\begin{proposition}[Resolution proves that Resolution is closed under restrictions]
    \label{prop:Res-closed-restrictions}
    There are polynomial-size $\P$-uniform Resolution derivations of the form
    \begin{equation*}
        \Reff_t(\varphi, \pi) \land (\pi' = \pi_{\restriction\rho}) \land (\varphi' =\varphi_{\restriction\rho}) \vdash_\Res \Reff_t(\varphi', \pi')\, ,
    \end{equation*}
    where, once $\rho$ is fixed, the extension variables defining the sets of clauses $(\pi' = \pi_{\restriction\rho}) \land (\varphi' =\varphi_{\restriction\rho})$ amount to constant-depth formulas of depth at most $2$.
\end{proposition}
The reader may find a proof in Section 6 of the full version of the paper.

We also use the following straightforward lemma; proof can be found in Section 7 of the full version.
\begin{lemma}
    \label{lemma:size-depth-increase}
    Let \(d \in \mathbb{N}\), and let \(\pi\) be a \(\PK_{d+1}\)-refutation of \(\varphi\) where every connective in \(\pi\) at depth \(d+1\) has arity at most \(r\). Let \(\psi\) be a formula of depth \(k\), whose top connective has arity \(\ell\), and whose bottom connectives have arity at most \(b\). It holds that the substitution of \(\psi\) into \(\pi\), denoted \(\pi(\psi)\), is a \(\PK_{d+k}\)-refutation of \(\varphi(\psi)\) where the connectives at depth \(d+k\) have arity \(b\) and $|\pi(\psi)| = O\bigl(|\pi| \cdot |\psi| \cdot \ell^r\bigr)$.
\end{lemma}
\fi

\subsection{Haken's lower bound in the polynomial-size regime}
\label{subsec:LLR-PHP}
Building on the formalizations of \citeauthor{LLR24} \autocite{LLR24} in $\LLR$, we show how for the particular case of the pigeonhole principle formulas, exponential lower bounds can be turned into fixed polynomial lower bounds with a mild increase in the depth of the proof.

Let us recall the definition of the CNF formulas encoding the \emph{Pigeonhole Principle} ($\PHP$). For every $m \in \bbN$ and $n\in \bbN$ such that $m > n$, the formula $\PHP^m_n$ stands for the CNF formula over variables $p_{i,j}$ for $i \in [m]$ and $j \in [n]$ consisting of the clauses
\ifcameraready
{\allowdisplaybreaks
\begin{axiomdef}{2}{PHP}
        &\!\!\!\! \bigvee_{j\in[n]} p_{i,j} &&\!\!\!\quad\text{ for }i\in[m], \label{php-1} \\
        &\!\!\!\!\neg p_{i,j} \lor \neg p_{i',j} &&\!\!\!\quad\text{ for all }i, i' \in [m], i \neq i' \text{ and } j\in [n] \label{php-2}
    \end{axiomdef}}
\else
\begin{axiomdef}{2}{PHP}
        & \bigvee_{j\in[n]} p_{i,j} &&\quad\text{ for all }i\in[m], \label{php-1} \\
        &\neg p_{i,j} \lor \neg p_{i',j} &&\quad\text{ for all }i, i' \in [m], i \neq i' \text{ and } j\in [n] \label{php-2}
\end{axiomdef}
\fi
We denote the formula $\PHP^{n+1}_n$ by $\PHP_n$ .

Haken \autocite{Haken85} proved that every Resolution refutation of $\PHP_n$ requires size $2^{\Omega(n)}$. The result can be formalized in $\LLR$, assuming $n\in\Log$.

\begin{theorem}[Haken's lower bound in the exponential-size regime {\autocite[Theorem 3.17]{LLR24}}]
    \label{thm:LLR-PHP-formal}
    There exists a positive $\varepsilon \in \bbQ$ such that $\BussTonetwo(\Pi) + \dWPHP(\PV_1(\Pi))$ proves the following statement: for every $t$ and every $n \in \Log$, if $\Pi$ encodes a length-$t$ Resolution refutation of $\PHP_n$, then $t > 2^{\varepsilon \cdot n}$.
\end{theorem}

By applying the Paris--Wilkie translation (\Cref{thm:Paris-Wilkie-translation}) to $\BussTtwotwo(\Pi) \supseteq \BussTonetwo(\Pi) + \dWPHP(\PV_1(\Pi))$, we immediately get the following upper bound in $\Res(\log)$.

\begin{corollary}
    \label{cor:PHP-in-Reslog}
    There exists a positive $\varepsilon \in \bbQ$ such that the sequence of formulas $\{ \Reff_{2^{\varepsilon n}}(\PHP_n) \}_{n\in\bbN}$ admits $\QP$-uniform quasipolynomial-size refutations in $\Res(\log)$.
\end{corollary}

The formulas $\Reff_{2^{\varepsilon n}}(\PHP_n)$ have size exponential in $n$, so it is not obvious how one could derive directly a formula of the form $\Reff_{n^k}(\PHP_n)$. The following theorem shows this is possible for the case of $\PHP_n$ if we go slightly above $\Res(\log)$.

\begin{theorem}
    \label{thm:RefPHP-poly-regime}
    For every $k \in \bbN$, the formulas $\{  \Reff_{n^k}(\PHP_n)\}_{n\in\bbN}$ admit $\QP$-uniform $\PK_{2}$-refutations in quasipolynomial size.
\end{theorem}

\begin{proof}
    Fix $k\in\bbN$ and let $\varepsilon\in \bbQ$ be the constant in \Cref{cor:PHP-in-Reslog}. Let $N \coloneq \lceil \log n^{k/\varepsilon} \rceil$ and consider a partial matching $\rho$ from $[n+1]$ to $[n]$, so that as a restriction over the variables of $\PHP_n$ it leaves $N+1$ pigeons and $N$ holes unassigned. That is, ${\PHP_{n}}{\restriction_\rho} \equiv \PHP_N$. The goal is to derive the same fact inside the proof system.

    Since $n^k \leq 2^{\varepsilon N}$, we have that by \Cref{prop:Res-closed-restrictions},
    \ifcameraready
    \begin{equation}
        \label{eq:restriction-PHP}
        \begin{multlined}[c][0.85\linewidth]
        \Reff_{n^k}(\PHP_n, \pi) \land (\pi' = \pi_{\restriction\rho}) \\ \vdash_\Res^{n^{O(1)}} \Reff_{2^{\varepsilon N}}(\PHP_N, \pi'),
        \end{multlined}
    \end{equation}
    \else
    \begin{equation}
        \label{eq:restriction-PHP}
        \Reff_{n^k}(\PHP_n, \pi) \land (\pi' = \pi_{\restriction\rho}) \vdash_\Res^{n^{O(1)}} \Reff_{2^{\varepsilon N}}(\PHP_N, \pi'),
    \end{equation}
    \fi
    and by \Cref{cor:PHP-in-Reslog},
    \begin{equation}
        \Reff_{2^{\varepsilon N}}(\PHP_N, \pi') \vdash_{\Res(\log)}^{n^{{(\log n)}^{O(1)}}} \bot .
    \end{equation}
    Together, these imply
    \begin{equation}
        \Reff_{n^k}(\PHP_n, \pi) \land (\pi' = \pi_{\restriction\rho}) \vdash_{\Res(\log)}^{n^{{(\log n)}^{O(1)}}} \bot.
    \end{equation}
    In order to get rid of the $\pi'$ variables we need to substitute them in the proof.
    \ifcameraready
    Since these come from the depth-2 extensions in \Cref{prop:Res-closed-restrictions}
    \else
    Since these come from the $\formDeltab^b_2(\Phi, \Pi)$-definition in \Cref{prop:closure-restrictions}
    \fi
    and $\rho$ is fixed, we get that the conjunct $(\pi' = \pi_{\restriction\rho})$ amounts to a depth-$2$ formula of polynomial size. By \Cref{lemma:size-depth-increase}, when substituting this into a $\Res(\log)$ refutation, we get a $\PK_{2}$-refutation in quasipolynomial size.
\end{proof}


This restriction trick for $\PHP_n$ works because we can send the formula to an exponentially smaller subinstance of itself. This appears problematic when dealing with the $\Reff$ formulas, as we cannot control the innermost formula $\varphi$. When looking at $\Reff_s(\varphi)$, we can disable blocks to send the formula to an exponentially smaller subformula $\Reff_{\log s}(\varphi)$, but we have no way of restricting $\varphi$.

More formally, let us suppose we want to obtain fixed-polynomial lower bounds, of the form encoded by $\neg\Reff_{(\nicefrac{s}{n})^k}(\Reff_s(\varphi))$. Then, let $\sigma \coloneq n \log (\nicefrac{s}{n})^{k/\varepsilon}$, so that $(\nicefrac{s}{n})^k = 2^{\varepsilon \sigma /n}$. By a restriction argument analogous to \cref{eq:restriction-PHP}, we can start from $\Reff_{(\nicefrac{s}{n})^k}(\Reff_s(\varphi))$ and derive $\Reff_{2^{\varepsilon \sigma / n}}(\Reff_\sigma(\varphi))$. To complete the proof we need a refutation of this last formula.
\ifcameraready
\Cref{thm:AM-LLR-regime-informal}
\else
\Cref{thm:exp-regime-ref-lwb}
\fi
yields such a refutation in quasipolynomial-size $\Res(\log)$, but this is no simplification: the formalization in the exponential-size regime uses the exact same proof as
\ifcameraready
\Cref{thm:AM-new-proof-informal} 
\else
\Cref{thm:poly-regime-ref-lwb-all-CNF} 
\fi
without really exploiting the fact that the formula parameters are in $\Log$ (that is, we still use limited independence and the more sophisticated restriction argument in \Cref{sec:restriction-formalized-APC}, so we can derive fixed-polynomial lower bounds directly from this argument without the restriction trick).

For the trick in \cref{eq:restriction-PHP} to constitute a simplification, we would need to obtain a simpler formalization of the restriction argument in \Cref{lemma:restriction} exploiting the fact that the parameters $n$ and $s$ are in $\Log$. The advantage in this regime is that we can carry out the counting arguments exactly. Since the disabling restrictions in \Cref{lemma:restriction} are strings of the form $\rho \in \{ 0,1\}^{2s}$, the space of restrictions is of size $2^{2s}$. For this to be quasipolynomial in the size of the formula $\Reff_{2^{\varepsilon s/n}}(\Reff_s(\varphi))$, we need $2^{2s} = \quasipoly(2^{s/n})$. Unfortunately, this latter requirement makes the restriction trick in \Cref{thm:RefPHP-poly-regime} impossible: we would restrict to a formula where $s$ becomes $\sigma = n \log (\nicefrac{s}{n})^{k/\varepsilon}$. If $s$ is, say, $n^{O(1)}$, then this gives $\sigma = \Theta(n \log n)$, and the requirement becomes $\Theta(n\log n) \leq \operatorname{poly}(\log n)$, which is impossible.

\subsection{The hardness reduction}
\label{subsec:hardness-PAP}
The $\NP$-hardness proof follows the ideas of \citeauthor{PAP}~\autocite{PAP} and the streamlined presentation sketched in \Cref{subsubsec:intro-techniques-PAP-NP-hard}. Formally, this is captured by the following methodology theorem. The statement below is a slight modification of the original one in \autocite{PAP}, tailored to constant-depth subsystems of $\PK$. For the sake of completeness, we give a formal proof here as well.

\begin{theorem}[General $\NP$-hardness of $\PAP$ {\autocite[Corollary 5.7]{PAP}}]
    \label{thm:general-NP-hardness}
    Let $\Res \leq_{\operatorname{p}}S_1 \leq_{\operatorname{p}}S_2$ be propositional proof systems such that substituting polynomial-size depth-$2$ formulas in an $S_1$-proof results in an $S_2$-proof with at most a polynomial blow-up. Suppose there exists a constant $c > 0$, a time-constructible function $t : \bbN \to \bbN$ and a $\TIME[t(n)]$-uniform sequence of unsatisfiable formulas $\{ \psi_n\}_{n\in \bbN}$ such that $\psi_n$ has at most $n$ variables and $\poly(n)$ clauses and there exists a $\TIME[t(n)]$-uniform sequence of $S_1$-proofs $\{ \pi_n\}_{n\in\bbN}$ such that $\pi_n :S_1 \vdash \neg\Reff_{n^{c+1}}(\psi_n)$. Then, the problem $\PAP_{S_2}[n^c]$ is $\NP$-hard under many-one reductions computable in time $\poly(t(n))$.
\end{theorem}

\newcommand{\disj}{\operatorname{disj}}
\begin{proof}
    For every two CNF formulas $\varphi = C_1 \land \cdots \land C_k$ and $\psi = D_1 \land \cdots \land D_m$  over disjoint variables, let us denote by $\disj(\varphi, \psi)$ the formula
    \begin{equation}
            \disj(\varphi, \psi) \coloneq \bigwedge_{i=1}^k (\neg y \lor C_i) \land \bigwedge_{i=1}^m (y \lor D_i),
    \end{equation}
    where $y$ is a fresh variable not present in $\varphi$ or $\psi$.

    By \Cref{prop:Res-closed-restrictions} and the assumption that $S_1$ proves Resolution lower bounds on $\psi$ we have that for every CNF formula $\varphi$ over $n$ variables,
    \ifcameraready
    \begin{equation}
        \begin{split}
        \!\!\!  \Reff_{n^c}(\disj(\varphi, \psi_n), \pi) \land (\pi' = \pi_{\restriction\rho}) &\vdash_{\Res} \Reff_{n^c}(\psi_n, \pi')\! \\ &\vdash_{S_1} \bot,
        \end{split}
    \end{equation}
    \else
    \begin{equation}
        \Reff_{n^{c+1}}(\disj(\varphi, \psi_n), \pi) \land (\pi' = \pi_{\restriction\rho}) \vdash_{\Res} \Reff_{n^{c+1}}(\psi_n, \pi') \vdash_{S_1} \bot,
    \end{equation}
    \fi
    where $\rho$ is fixed to be the restriction that sets $\rho : y \mapsto 0$ and sets all variables of $\varphi$ to $1$. Note that the first derivation has size $\poly(n)$ and the second derivation has size $O(t(n))$.

    \ifcameraready
    Since the conjunct $(\pi' = \pi_{\restriction\rho})$ comes from the depth-2 extension variables in \Cref{prop:Res-closed-restrictions} and $\rho$ is fixed, we get that these can be substituted to get an $S_2$-refutation of $\Reff_{n^c}(\disj(\varphi, \psi_n), \pi)$ in size $\poly(T(n))$.
    \else
    Since the conjunct $(\pi' = \pi_{\restriction\rho})$  comes from the $\formDeltab^b_2(\Phi, \Pi)$-definition in \Cref{prop:closure-restrictions} and $\rho$ is fixed, we get that this amounts to a depth-$2$ formula of polynomial size, which can be substituted to get an $S_2$-refutation of $\Reff_{n^{c+1}}(\disj(\varphi, \psi_n), \pi)$ in size $\poly(t(n))$.
    \fi

    The $\NP$-hardness reduction is now from $\SAT$ and maps a CNF formula over $n$ variables to the instance $(\disj(\varphi, \psi_n), \tau_n, 1^{{N}^c})$ of $\PAP_{S_2}[N^c]$, where $\tau_n : S_2\vdash\neg \Reff_{N^c}(\disj(\varphi, \psi_n), \pi)$, $N$ is the number of variables of $\disj(\varphi, \psi_n)$. Note that for large enough $n$, we have $N^c = (2n + 1)^c \leq n^{c+1}$, so the $n^{c+1}$ Resolution lower bound at hand is good enough. It is easy to see that $\varphi$ is satisfiable if and only if $\disj(\varphi, \psi_n)$ is, and $\tau_n$ is always a correct $S_2$-proof. This completes the proof.
\end{proof}

 This yields the desired $\NP$-hardness reduction.

\begin{theorem}
    \label{thm:NP-hardness-PAP-PK}
   Let $Q$ be a propositional proof system that p-simulates $\PK_{2}$. For every $c > 0$, the problem $\PAP_Q[n^c]$ is $\NP$-hard under quasipolynomial-time many-one reductions.
\end{theorem}

\begin{proof}
    Consider $m \coloneq \lceil \nicefrac{c+1}{\varepsilon} \log n\rceil$ and the formulas $\PHP_m$. By \Cref{cor:PHP-in-Reslog}, we have $\Res(\log)$ refutations of $\Reff_{2^{\varepsilon m}}(\PHP_m)$, which is precisely $\Reff_{n^{c+1}}(\PHP_m)$.

    Since \Cref{lemma:size-depth-increase} guarantees that substituting depth-2 formulas in $\PK_{0+\nicefrac{1}{2}}$-proofs results in $\PK_{2}$-proofs with a quasipolynomial increase in size, we can take $S_1 \coloneq \PK_{0+\nicefrac{1}{2}}$ and $S_2\coloneq \PK_{2}$ and invoke  \Cref{thm:general-NP-hardness} to conclude that $\PAP_{\PK_2}[n^c]$ is $\NP$-hard under quasipolynomial-time many-one reductions. Since $Q \psim \PK_2$, we immediately extend the $\NP$-hardness reduction from $\PAP_{\PK_2}[n^c]$ to $\PAP_Q[n^c]$. \qedhere
    
\end{proof}

\section*{Acknowledgements}
\label{sec:acknowledgements}
\addcontentsline{toc}{section}{\nameref{sec:acknowledgements}}

We are grateful to Albert Atserias for useful initial discussions, and we thank Jonas Conneryd for helpful comments and pointers. We thank Sam Buss and Jan Krajíček for kindly answering technical questions about details of the Paris--Wilkie translation. We also thank the anonymous FOCS reviewers for helpful comments that improved the clarity of the presentation.

Noel Arteche was supported by the Wallenberg AI, Autonomous Systems and Software Program (WASP) funded by the Knut and Alice Wallenberg Foundation. Susanna F. de Rezende received funding from the Knut and Alice Wallenberg grant \mbox{KAW 2023.0116}, ELLIIT, and the Swedish Research Council grant \mbox{2021-05104}. Erfan Khaniki was supported by the Royal Society University Research Fellowship URF$\backslash$R1$\backslash$211106 \say{Proof complexity and circuit complexity: a unified approach}. Kilian Risse was supported by the Swiss National Science Foundation Postdoc.Mobility fellowship \mbox{P500-2\_235298}. We also gratefully acknowledge that we have benefited greatly from being part of the Basic Algorithms Research Copenhagen (BARC) center financed by the Villum Investigator grant~54451. The authors would also like to thank the Isaac Newton Institute for Mathematical Sciences, Cambridge, for support and hospitality during the programme \emph{Logical Foundations of Complexity Theory}, where part of the final work on this paper was undertaken; this work was supported by EPSRC grant EP/Z000580/1.

\section*{AI Disclosure}
The authors used GPT 5.6-Sol only to assist with typesetting and generating initial versions of \Cref{fig:system-metro-line} and \Cref{fig:prove-delayer-example}, as well as polishing the presentation of some of the proofs and finding small errors. 


\printbibliography[
        heading=bibintoc
]


\end{document}